
%
\input harvmac.tex
\input epsf
%
%
\lref\Ahl{L. V.~ Ahlfors and L.~ Sario, {\it Riemann Surfaces},
Princeton University Press, 1960.}
\lref\Al{ O.~ Alvarez, ``Theory of Strings with Boundaries:
 Fluctuations, Topology and
Quantum Gravity", Nucl. Phys. {\bf B216} (1983 ) 125.}
\lref\Ar{See V.~ Arnold, ``Remarks on Enumeration of Plane Purves'';
``Plane Curves, their Invariants, Perestroikas and Classifications".}
\lref\BaTa{J.~ Baez and W.~ Taylor, "Strings and Two-dimensional
QCD for Finite N", MIT-CTP-2266, hep-th/9401041.}
\lref\demeterfi{J.L.F.~ Barbon and K.~ Demetrfi,
 ``Effective Hamiltonians for
$1/N$ Expansion in Two-Dimensional QCD",
 hep-th/9406046.}
\lref\Ba{I.~ Bars, ``QCD and Strings in 2d",
Talk given at International
Conference on Strings, Berkeley, CA, 24-29
May 1993, hep-th/9312018.}
\lref\BaHa{I.~  Bars and A.~  Hanson, ``Quarks
at the Ends of the String",
Phys. Rev. {\bf D13} (1976)1744;
``A Quantum String Theory of Hadrons and its Relation to
Quantum Chromodynamics
in Two Dimensions", Nucl. Phys. {\bf B111} (1976) 413.}
\lref\BeEd{I.~ Berstein and A.L.~ Edmonds, ``On the Classification of
Generic Branched Coverings of Surfaces", Illinois Journal
of Mathematics, Vol {\bf 28}, number {\bf 1}, (1984 ).}
\lref\Bir{J. S.~ Birman, {\it Braids, Links and Mapping Class groups},
pages 11, 12, Princeton University Press, 1975.}
\lref\BogSh{
N. N.~ Bogoliubov D.V.~ Shirkov, {\it Introduction to the
 Theory of Quantized Fields},
Interscience Publishers, Inc., New York}
\lref\Bo{R.~ Bott, ``On the Chern-Weil Homomorphism and
the Continuous Cohomology of Lie Groups,'' Adv. in Math.
{\bf 11} (1973) 289-303.}
\lref\BoTu{R.~ Bott and L.~ Tu, {\it Differential Forms in
Algebraic Topology}, Springer Verlag, New York, 1982.}
\lref\Boul{ D.V. Boulatov : `` q-deformed lattice gauge theory
and $3$ manifold invariants, '' Hepth 9210032, Int. J. Mod.Phys. A8: 3139-3162,
1993. }
\lref\Br{N.~ Bralic, "Exact Computation of Loop Averages
 in Two-Dimensional
Yang-Mills Theory", Phys. Rev. {\bf D22} (1980) 3090.}
\lref\Bthom{ M. Blau and G. Thompson, ``Derivation of the Verlinde
 formula from Chern-Simons Theory and the G/G model,'' Nucl. Phys. B408  (1993)
 345. }
\lref\CreTay{ M. Crescimanno and W. Taylor,
`` Large N phases of Chiral  $QCD_2$, '' hep-th
9408115. }
\lref\Creu{ M. Creutz, `` On invariant
 integration over $SU(N)$, '' J. Math. Phys. {\bf 19} (10), October 1978.  }
\lref\CMROLD{S.~ Cordes, G.~ Moore, and S.~ Ramgoolam,
``Large N 2D Yang-Mills Theory and Topological String Theory",
hep-th/9402107. }
\lref\CMRlsch{ S.~ Cordes, G.~ Moore, and S.~ Ramgoolam,
in Les Houches Session LXII on http://xxx.lanl.gov/lh94.}
\lref\wadia{A.~ Dhar, P.~ Lakdawala, G.~ Mandal and S.~ Wadia,
``String Field Theory of Two-Dimensional QCD on a Cylinder:
 A Realization of
$W ( \infty )$ Current Algebra", Phys. Lett. {\bf B350} (1994);
 hep-th/9403050.}
\lref\DiRu{R.~ Dijkgraaf and R.~ Rudd, unpublished.}
\lref\dgcrg{M.R.~ Douglas, ``Conformal Field
Theory Techniques for Large $N$
Group Theory,'' hep-th/9303159;
M.R.~ Douglas, ``Conformal Field Theory
Techniques in Large $N$ Yang-Mills Theory",
hep-th/9311130,to be published in the
 proceedings of the May 1993 Carg\`ese
workshop on Strings,  Conformal Models
 and Topological Field Theories.}
\lref\Dosc{M. R.~ Douglas, ``Some Comments
on QCD String,''  RU-94-09,
hep-th/9401073, to appear in Proceedings of the Strings '93
 Berkeley conference.}
\lref\Docft{M.R.~ Douglas, ``Conformal Field Theory Techniques
 in
Large N Yang-Mills Theory,'' hep-th 9303159. }
\lref\DoKa{M.~ Douglas and V.A.~ Kazakov, ``Large $N$ Phase
Transition in Continuum
QCD$_2$", Phys. Lett. {\bf B319} (1993) 219, hep-th/9305047 }
\lref\Doug{ M. R. Douglas,  `` Large N gauge theory- Expansions
 and transitions, ''
RU-94-72, Hepth 9409098.  }
\lref\DrZu{ J.M.~ Drouffe and J. B.~ Zuber, ``Strong Coupling and
 Mean Field Methods
in Lattice Gauge theories", Phys. Rept {\bf 102} (1983) 1. }
\lref\Du{J. J.~ Duistermaat and G. J.~ Heckman, ``On The Variation
In The Cohomology In The Symplectic Form Of The Reduced
Phase Space,''
Invent. Math. {\bf 69} (1982) 259.}
\lref\Ed{A. L.~ Edmonds, ``Deformation of Maps to
Branched Coverings in Dimension Two", Annals of Mathematics
{\bf 110} (1979), 113-125. }
\lref\egh{T.~ Eguchi, P.B.~ Gilkey, and A.J.~ Hanson,
``Gravitation, Gauge Theories, and  Differential Geometry",
 Phys. Rep. {\bf 66}(1980) 214.}
\lref\Ez{C. L.~ Ezell, ``Branch Point Structure of Covering
Maps onto
Nonorientable Surfaces", Transactions of the American
 Mathematical
society, vol {\bf 243}, 1978. }
\lref\Fi{D.~ Fine, ``Quantum Yang-Mills On The
 Two-Sphere", Commun. Math.
Phys. {\bf 134} (1990) 273;
``Quantum Yang-Mills On A Riemann Surface,''
Commun. Math. Phys. {\bf 140} (1991) 321.}
\lref\Fo{R.~ Forman, ``Small Volume Limits of $2-d$
Yang-Mills,'' preprint,
Rice University (1991).}
\lref\Freed{D.~ Freed and K.~ Uhlenbeck, {\it
Instantons and 4-Manifolds},
Springer 1984.}
\lref\FrieWind{D.~ Friedan and P.~ Windey,
 ``Supersymmetric Derivation
of the Atiyah-Singer Index and the Chiral
 Anomaly", Nucl. Phys. {\bf B235}
(1984) 395.}
\lref\Fu{W.~ Fulton, ``Hurwitz Schemes
and Irreducibility
of Moduli of Algebraic Curves", Annals of Math. {\bf 90}
(1969) 542.}
\lref\GlJa{See, e.g., J.~ Glimm and A.~ Jaffe,
{\it Quantum Physics},
Springer 1981.}
\lref\GaYaSo{ O. Ganor, J. Sonnenschein, S. Yankielowicz,
 " The string theory approach to 2D
Yang Mills Theory, " Jul 1994,  hep-th 9407114}
\lref\Grtalk{D.~ Gross, ``Some New/Old Approaches to QCD,''
Int. Workshop on String Theory, Quantum Gravity and the
Unification of Fundamental
Interactions, Rome, Italy, 21-26 Sept. 1992, hep-th/9212148.}
\lref\Gera{ A. Gerasimov, ``Localization in GWZW and Verlinde formula, ''
hep-th 9305090.  }
\lref\GrHa{P.~ Griffiths and J.~ Harris, {\it Principles
 of Algebraic geometry},
p. 445, J.Wiley and Sons, 1978. }
\lref\GrTa{D.~ Gross, ``Two Dimensional QCD as a
 String Theory",
Nucl. Phys. {\bf B400}  (1993) 161, hep-th/9212149.;
D.~ Gross and W.~ Taylor, ``Two-dimensional QCD is
 a String Theory",
Nucl. Phys. {\bf  B400}  (1993) 181, hep-th/93011068;
D.~ Gross and W.~ Taylor, ``Twists and Loops in the String
Theory of Two Dimensional QCD", Nucl. Phys. {\bf 403}
(1993) 395, hep-th/9303046.}
\lref\GrTatalk{D.J.~ Gross and W.~ Taylor, ``Two-Dimensional
QCD
and Strings", Talk presented at  Int. Conf. on Strings '93,
Berkeley CA,  24-29 May 1993;
hep-th/9311072.}
\lref\GrWi{D.~ Gross and E.~ Witten, ``A Possible Third
Order Phase Transition on the
Large $N$ Lattice Gauge Theory", Phys. Rev.
{\bf D21} (1980) 446.}
\lref\GrKiSe{L. ~ Gross, C.~ King and A.~
Sengupta, ``Two-Dimensional Yang-Mills via
Stochastic Differential Equations", Ann. of Phys. {\bf 194} (1989) 65.}
\lref\hamermesh{M.~ Hamermesh, {\it Group Theory and its
applications to physical problems}, Addison-Wesley 1962}
\lref\HaMo{ J.~ Harris and I.~ Morrison, ``Slopes
 of Effective Divisors on the
Moduli Space of Stable Curves", Invent. Math.
{\bf 99}  (1990) 321-355.}
\lref\HaMu{ J.~ Harris and D.~ Mumford, ``On
 the Kodaira Dimension of
the Moduli Space of Curves", Invent. Math.
 {\bf  67} (1982) 23-86.}
\lref\Hi{F.~ Hirzebruch, {\it Topological Methods
in Algebraic Geometry},
Springer Verlag, 1966.}
\lref\Hitchin{N.~ Hitchin, ``The Geometry and
Topology of
Moduli Spaces", in {\it Global Geometry and
 Mathematical Physics},
LNM 1451, L.~ Alvarez-Gaum\'e et. al.,  eds.}
\lref\Hora{P.~ Horava, ``Topological Strings and
QCD in Two Dimensions",
to appear in Proc. of The Cargese Workshop,1993;
hep-th/9311156.}
\lref\Hott{ P.~ Horava, ``Two-Dimensional String
 Theory and the
Topological Torus", Nucl. Phys. {\bf B386} ( 1992 ) 383-404.}
\lref\Hori{E.~ Horikawa, ``On Deformations of Holomorphic Maps,
I, II, III", J. Math. Soc. Japan, {\bf 25} (1973) 647; ibid {\bf 26}
(1974) 372; Math. Ann. {\bf 222} (1976) 275.}
\lref\husemoller{D.~ Husemoller, {\it Fiber Bundles},
3$^{{\rm rd}}$ edition, Springer, 1993.}
\lref\Ing{R.E.  Ingram , ``Some Characters of the
Symmetric Group",
Proc. Amer. Math. Soc. (1950), 358-369}
\lref\KaKo{V.A.~ Kazakov and I.~ Kostov,
 ``Non-linear Strings in
Two-Dimensional $U(\infty)$ Gauge Theory,''
 Nucl. Phys. {\bf B176}
(1980) 199-215;
``Computation of the Wilson Loop Functional in
Two-Dimensional $U ( \infty )$
Lattice Gauge Theory", Phys. Lett. {\bf B105}
(1981) 453;
V.A.~ Kazakov, ``Wilson Loop Average for an
Arbitrary Contour in
Two-Dimensional $U(N)$ Gauge Theory",
 Nuc. Phys. {\bf B179} (1981) 283-292.}
\lref\Kazkos{V.A. Kazakov, ``$U ( \infty )$ Lattice Gauge
Theory as a Free Lattice String
Theory", Phys. Lett. {\bf B128} (1983) 316;
V.I.~ Kostov, ``Multicolour QCD in Terms of Random
Surfaces",
Phys. Lett. {\bf B138} (1984) 191.}
\lref\Ki{F.~ Kirwan, {\it Cohomology of Quotients In
Symplectic
And Algebraic Geometry},  Princeton University Press.}
\lref\Kauff{ L.H. Kauffmann, ``Knots and Physics, '' World Scientific 1991. }
\lref\jeffkir{L.~ Jeffrey and F.~ Kirwan, ``Localization for
 Nonabelian
Group Actions",  alg-geom/9307001.}
\lref\Kn{F.~ Knudsen, ``The Projectivity of the Moduli
Space of Stable
Curves",  Math. Scand. {\bf 52} (1983) 161.}
\lref\kobnom{S.~ Kobayashi and K.~ Nomizu, {\it
Foundations of
Differential Geometry I,II}, Interscience Publishers,
New York, 1963.}
\lref\Kon{M.~ Kontsevich, ``Intersection Theory On
The Moduli Space
Of Curves And The Matrix Airy Function", Commun.
Math. Phys. {\bf 147} (1992) 1.}
\lref\Kos{I.K.~ Kostov, ``Continuum QCD2 in Terms of
 Discretized Random
Surfaces with Local Weights", Saclay-SPht-93-050;
 hep-th/9306110.}
\lref \Ma{ W.S.~ Massey, {\it A Basic Course in Algebraic Topology},
Springer-Verlag, 1991.}
\lref\menof{ P. Menotti and E. Onofri, ``The action of $SU(N)$ lattice gauge
theory
in terms of the eat kernal on the group manifold, ''  NPB190: 288. 1981.}
\lref\Mig{A.~ Migdal, ``Recursion Relations in Gauge Theories",
Zh. Eksp. Teor. Fiz. {\bf 69} (1975) 810 (Sov. Phys. Jetp. {\bf 42} 413).}
\lref\MitH{A. A.~ Migdal, `` Loop Equations and 1/N Expansion",
Phys. Rep. {\bf 102} (1983)199-290;
G.~ `t Hooft, ``A Planar Diagram Theory for Strong
Interactions, Nucl. Phys. {\bf B72} (1974) 461.}
\lref\Mij{M.~ Mijayima,
``On the Existence of Kuranishi Family for Deformations of Holomorphic
Maps", Science Rep. Kagoshima Univ., {\bf 27} (1978) 43.}
\lref\miller{E.~ Miller, ``The Homology of the Mapping
Class Group", J. Diff. Geom. {\bf 24} (1986)1}
\lref\Min{J. Minahan, ``Summing over Inequivalent Maps in the String
Theory of QCD", Phys. Rev. {\bf D47} (1993) 3430. }
\lref\Minclas{ J.~ Minahan and A.~ Polychronakos,
 ``Classical Solutions for two
dimensional QCD on the sphere", Nucl. Phys.
 {\bf B422} (1994) 172; hep-th/9309119 }
\lref\MiPoetd{J.~ Minahan and A.~ Polychronakos,
``Equivalence of Two
Dimensional QCD and the $c=1$ Matrix Model",
Phys. Lett. {\bf B312} (1993) 155; hep-th/9303153.}
\lref\MiPoifs{J.~ Minahan and A.~ Polychronakos,
"Interacting Fermion
Systems from Two Dimensional QCD", Phys. Lett.
{\bf 326} (1994) 288;
hep-th/9309044.}
\lref\mrsb{G.~ Moore and N.~ Seiberg,
 ``Polynomial Equations for Rational Conformal Field
Theories,'' Phys. Lett.  {\bf 212B}(1988)451;
``Classical and Quantum Conformal Field Theory",
Commun. Math. Phys. {\bf 123}(1989)177;
``Lectures on Rational Conformal Field Theory",
 in {\it Strings 90}, the proceedings
of the 1990 Trieste Spring School on Superstrings.}
\lref\MoSeSt{G.~ Moore, N.~ Seiberg, and M.~ Staudacher,
``From Loops to States in $2D$ Quantum Gravity",  Nucl. Phys.
{\bf B362}(1991)665}
\lref\Mu{ D.~ Mumford, ``Towards an Enumerative Geometry
of the Moduli Space of Curves", in {\it Arithmetic and geometry}, M.~ Artin and
J.~ Tate eds., Birkhauser, 1983. }
\lref\nrswils{ S.G.~ Naculich, H.A.~ Riggs, H.J.~ Schnitzer,
`The String Calculation of Wilson Loops in Two-Dimensional
Yang-Mills Theory, '  hep-th/9406100.}
\lref\NaRiSc{S.G.~ Naculich, H. A.~ Riggs and H.G.~ Schnitzer,
 ``2D Yang Mills
Theories are String Theories,'' Mod. Phys. Lett. {\bf A8} (1993)
2223; hep-th/9305097.}
\lref\NR{ S.G.Naculich, H.A.Riggs, ``The String Calculation of QCD Wilson loops
on arbitrary surfaces,'' Hep-th 9411143.  }
\lref\ObZub{K.H.~ O'Brien and J.-B.~ Zuber, ``Strong Coupling
Expansion of Large
$N$ QCD and Surfaces", Nucl. Phys. {\bf B253} (1985) 621.}
\lref\Okubo{ S.~ Okubo,  J. Math. Phys. {\bf 18} (1977)  2382 }
\lref\Pe{ R. C.~ Penner, ``Perturbative Series and the Moduli Space
of Riemann Surfaces, J. Diff. Geo. {\bf 27} (1988) 35-53. }
\lref\PP{A. M.~ Perelomov and V. S.~ Popov, Sov. J. of
Nuc. Phys. {\bf 3} (1966) 676}
\lref\Pofs{A.M~ Polyakov, ``Fine Structure of Strings",
Nucl. Phys. {\bf B268} (1986) 406.}
\lref\Pofp{A.M.~ Polyakov, ``A Few Projects in String Theory", PUPT-1394;
hep-th/9304146.}
\lref\PrSe{A.~ Pressley and G.~ Segal, {\it Loop Groups},
Oxford Math. Monographs,
Clarendon, NY, 1986.}
\lref\zuckerman{The following construction goes back to
D.~ Quillen, ``Rational Homotopy Theory,'' Ann. of  Math.
 {\bf 90} (1969) 205.
The application to the present case was developed with G.~ Zuckerman.}
\lref\rajeev{S.~ Rajeev, ``Quantum Hadrodynamics in Two-Dimensions";
hep-th/9401115.}
\lref\Ra{S.~ Ramgoolam, ``Comment on Two-Dimensional $O(N)$ and $Sp(N)$
Yang Mills Theories as String Theories", Nuc. Phys. {\bf B418}, (1994) 30;
hep-th/9307085. }
\lref\Hcas{S.~ Ramgoolam,  Higher Casimir
 Perturbations of Two Dimensional
Yang Mills Theories}
\lref\RT { N. Reshetikhin and V.G. Turaev, ``Invariants of three-manifolds via
link polynomials
and quantum groups,'' Invent. math. 103 (1991).     }
\lref\Resh{ N. Reshetikhin, `` Quasitriangular Hopf algebras and invariants of
links,'' Algebra
and Analysis, v.1, N2, (1990).   }
\lref\Rudd{R.~ Rudd, ``The String Partition Function
for QCD on the Torus",
hep-th/9407176.}
\lref\Ru{B.~ Rusakov, ``Loop Averages And Partition Functions in $U(N)$
Gauge Theory On Two-Dimensional Manifolds", Mod. Phys. Lett. {\bf A5}
(1990) 693.}
\lref\Schw{A.~ Schwarz, ``The Partition Function Of Degenerate Quadratic
Functional And Ray-Singer Invariants", Lett. Math. Phys. {\bf 2} (1978) 247.}
\lref\She{S.~ Shenker, "The Strength of Non-Perturbative Effects in String
Theory", in Carg/`ese Workshop on {\it Random Surfaces, Quantum Gravity
and Strings}, Garg`ese, France, May 28 - Jun 1 1990, eds. O.~ Alvarez,
E.~ Marinari, and P.~ Windey (Plenum, 1991)  }
\lref\Se{A.~ Sengupta, ``The Yang-Mills Measure For $S^2$,'' to appear in
J. Funct. Anal., ``Quantum Gauge Theory On Compact Surfaces,'' preprint.}
\lref\singgrib{I.~ Singer, ``Some Remarks on the Gribov Ambiguity",
Commun. Math Phys. {\bf 60} (1978) 7.}
\lref\seif{ H. Seifert and W. Threlfall, ``A textbook of Topology, ''
translated by M.A. Goldman, Academic press 1980. }
\lref\SR{ S. Ramgoolam, ``Coupled
non-intersecting Loops in 2D Yang Mills and free fields, '' YCTP-P18-94.}
\lref\Sp{E. H.~ Spanier, {\it Algebraic Topology},
 Mc. Graw-Hill, Inc. 1966.}
\lref\steenrod{N.~ Steenrod, {\it The topology of
 Fiber Bundles}, Princeton Univ. Press}
\lref\St{A.~ Strominger,  ``Loop Space Solution of Two-Dimensional
QCD", Phys. Lett. {\bf 101B} (1981) 271.}
\lref\PhaTay{ W.~ Taylor, ``Counting strings and
Phase transitions in 2d QCD",
MIT-CTP-2297; hep-th/9404175}
\lref\Ve{G.~ Veneziano, ``Construction of a Crossing-Symmetric,
 Regge-Behaved
Amplitude for Linearly Rising Trajectories,''
Nuovo Cim. {\bf 57A} 190, (1968).}
\lref\DiVeVe{E.~ Verlinde and H.~ Verlinde ,
``A Solution of Two-Dimensional Topological
Quantum Gravity", Nucl. Phys. {\bf B348} (1991) 457,
R.~ Dijkgraaf, E.~ Verlinde and H.~ Verlinde, ``Loop
Equations and Virasoro Constraints
in Nonperturbative 2d Quantum Gravity",
Nucl. Phys. {\bf B348} (1991) 435;
``Topological Strings in $d < 1$", Nucl. Phys.
{\bf B352} (1991) 59;
``Notes on Topological String Theory and 2-d
Quantum Gravity", in {\it String Theory
and Quantum Gravity}, Proc. Trieste Spring School,
April 24 - May 2,  1990 (World Scientific,
Singapore, 1991).}
\lref\verlinde{E.~ Verlinde, ``Fusion Rules
and Modular Transformations
in 2d Conformal Field Theory",  Nucl. Phys. {\bf B300} (1988) 360.}
\lref\Vi{J.~ Vick, {\it Homology Theory}, Academic Press, 1973.}
\lref\Wi{K.G.~ Wilson, ``Confinement of Quarks",
Phys. Rev. {\bf D10} (1974) 2445.}
\lref\Witdgt{ E.~ Witten, ``On Quantum gauge theories in two dimensions,''
Commun. Math. Phys. {\bf  141}  (1991) 153.}
\lref\Witdgtr{E.~ Witten, ``Two Dimensional Gauge Theories Revisited",
J. Geom. Phys. {\bf G9} (1992) 303; hep-th/9204083.}
\lref\Witcs{ E. Witten, `` Chern Simons gauge  theory as a string theory,''
Hep-th 9207094 }
\lref\Zelo{D. P.~ Zhelobenko, Translations of American Math.
Monographs, {\bf 40}.  }


\def\boxit#1{\vbox{\hrule\hbox{\vrule\kern8pt
\vbox{\hbox{\kern8pt}\hbox{\vbox{#1}}\hbox{\kern8pt}}
\kern8pt\vrule}\hrule}}
\def\mathboxit#1{\vbox{\hrule\hbox{\vrule\kern8pt\vbox{\kern8pt
\hbox{$\displaystyle #1$}\kern8pt}\kern8pt\vrule}\hrule}}


\def\inbar{\,\vrule height1.5ex width.4pt depth0pt}

\def\CF {{\cal F}}

\def\cgp {c_\Gamma^l}
\def\cgm {c_\Gamma^r }

\def\CY{{\cal Y}}

\def\G {\Gamma}

\def\IB{\relax{\rm I\kern-.18em B}}
\def\IC{\relax\hbox{$\inbar\kern-.3em{\rm C}$}}
\def\ID{\relax{\rm I\kern-.18em D}}
\def\IE{\relax{\rm I\kern-.18em E}}
\def\IF{\relax{\rm I\kern-.18em F}}
\def\IG{\relax\hbox{$\inbar\kern-.3em{\rm G}$}}
\def\IGa{\relax\hbox{${\rm I}\kern-.18em\Gamma$}}
\def\IH{\relax{\rm I\kern-.18em H}}
\def\II{\relax{\rm I\kern-.18em I}}
\def\IK{\relax{\rm I\kern-.18em K}}
\def\IL{\relax{\rm I\kern-.18em L}}
\def\IM{\relax{\rm I\kern-.18em M}}
\def\IN{\relax{\rm I\kern-.18em N}}
\def\IO{\relax\hbox{$\inbar\kern-.3em{\rm O}$}}
\def\Iom{{\inbar\kern-3.00pt\Omega}}
\def\IOm{\relax\hbox{$\inbar\kern-3.00pt\Omega$}}
\def\IP{\relax{\rm I\kern-.18em P}}
\def\IPi{\relax\hbox{${\rm I}\kern-.18em\Pi$}}

\def\IQ{\relax\hbox{$\inbar\kern-.3em{\rm Q}$}}
\def\IR{\relax{\rm I\kern-.18em R}}
\def\ITh{\relax\hbox{$\inbar\kern-.3em\Theta$}}

\font\cmss=cmss10 \font\cmsss=cmss10 at 7pt
\def\IZ{\relax\ifmmode\mathchoice
{\hbox{\cmss Z\kern-.4em Z}}{\hbox{\cmss Z\kern-.4em Z}}
{\lower.9pt\hbox{\cmsss Z\kern-.4em Z}}
{\lower1.2pt\hbox{\cmsss Z\kern-.4em Z}}\else{\cmss Z\kern-.4em
Z}\fi}
\def\kG {\vec{k}_\Gamma}

\def\p {\partial}

\def\Sc {\Sigma_T^c}

\def\SG{{\Sigma_T}}
\def\Sh{{\Sigma_W}}
\def\Sw{{\Sh}}
\def\ST{{\SG}}

\def\ymt{$YM_2$}

%
%

\def\dim{\mathop{\rm dim}}



%
\def\exercise#1{\bgroup\narrower\footnotefont
\baselineskip\footskip\bigbreak
\hrule\medskip\nobreak\noindent {\bf Exercise}. {\it #1\/}\par\nobreak}
\def\endexercise{\medskip\nobreak\hrule\bigbreak\egroup}
%
%
%
\message{S-Tables Macro v1.0, ACS,
 TAMU (RANHELP@VENUS.TAMU.EDU)}
%
%
\newhelp\stablestylehelp{You must choose a style between 0 and 3.}%
\newhelp\stablelinehelp{You should not use special hrules when
stretching
a table.}%
\newhelp\stablesmultiplehelp{You have tried to place an S-Table
inside another
S-Table.  I would recommend not going on.}%
%
%
\newdimen\stablesthinline
\stablesthinline=0.4pt
\newdimen\stablesthickline
\stablesthickline=1pt
%
%
\newif\ifstablesborderthin
\stablesborderthinfalse
\newif\ifstablesinternalthin
\stablesinternalthintrue
\newif\ifstablesomit
\newif\ifstablemode
\newif\ifstablesright
\stablesrightfalse
%
%
\newdimen\stablesbaselineskip
\newdimen\stableslineskip
\newdimen\stableslineskiplimit
%
%
\newcount\stablesmode
\newcount\stableslines
\newcount\stablestemp
\stablestemp=3
\newcount\stablescount
\stablescount=0
\newcount\stableslinet
\stableslinet=0
%
%
%
\newcount\stablestyle
\stablestyle=0
%
%
\def\stablesleft{\quad\hfil}%
\def\stablesright{\hfil\quad}%
%
%
\catcode`\|=\active%
%
%
\newcount\stablestrutsize
\newbox\stablestrutbox
\setbox\stablestrutbox=\hbox{\vrule height10pt depth5pt width0pt}
\def\stablestrut{\relax\ifmmode%
                         \copy\stablestrutbox%
                       \else%
                         \unhcopy\stablestrutbox%
                       \fi}%
%
%
\newdimen\stablesborderwidth
\newdimen\stablesinternalwidth
\newdimen\stablesdummy
\newcount\stablesdummyc
\newif\ifstablesin
\stablesinfalse
%
%
%
%
%
\def\stablesadj{%
  \ifcase\stablestyle%
    \hbox to \hsize\bgroup\hss\vbox\bgroup%
  \or%
    \hbox to \hsize\bgroup\vbox\bgroup%
  \or%
    \hbox to \hsize\bgroup\hss\vbox\bgroup%
  \or%
    \hbox\bgroup\vbox\bgroup%
  \else%
    \errhelp=\stablestylehelp%
    \errmessage{Invalid style selected, using default}%
    \hbox to \hsize\bgroup\hss\vbox\bgroup%
  \fi}%
\def\stablesend{\egroup%
  \ifcase\stablestyle%
    \hss\egroup%
  \or%
    \hss\egroup%
  \or%
    \egroup%
  \or%
    \egroup%
  \else%
    \hss\egroup%
  \fi}%
\def\stablestart{%
  \ifstablesin%
    \errhelp=\stablesmultiplehelp%
    \errmessage{An S-Table cannot be placed within an S-Table!}%
  \fi
  \global\stablesintrue%
  \global\advance\stablescount by 1%
  \message{<S-Tables Generating Table \number\stablescount}%
  \begingroup%
  \stablestrutsize=\ht\stablestrutbox%
  \advance\stablestrutsize by \dp\stablestrutbox%
  \ifstablesborderthin%
    \stablesborderwidth=\stablesthinline%
  \else%
    \stablesborderwidth=\stablesthickline%
  \fi%
  \ifstablesinternalthin%
    \stablesinternalwidth=\stablesthinline%
  \else%
    \stablesinternalwidth=\stablesthickline%
  \fi%
  \tabskip=0pt%
  \stablesbaselineskip=\baselineskip%
  \stableslineskip=\lineskip%
  \stableslineskiplimit=\lineskiplimit%
  \offinterlineskip%
  \def\borderrule{\vrule width \stablesborderwidth}%
  \def\internalrule{\vrule width \stablesinternalwidth}%
  \def\thinline{\noalign{\hrule height \stablesthinline}}%
  \def\thickline{\noalign{\hrule height \stablesthickline}}%
  \def\trule{\omit\leaders\hrule height \stablesthinline\hfill}%
  \def\ttrule{\omit\leaders\hrule height \stablesthickline\hfill}%
  \def\tttrule##1{\omit\leaders\hrule height ##1\hfill}%
  \def\stablesel{&\omit\global\stablesmode=0%
    \global\advance\stableslines by 1\borderrule\hfil\cr}%
  \def\el{\stablesel&}%
  \def\elt{\stablesel\thinline&}%
  \def\eltt{\stablesel\thickline&}%
  \def\elttt##1{\stablesel\noalign{\hrule height ##1}&}%
  \def\elspec{&\omit\hfil\borderrule\cr\omit\borderrule&%
              \ifstablemode%
              \else%
                \errhelp=\stablelinehelp%
                \errmessage{Special ruling will not display
properly}%
              \fi}%
  \def\stmultispan##1{\mscount=##1 \loop\ifnum\mscount>3
\stspan\repeat}%
  \def\stspan{\span\omit \advance\mscount by -1}%
  \def\multicolumn##1{\omit\multiply\stablestemp by ##1%
     \stmultispan{\stablestemp}%
     \advance\stablesmode by ##1%
     \advance\stablesmode by -1%
     \stablestemp=3}%

\def\multirow##1{\stablesdummyc=##1\parindent=0pt\setbox0\hbox\bgroup%

    \aftergroup\emultirow\let\temp=}
  \def\emultirow{\setbox1\vbox to\stablesdummyc\stablestrutsize%
    {\hsize\wd0\vfil\box0\vfil}%
    \ht1=\ht\stablestrutbox%
    \dp1=\dp\stablestrutbox%
    \box1}%
  \def\stpar##1{\vtop\bgroup\hsize ##1%
     \baselineskip=\stablesbaselineskip%
     \lineskip=\stableslineskip%

\lineskiplimit=\stableslineskiplimit\bgroup\aftergroup\estpar\let\temp
=}%
  \def\estpar{\vskip 6pt\egroup}%
  \def\stparrow##1##2{\stablesdummy=##2%
     \setbox0=\vtop to ##1\stablestrutsize\bgroup%
     \hsize\stablesdummy%
     \baselineskip=\stablesbaselineskip%
     \lineskip=\stableslineskip%
     \lineskiplimit=\stableslineskiplimit%
     \bgroup\vfil\aftergroup\estparrow%
     \let\temp=}%
  \def\estparrow{\vfil\egroup%
     \ht0=\ht\stablestrutbox%
     \dp0=\dp\stablestrutbox%
     \wd0=\stablesdummy%
     \box0}%
  \def|{\global\advance\stablesmode by 1&&&}%
  \def\|{\global\advance\stablesmode by 1&\omit\vrule width 0pt%
         \hfil&&}%
  \def\vt{\global\advance\stablesmode by 1&\omit\vrule width
\stablesthinline%
          \hfil&&}%
  \def\vtt{\global\advance\stablesmode by 1&\omit\vrule width
\stablesthickline%
          \hfil&&}%
  \def\vttt##1{\global\advance\stablesmode by 1&\omit\vrule width
##1%
          \hfil&&}%
  \def\vtr{\global\advance\stablesmode by 1&\omit\hfil\vrule width%
           \stablesthinline&&}%
  \def\vttr{\global\advance\stablesmode by 1&\omit\hfil\vrule width%
            \stablesthickline&&}%
  \def\vtttr##1{\global\advance\stablesmode by 1&\omit\hfil\vrule
width ##1&&}%
  \stableslines=0%
  \stablesomitfalse}
\def\stablesdef{\bgroup\stablestrut\borderrule##\tabskip=0pt plus
1fil%
  &\stablesleft##\stablesright%

&##\ifstablesright\hfill\fi\internalrule\ifstablesright\else\hfill\fi%

  \tabskip 0pt&&##\hfil\tabskip=0pt plus 1fil%
  &\stablesleft##\stablesright%

&##\ifstablesright\hfill\fi\internalrule\ifstablesright\else\hfill\fi%

  \tabskip=0pt\cr%
  \ifstablesborderthin%
    \thinline%
  \else%
    \thickline%
  \fi&%
}%
\def\endtable{\advance\stableslines by 1\advance\stablesmode by 1%
   \message{- Rows: \number\stableslines, Columns:
\number\stablesmode>}%
   \stablesel%
   \ifstablesborderthin%
     \thinline%
   \else%
     \thickline%
   \fi%
   \egroup\stablesend%
\endgroup%
\global\stablesinfalse}
%
%


\def\subsubseclab#1{\DefWarn#1\xdef #1{\noexpand\hyperref{}{subsubsection}%
{\secsym\the\subsecno.\the\subsubsecno}%
{\secsym\the\subsecno.\the\subsubsecno}}%
\writedef{#1\leftbracket#1}\wrlabeL{#1=#1}}

\def\figin{\epsfcheck\figin}\def\figins{\epsfcheck\figins}
\def\epsfcheck{\ifx\epsfbox\UnDeFiNeD
\message{(NO epsf.tex, FIGURES WILL BE IGNORED)}
\gdef\figin##1{\vskip2in}\gdef\figins##1{\hskip.5in}
\else\message{(FIGURES WILL BE INCLUDED)}%
\gdef\figin##1{##1}\gdef\figins##1{##1}\fi}
\def\DefWarn#1{}
\def\figinsert{\goodbreak\midinsert}
\def\ifig#1#2#3{\DefWarn#1\xdef#1{fig.~\the\figno}
\writedef{#1\leftbracket fig.\noexpand~\the\figno}%
\figinsert\figin{\centerline{#3}}\medskip\centerline{\vbox{\baselineskip12pt
\advance\hsize by -1truein\noindent\footnotefont{\bf Fig.~\the\figno:} #2}}
\bigskip\endinsert\global\advance\figno by1}

\Title{
\vbox{\baselineskip12pt\hbox{hep-th/9412110}\hbox{YCTP-P17-94}}}
 { \vbox{
\centerline{ Wilson loops in 2D Yang Mills: }
\centerline{ Euler Characters and Loop equations} }}
\bigskip
\centerline{ Sanjaye Ramgoolam}
\bigskip
\centerline{skr@genesis2.physics.yale.edu}
\smallskip\centerline{Dept.\ of Physics}
\centerline{Yale University}
\centerline{New Haven, CT \ 06511}
\bigskip
\bigskip
\bigskip
\centerline{\bf Abstract}
\noindent
We give a simple diagrammatic algorithm for writing the
chiral large $N$ expansion of  intersecting  Wilson loops
in  $2D$  $SU(N)$ and $U(N)$Yang Mills theory
 in terms of symmetric groups, generalizing
the result of Gross and Taylor for  partition functions.
We prove that these expansions  compute Euler characters
of a space of holomorphic maps from string worldsheets with boundaries.
We prove that the Migdal-Makeenko equations  hold
for the chiral theory  and show
that they can be expressed  as  linear constraints on perturbations
of the chiral $YM2$ partition functions.   We briefly  discuss
 finite $N$ , the non-chiral expansion, and higher  dimensional  lattice
models.

\Date{ 10/94}
\newsec{ Introduction}

We develop formulae
for    the chiral large $N$ expansion
of  Wilson averages in $2D$ Yang Mills. This
 generalizes the
results of Gross and Taylor \GrTa\ who expressed
the chiral expansion
of partition functions in terms of a sum of delta functions
 over symmetric groups.
The generalisation of their result is best expressed in
a diagrammatic form.
This diagrammatic expression  has  a direct interpretation
in terms of branched covers. The general structure of
the diagram is understood
in terms of elementary properties  of  the appropriate
space of
branched covers :

 (a) degrees of the
maps in various regions of the target space and,

(b)  the fundamental groups
of  the target space and of the Wilson graph (see definition  in section 4.1).

The detailed nature of the
sums over symmetric group elements  appearing is  understood
in terms of  the counting of equivalence classes (definition in section 4.2) of
the
 relevant class of branched covers, and from the conjecture (section 9 of  ref
\CMROLD )
 that Wilson averages
compute Euler characters of the space of branched covers.

In sections 2 and 3 we explain these ideas in the
familiar cases of manifolds with and without boundary.
In section 4 we introduce Wilson observables, associated to
curves in the target $\ST$, and state the conjecture
from \CMROLD\ for the geometrical interpretation
of their expectation values. We recall  the proof of the conjecture
for non-intersecting Wilson loops in section 5. Readers familiar with
the string theory of 2D Yang Mills may prefer to move on to section 4
after reading the introduction. The main
 results  of this paper are in sections  6, 8, 9, 11 and  12.

In section 6, we give the general algorithm for writing
 the chiral large $N$ expansion. It uses words in
 the generators of
the  fundamental group
of the Wilson graph,  arranges them along strands
 labelled with positive integers,
manipulates the strands via a simple diagrammatic
 move (see fig. 12), and constructs
the final answer by replacing the generators with
certain sums over permutations.

In section 7, this algorithm is derived from exact
answers for 2D Yang Mills in terms of
group integrals which were derived in
 \refs {\Witdgt , \Ru, \Mig }. In sections 8-10, we
give  the geometrical interpretation in terms of branched
 covers and Euler characters, using many ideas
 from \GrTa\  and \CMROLD .
We have adopted the words  `braid diagrams'
 for the diagrams
which express the chiral large $N$
 expansions of Wilson averages.
This terminology will be explained in
section 7.1  and appendix A.

In this diagrammatic expression of Wilson averages,
some surprising  properties of Wilson averages
become  transparent.
First we show that in the chiral theory the Wilson
averages can be rather
directly related to closed strings. We explain this
in remarks in sections 5 and 8.
Second we show in section 11 that the { \it chiral
 Wilson averages}
 satisfy the Migdal-Makeenko loop equations.
Recall that the chiral
expansion only gives part of the full $1/N$
expansion of $2D$ Yang Mills \GrTa :
however from the string point of view it is an
 interesting theory. In fact, ref \CMROLD\ constructs
a string action for the chiral theory by localizing to
spaces of holomorphic maps
(as well as one for the non-chiral theory by localising
to spaces of  holomorphic + anti-holomorphic maps).
The full  large $N$ expansion of \ymt\ is the
non-chiral (coupled) expansion.
Standard proofs of Migdal-Makeenko imply
that the coupled expansion
for Wilson averages should satisfy the loop
 equations.
The fact  that chiral Wilson
averages satisfy loop equations may have
 some mathematical interest since
the chiral theory is the worldsheet field  theory
 for  classical
 Hurwitz spaces of branched covers.
The third property  concerns an analogy to 2D gravity.
The loop equations are non-linear equations,
but as in  2D gravity \refs{  \DiVeVe }: The  non-linear  loop equations
 can be used to derive  linear constraints
on perturbations of the partition function.

We will generally be dealing with
 the chiral $SU(N)$ theory. Occasionally,  we will
deal with  the chiral zero charge sector
 of the $U(N)$ theory, which has the same structure as the chiral $SU(N)$
theory
except for a small difference in the area
dependent exponentials, and which has been studied,  for example,  in \refs{
\GrTa,
 \DoKa , \Rudd, \CreTay } .
 This will simplify some formulae in section 12.

In the final section,  we have  some general
 remarks on  the application
 of this diagrammatic formalism for the geometry of
large $N$ group integrals to
finite $N$ \ymt, non-chiral \ymt\ and higher dimensional
 lattice  models.

\newsec{ Partition function}

The basic tool that leads to an interpretation
of the chiral expansion
of \ymt , in terms of covering spaces and
 Euler characters
of spaces of holomorphic  maps ( section 5 of \CMROLD )  is its expression
 in terms
 of delta functions over symmetric groups, which
are related,  by Riemann's theorem, to
the counting of branched covers \GrTa ( for a review see \CMRlsch ).

The main ingredient in the derivation of the
chiral expansion from exact results for
 the partition
function  \refs{  \Mig ,\Witdgt , \Ru }
 is Schur-Weyl duality, which we describe briefly.
All unitary irreps of  $SU(N)$ are
 obtained from reducing tensor
products of the fundamental rep $V$
 (for $U(N)$ we need tensor products of both
the fundamental
 and its complex conjugate).
Consider the action of $SU(N)$ on $V^{\otimes n}$.
Let $\rho(U)$ represent the action of $U$ in $V^{\otimes n}$.
The symmetric group $S_n$ acts on $V^{\otimes n} $ by
 permuting the factors.
Let $\tilde{\rho}$ be the map from $S_n$ to $End(V^{\otimes n})$.
So
\eqn\rhoact{ \tilde{\rho} (\sigma) \vert e_{i_1}>  \otimes \vert e_{i_2}>
 \otimes \cdots \vert e_{i_n}>
  =  \vert e_{i_{\sigma(1)} }>  \otimes \vert e_{i_{\sigma (2)} }>  \otimes
\cdots \vert e_{i_{\sigma(n)} } >. }

\bigskip

\boxit{{\bf Schur-Weyl duality}: The commutant of
$\rho (SU(N))$ in $V^{\otimes n}$ is $\tilde{\rho}( \IC(S_n) )$,
and
$V^{\otimes n}$ is completely reducible, as a representation
 of $SU(N) \times S_n$,  to $V^{\otimes n}\cong
 \sum_{Y}  R(Y)\otimes r(Y). $}

In the $SU(N)$ theory irreps can be
labelled by Young diagrams with fewer than $N$ rows:
$Y\in \CY_n^{(N)}$, where
$\CY_n^{(N)}$ is the set of valid Young
diagrams for $SU(N)$.
For $N > n$ (which is relevant  in the large $N$
 expansions), $\CY_n^{(N)}=\CY_n$, the set of all
Young diagrams with $n$ boxes.
We denote by $R(Y)$ ( $r(Y)$ ) the irrep of  $SU(N)$ ($S_n$ ) corresponding to
$Y$.
So if $P_Y$ is the Young projector for the rep of $S_n$
associated with $Y$,
we have \refs{\Zelo}
\eqn\pyeq{   P_Y V^{\otimes n} \cong R(Y)\otimes r(Y) . }

We recall the result for the chiral partition function,
 for a target Riemann surface $\ST$ of genus $G$ and  area $A$,
as an expansion
in terms of  symmetric group quantities :
\eqn\Zchir{
\eqalign{
&Z(G,A, N)\cr
& =~\sum_{n=0}^{\infty}
\biggl[ {1\over {n!}}
\delta_{n} (e^{-{nA\over 2} -A {T_2^{(n)}\over N} + {A (n^2 )\over 2N^2} }
N^{n(2-2G)}  \Omega_n^{2-2G}
\prod_{j=1}^G \sum_{s_j,t_j \in S_n}  s_jt_js_j^{-1}t_j^{-1}) \biggr]
.\cr
 }}
Here
\eqn\omeg{ \Omega_n = \sum_{v \in S_n} v
N^{K_v - n }  = N^{-n} \sum_{v \in S_n} v tr_n( \tilde\rho (v) ), }
where $tr_n$ is the trace in $V^{\otimes n}$.
The $n=0$ term in \Zchir\ is defined to be $1$.
$T_2^{(n)}$ is an element in the group algebra of $S_n$,
 equal to the sum of all the transpositions.
 The delta function $\delta_n$
 acting on the
group algebra of $S_n$ evaluates the coefficient of the identity
permutation.
 The omega factors
 were expanded out in \CMROLD\
producing words in the  free algebra generated by
elements of $S_n$, $  \CF( S_n) $.
Each permutation comes with a power of $N$
 which is $N^{K_{\sigma}-n}$.
All words of length $L$ are  weighted by
the  Euler character of the configuration space of $L$ points on $\ST$.
This uses the fact  the power of $\Omega_n$ is
exactly the Euler character of the target space.
By Riemann's theorem,  the sums
 over symmetric groups count branched covers (holomorphic maps)
of degree $n$,
weighted by the inverse of the order of the automorphism group.
Using the fibration of Hurwitz space over configuration spaces
leads to the result that  the zero area partition function is
 the generating function of Euler characters of  the
space of holomorphic maps  from a smooth
worldsheet $\Sw$ onto the target $\ST$. The detailed discussion
is in  \CMROLD .
For finite area we have a similar
 structure:   elements in  $  \CF( S_n) $,
 weighted by $A$ dependent  { \it power series } in $1/N$.
For the geometrical interpretation in terms of branched covers,
the factor $e^{-A {T_2^{(n)}\over N}} $ is expanded out to
give sums of transpositions. Terms with these sums
  are interpreted as contributions
 from maps which have some number of simple
branch points weighted by factors of the area.
 The term $e^{-n^2A\over {2N^2}}$ is   also expanded out and  interpreted
as contributions from maps
involving worldsheets with double points and
collapsed handles  \refs{  \Min , \GrTa }.

\ifig\fonpf{ Pictorial representation of delta function for partition function
}
{\epsfxsize4.5in\epsfbox{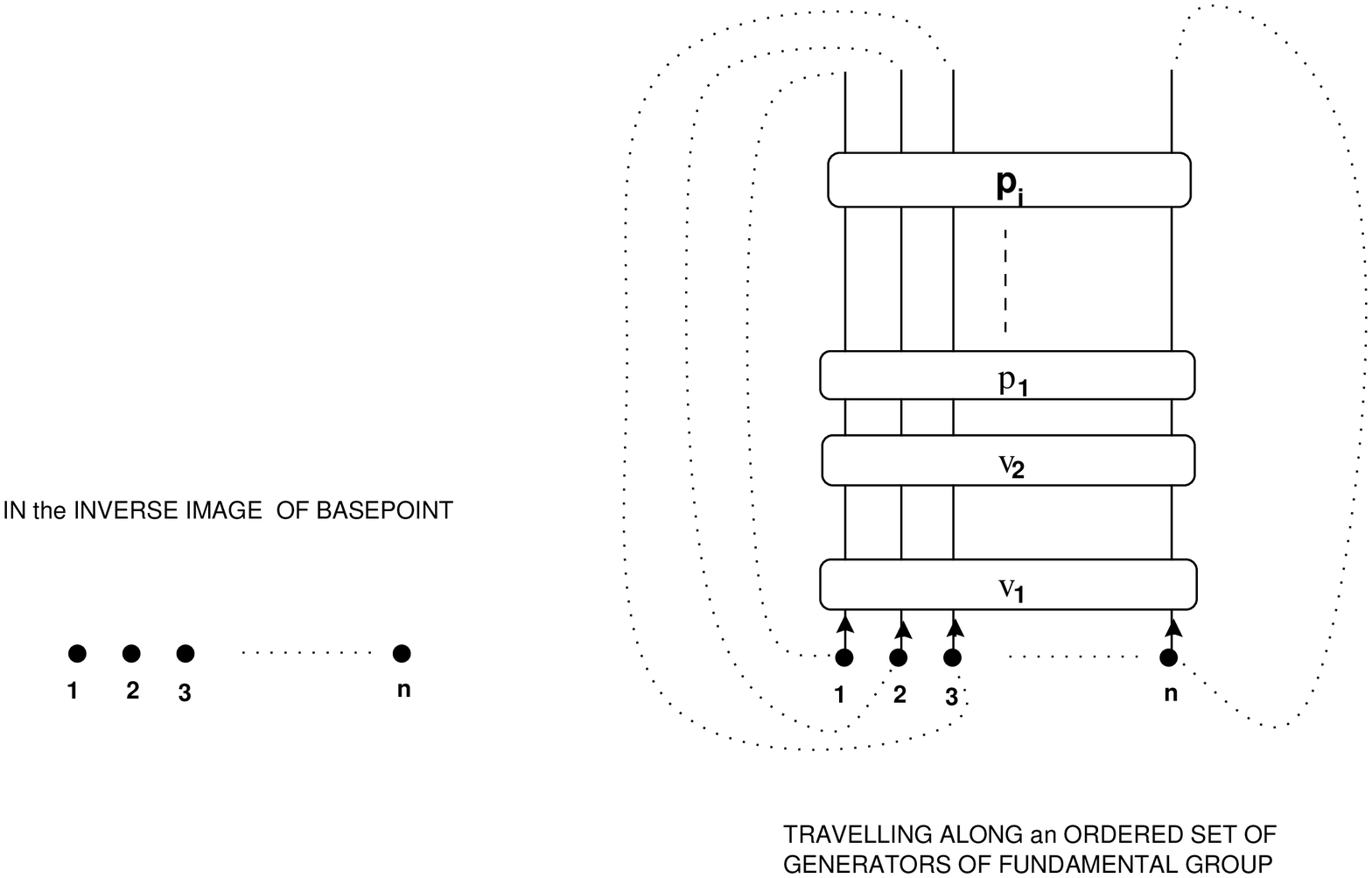}}

\ifig\ftwpf{ Simplified pictorial representation for partition function}
{\epsfxsize3.5in\epsfbox{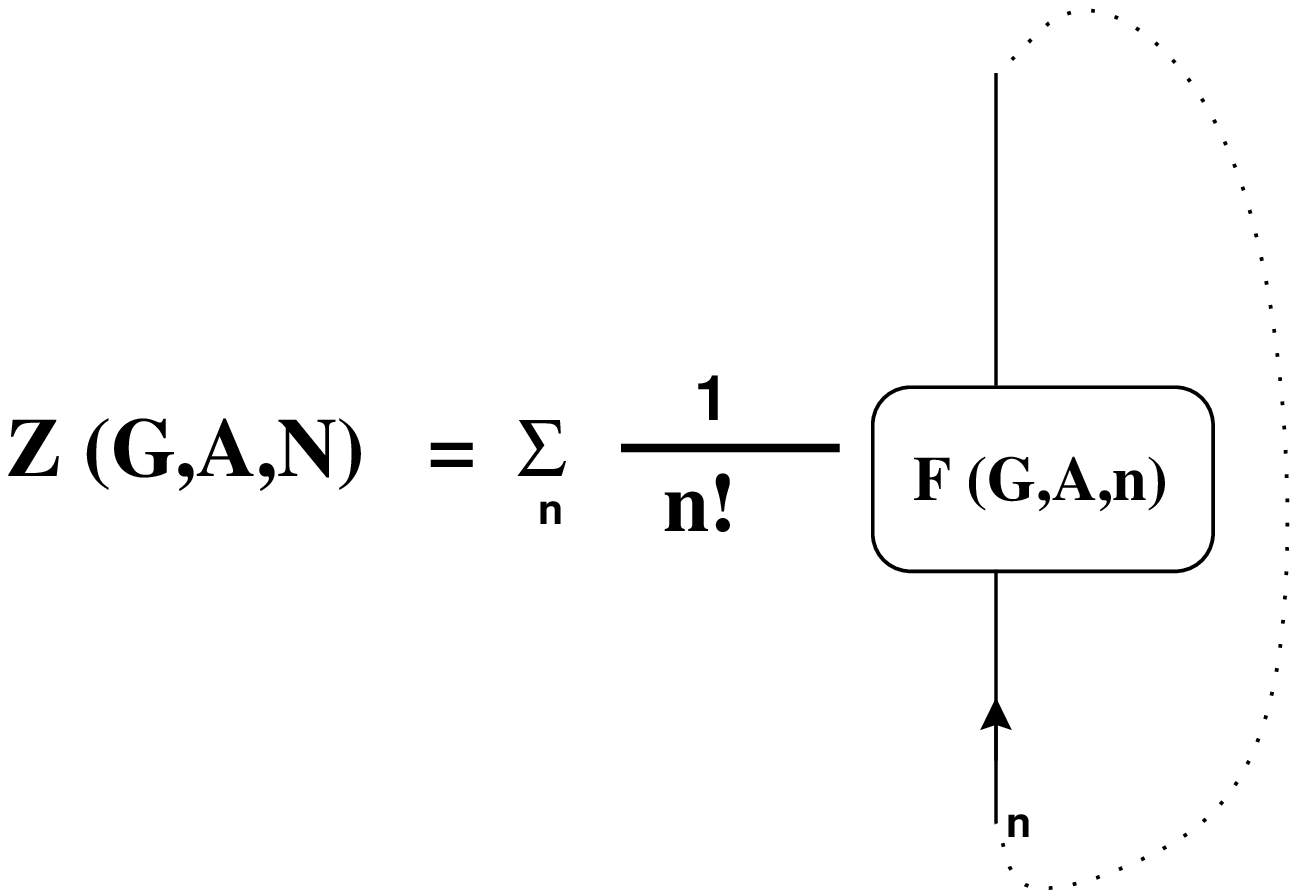}}

Each term contributing to the partition
 function is a sequence of permutations which multiply to $1$.
Such a sequence can be expressed as in \fonpf,
 where we have drawn a typical term for $G=0$.
The $v$'s come from the
 $\Omega_n$ factors and the $p_i$'s come from the $AT_2^{(n)}$ term in
the exponential.
 At the base of the diagram are $n$ points   labelled from $1$ to $n$.
They correspond to the $n$ points in the inverse
image of an arbitrarily chosen  basepoint, $y_0$,
on the target $\ST$. For each generator of the
 fundamental group  $\ST - \{  {\hbox{branch points} }\}$  the diagram contains
a permutation,
which  may be thought more pictorially in terms of the strands
winding around each other according to the permutation.
The product of the permutations
read from top to bottom is equal to the identity.
This is expressed by the dotted lines which join top
 to bottom with no further winding.
 In \ftwpf , we draw a simplified version of the
diagram with the whole sequence of permutations
written as one $F$ factor.
\eqn\Ffact{\eqalign{
F(G,A,n) &
  \in  \CF( S_n) \otimes \IR ( 1/N ) \cr
& =   e^{-{nA\over 2} -
A {T_2^{(n)}\over N} + {n^2A\over 2N^2} }   N^{n(2-2G)} \Omega_n^{2-2G}
  \prod_{i=1}^{G} \sum_{s_i,t_i \in S_n}
s_it_is_i^{-1}t_i^{-1}. \cr
} }

After  multiplying out the permutations, $F$ determines an element
of $ \IZ (S_n) \otimes \IR (1/N)$.

\ifig\fpfunthr{ Partition function as a trace in tensor space. }
{\epsfxsize3.5in\epsfbox{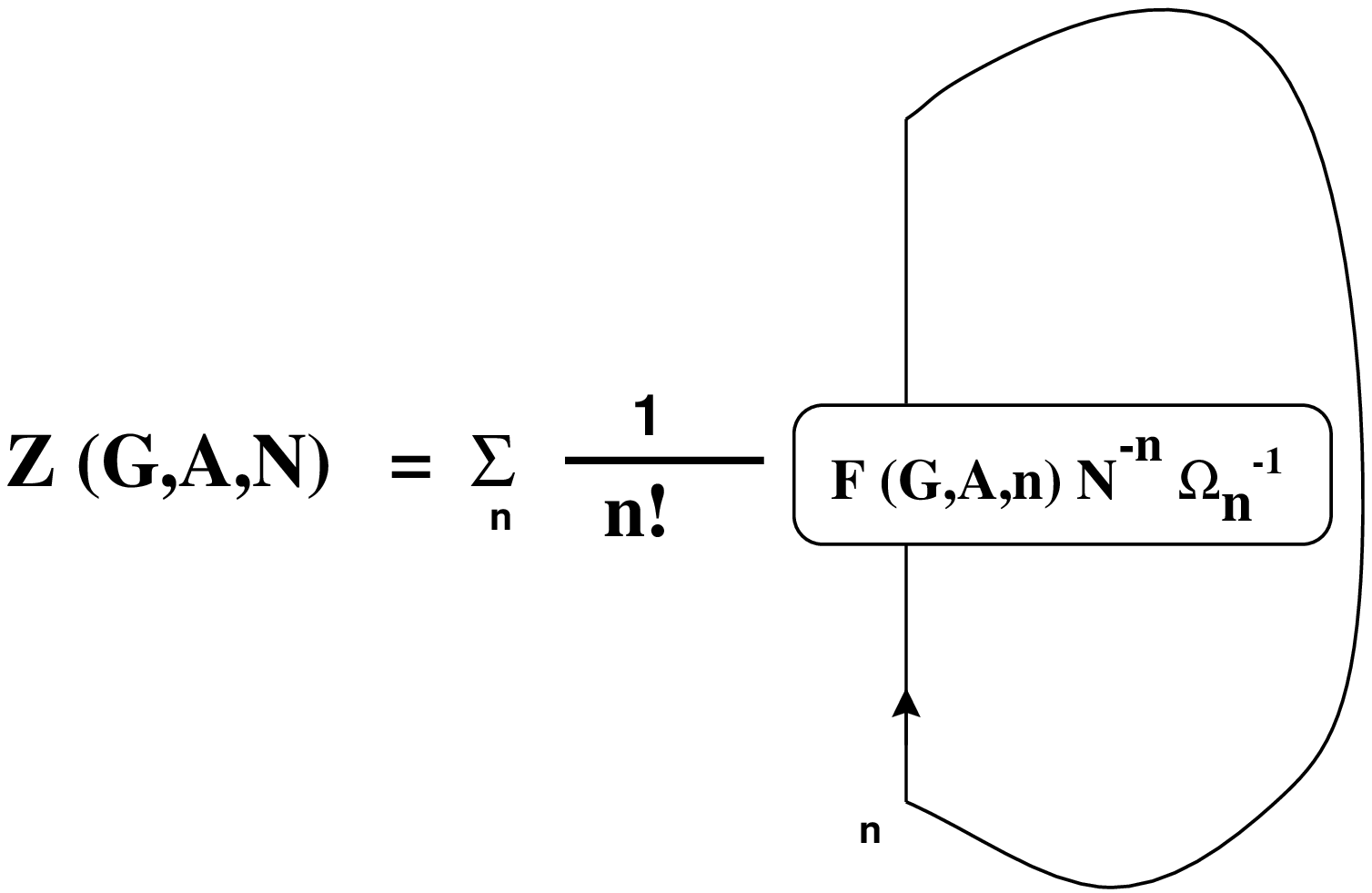}}

The permutations can be represented in the
standard way in tensor space,
as operators in $End ( V^{\otimes n} ) $.
The identity pemutation maps to the  identity operator,
so the delta function is equal to a
delta function on the corresponding operators
in tensor space.
Observing that
\eqn\deltatr{tr_n( \tilde \rho (\sigma) ) \equiv
  tr_n (\sigma) = N^{K_{\sigma}} = \delta_n ( \sigma N^{n} \Omega_n ), }
we can also write $Z$ very simply as a sum over traces in $V^{\otimes n}$ :
\eqn\Zchirt{
Z (G,A,N) = \sum_{n} {1\over {n!}} tr_{n} (F_n ( G,A,N) \Omega_n^{-1} N^{-n} )
}

This can be represented pictorially by replacing the
 dotted lines with solid lines,
which indicate contracted indices, as in \fpfunthr .

\newsec{Manifolds with boundary. }
\subsec{ Hurwitz spaces for manifolds with boundary. }

For target spaces with boundary, $(\ST , \p\ST)$,
chiral \ymt\
partition functions are generating functions for
Euler characters of spaces of branched covers  from
manifolds with boundary,  $\bigl(\Sw,\p\Sw\bigr) $,   to the target space.

\noindent
{\bf Definition 3.1}. A {\it boundary-preserving branched
covering} is a map
\eqn\bpbc{
f:\bigl(\Sw,\p\Sw\bigr)\rightarrow \bigl(\ST,\p\ST\bigr)
}
such that

\noindent
1. $f:\p\Sw\rightarrow\p\ST$ is a covering map.

 \noindent
2. $f:\Sw - \p\Sw \rightarrow \ST- \p\ST$ is an
 orientation preserving branched covering.

\noindent
{ \bf Definition 3.2 }
Two maps $f_1$ and $f_2$ are said to be { \it equivalent } if
there is a homeomorphism $\phi : \Sw \rightarrow \Sw$,
  such that $f_1= f_2\circ \phi$.
An { \it automorphism}   of a boundary-preserving branched cover
is a homeomorphism $\phi$ such that $f= f\circ \phi$.

By (1) $f$ determines a class $\kG$ for each component
$\G$
of $\p\ST$:  $\kG^{(j)}$ is the number of circles of winding number $j$ in the
inverse image of $\G$.
Let us assume that $\ST- \p\ST$ is connected.
Then, by  (2) $f$ determines a branch locus
$S(f)\subset \ST- \p\ST$, an index $n$,
 and an equivalence class of a homomorphism
$\psi_f: \pi_1(\ST-S(f), y_0) \to S_n$.  We have the direct
analog of  Riemann's   theorem:

\noindent
{\bf Proposition 3.3}. Let $\ST$ be a connected,
closed surface with
boundary. Let $S\subset \ST- \p\ST$ be a finite set,
 and let $n$ be a positive integer.
There is a one-one
correspondence between equivalence classes
of boundary-preserving
branched covers \bpbc\ with branch locus $S$ and
equivalence classes of homomorphisms
$\psi: \pi_1(\ST-S(f), y_0) \to S_n$.

The proof is in section 9 of \CMROLD. Given the homomorphism we can
construct a branched cover, and given a branched cover,
we can obtain a homomorphism by lifting paths
which generate  the fundamental group \Ez .
The boundaries can be shrunk to give a surface with deleted points.
So the counting of boundary preserving branched
 covers is related to a classical
problem:  that of counting branched covers with
specified monodromies at the deleted points.

 \ifig\fgnsmb{ Generators of fundamental group for manifold with boundary.}
{\epsfxsize3.5in\epsfbox{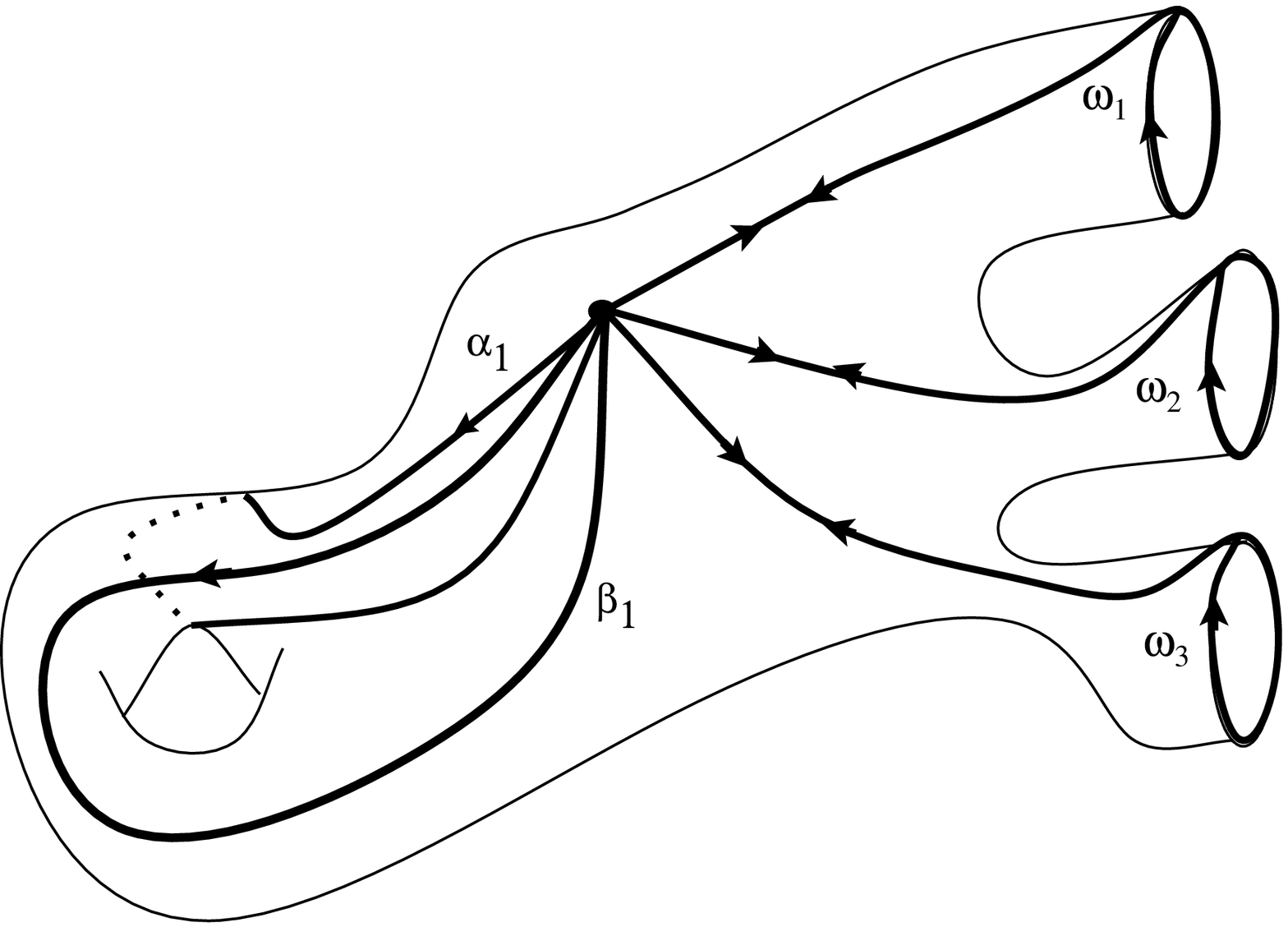}}

\subsec{ \ymt\ partition functions for Manifolds with boundary}

The partition function for a manifold of genus
$G$, area $A$,  with  $b$  boundaries, with
specified holonomies $U_1, \cdots
U_b$, is given by \Witdgt, \Ru,
\eqn\mbdy{ Z(G, A, b; U_1, U_2, \cdots U_b) =
 \sum_{R} (dim R)^{2-2G-b} e^{-{A\over 2N}C_2(R) }
\chi_R (U_1)  \cdots  \chi_R (U_b).  }
We can consider expectation values of
$\chi_{R_i} (U_i ^{\dagger})$ inserted at the $i'th$ boundary.
They are given by
\eqn\mbdyi{\eqalign{
 Z(G,A, b; R_1, R_2, \cdots R_b) &=\prod_{i=1}^{b} \int dU_i
Z(G,A, b; U_1, U_2, \cdots U_b)
\prod_{i=1}^{b}  \chi_{R_i } (U_i^\dagger) \cr
&= \sum_{R}  (dim R)^{2-2G-b}
 e^{-{AC_2 (R)\over 2 N} }  \delta_{R_1 , R}\delta_{R_2 , R} \cdots \delta_{R_b
, R}.   \cr  } }
A more direct  string interpretation is
obtained for  insertions of
certain linear combinations of the
 irreducible characters, called {\it loop functions},  $\Upsilon_{\vec k} (U)$,
which depend on a {\it winding vector}
$\vec k$. $\vec k$ is
a vector with an infinite number of
components, most of which are zero.
The vector $\vec k$ determines a conjugacy
 class $[\vec k]$ of permutations
with $k_1$ cycles of length $1$,  $k_2$ cycles
 of length $2$ etc. ,
in $S_{n(\vec k )} $, where $n(\vec k) = \sum_{j} jk_j $.
 If  $\sigma$ is  a permutation in the class $[\vec k ]$,
\eqn\upsi{\eqalign{
\Upsilon_{\vec k} (U) & = \Upsilon_{\sigma} (U) \cr
                                       & = tr_n ( \sigma U) \cr
& = \sum_{ Y \in Y_n} \chi_{r(Y)} (\sigma) \chi_{R(Y)} (U), \cr} }
where the last equality follows directly from Schur-Weyl duality.

Insertion of the loop  function into the path integral yields:
\eqn\mbdyupsi{ \eqalign{
 &Z(G,A,b;\vec  k_1, \vec k_2, \cdots \vec k_b)\cr
&= \int dU_1 dU_2 \cdots dU_b Z(G,A,b; U_1, U_2, \cdots U_b)
 \prod_{i=1}^{b}
\sum_{u_i \in [\vec k_i]} {1 \over {n_i!} }
 \Upsilon_{ u_i}  (U_i^\dagger ) \cr
&=\sum_{n}\sum_{u_1, \cdots u_b} {1\over n!} \delta (F (G,A,b,n) u_1\cdots u_b)
 \delta_{n_1,n} \delta_{n_2,n} \cdots
\delta_{n_k,n} \quad , \cr}}
where   $F(G,A,b,n) $ is given by
\eqn\Fbdy{\eqalign { & F (G,A,b,n)
= \cr
& e^{-n A/2 - AT_2^{(n)} + {n^2A\over 2N^2} }
 N^{n(2-2G-b)} \Omega_n^{2-2G-b}
\prod_{j=1}^G \sum_{s_j,t_j \in S_n}
s_jt_js_j^{-1}t_j^{-1}
.\cr }}

This formula can be derived by manipulations
similar to those leading to the
derivation of \Zchir\ \GrTa. Note that we need the insertion
 $ \sum_{\sigma_i \in [\vec k_i]} {1 \over {n_i!} }
\Upsilon_{ \sigma_i}  (U_i^\dagger )=
 {  \vert {[ \vec k_i] } \vert \over {n (\vec k_i) !} }$
( where $\vert [ \vec k_i]  \vert )$ is the
number of elements in the conjugacy class  $[\vec k_i]$ ),
in order  to make sure that we are counting
equivalence classes of branched covers weighted
by the inverse of the automorphism.

\ifig\fgnsmb{  Delta function for Manifold with Boundary}
{\epsfxsize4.5in\epsfbox{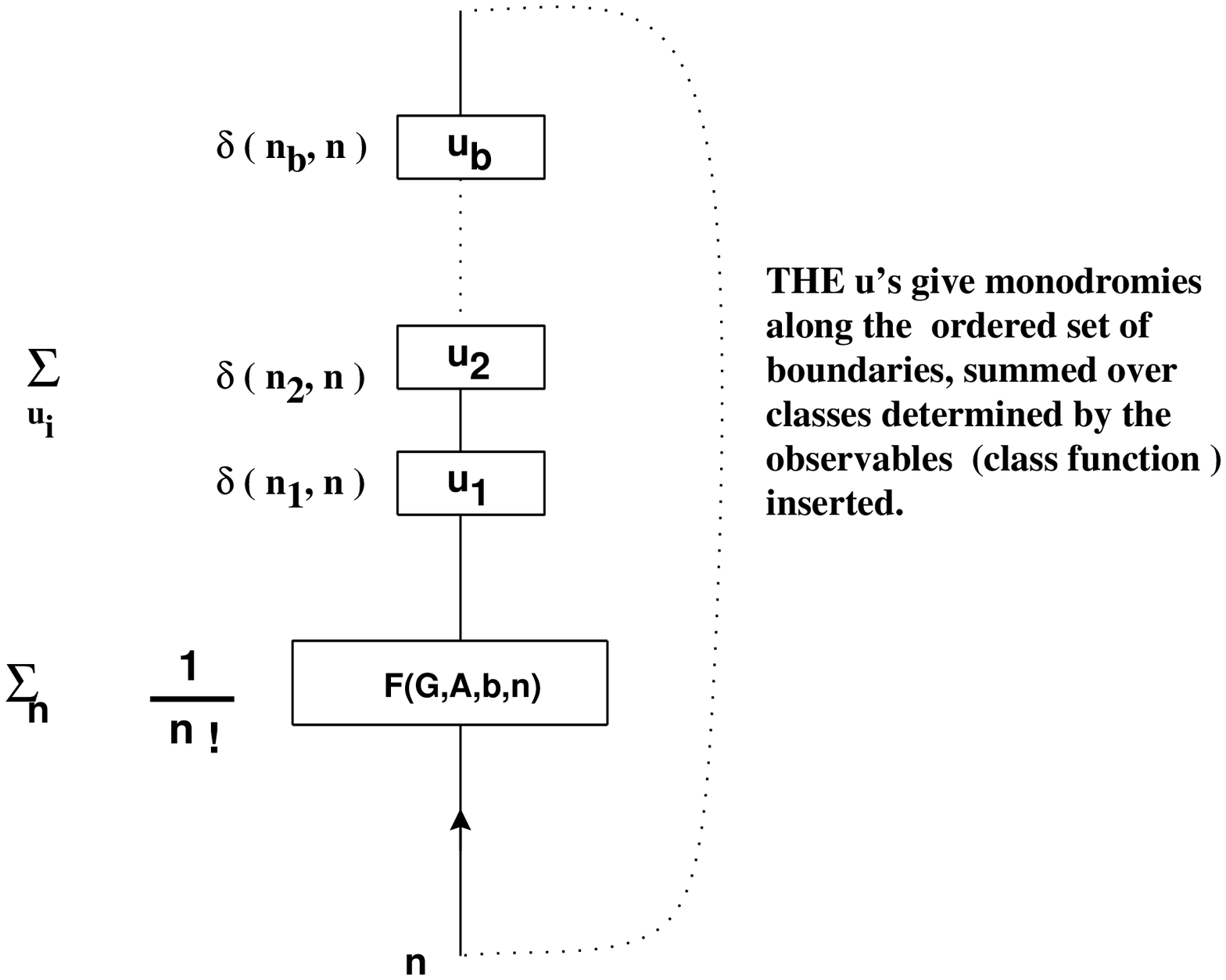}}
Again the delta function can be presented pictorially as in \fgnsmb  .
It expresses the fact that each term is obtained from a
sequence of permutations obtained by travelling around
cycles in the interior of the manifold ( collected in the $F$ factor),
and around the boundaries, which all multiply to $1$.

At $A=0$ the partition function  counts Euler characters of
the space of branched covers
where  the map from worldsheet boundaries  to the $i'th$
target boundary is in  the
homotopy class   given by
$\vec k_i$.  To see this we expand the omega factors to
generate arbitrary numbers of branch points, as for the
chiral partition function. Note that the power of omega is
 appropriate for producing the Euler character
of configuration spaces of points on $\Sigma_T $.

\newsec{  Wilson loops ---observables and maps}

\subsec{ Observables and exact answers }
Wilson averages are computed by inserting in the path integral,
traces of holonomies around specified paths.
Each path $\Gamma$ is a map from $S^1$ to the  target space $\ST$.
The image of such a map is called $\CG$. In general we have
$ \amalg S^1 \rightarrow \amalg \CG $. We will call
 $\amalg \CG $ the  {\it Wilson graph}, which
 is made of a union of open edges and vertices.
 It may be connected or disconnected.
For non-intersecting Wilson loops,  the Wilson
 graph is a disjoint union of circles.
The observable associated  with it is
$$W_{R_\G} = tr_{R_\G}  (Pexp { \int_{\Gamma} A} ),
 $$ the trace in rep
$R_{\G}$ of the holonomy of the gauge field along $\G$.
As in the case of manifolds with boundary,
insertions of the appropriately normalised
$\Upsilon_{\vec k_{\G}} (U_\G ^\dagger) $
will have a more direct
geometrical interpretation.
\ifig\useorie{Using the orientation of the surface and of the
Wilson line we can define two infinitesimal deformations of
the Wilson line $\G^l$ and $\G^r$ .}
{\epsfxsize2.0in\epsfbox{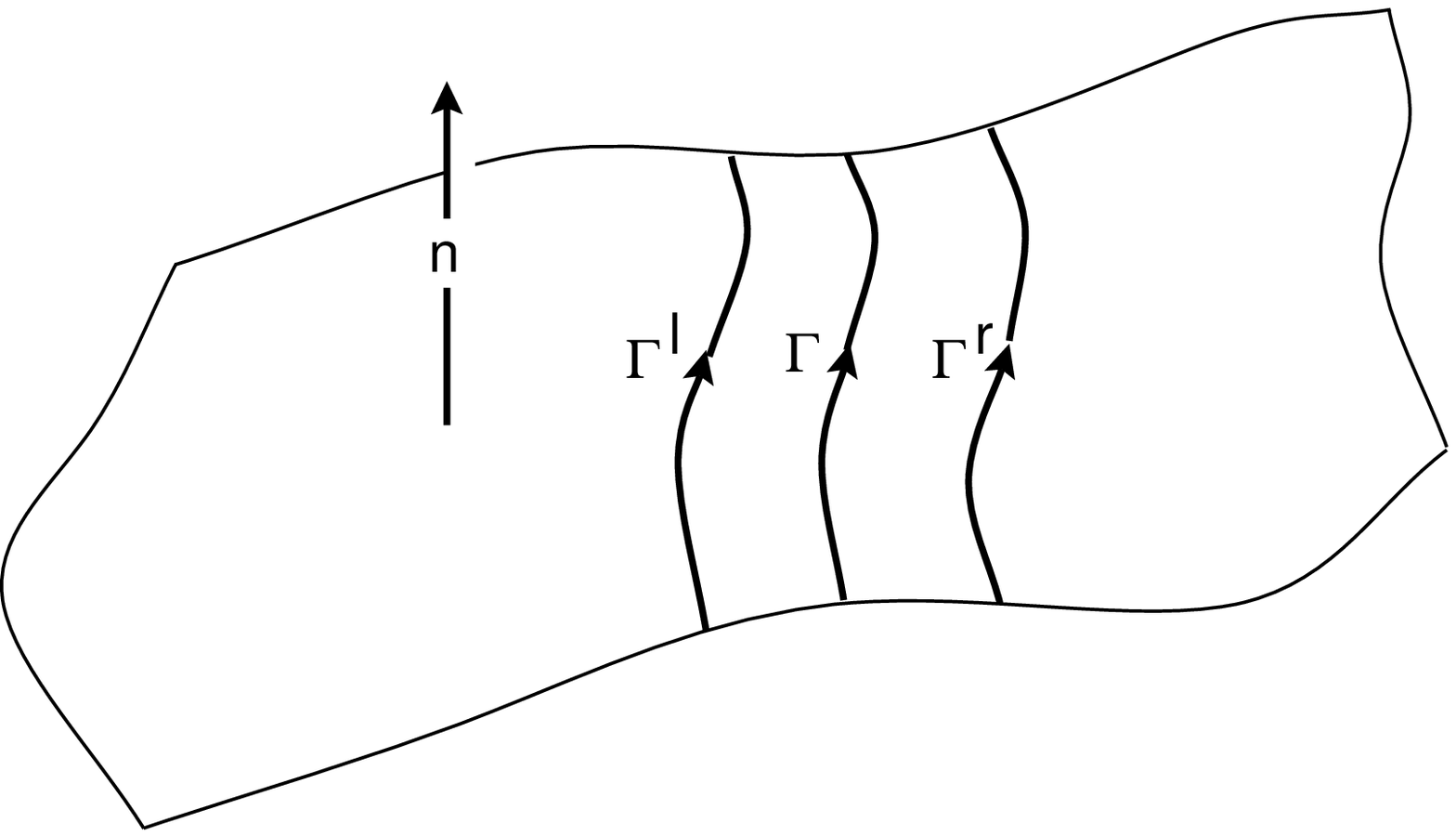}}
Using the orientation of $\ST$ we can perturb each oriented Wilson loop
$\G$ to its left or right, into loops $\G^l$ and $\G^r$.

 Witten \Witdgt\ and Rusakov  \Ru\
 wrote exact answers for the Wilson averages in terms
of  sums over representations and
group integrals.
Given a collection of curves $\{ \G \} $ on $\ST$,
we have a decomposition into disjoint connected open components:
\eqn\ccomp{
\ST-\amalg~ \CG = \amalg_c~ \Sc .}
The component $\Sc$ has $G_c$ handles, area $A_c$ and $b_c$ boundaries.
A choice of orientation  of the manifold $\ST$ induces an orientation
on each component.  Assign to each oriented edge $e$ of the
 Wilson graph a group element $U_e$. Each region induces
on its boundaries an orientation. Any oriented boundary of a
 region can be expressed as a word in  the edge paths  and
 is assigned a group element equal to the corresponding
product  of edge variables. For each region there is a local factor
equal to :
\eqn\Zc{ Z_c = \sum_{R_c} \dim~ R_c~
e^{- A_cC_2(R_c)/2}    \prod_{i=1}^{b_c}
 \chi_{R_{c}} ( U_c^{ {(i)} } )   , }
where $U_c^{  {(i)} } $ is the holonomy
around the $i'th$ oriented boundary
of  $\Sc$. The full answer is
\eqn\Wilsav{\langle \prod_{ \{ \G \}}
W_{R_{\G}} \rangle =  \sum_{R_c} \int \prod_{e}
dU_{e} \prod_{c} Z_c \prod_{\Gamma}
\chi_{R_\Gamma}(U_{\Gamma}^\dagger ). }
If we insert $\Upsilon_{\vec k_{\G}} (U_\G ^\dagger) $,
 then we have instead of irreducible characters, the
 loop functions in the product over $\G$.

Witten \Witdgt\  also showed that these exact
 answers can be written as a sum over
 reps associated with each region, together
with six $j$
symbols associated with the vertices. In the
 case of
non-intersecting Wilson averages,
everything can be written in terms of  sums over reps and
fusion numbers.

\subsec{Geometrical conjecture }
In \CMROLD\ there was a simple conjecture,
based on results of \GrTa ,
 for interpreting the  expectation values of all
Wilson averages in the zero area limit
in terms of Euler characters.
Consider the insertion of $\prod_{\G} \Upsilon_{\vec k_\G}$ .
Elaborating the  conjecture of  \CMROLD , we have,

\medskip
\noindent
{\bf   CONJECTURE 4.1  }. The string interpretation
 for chiral  Wilson loop amplitudes
on $\ST$
is obtained by imposing on branched covers
$f : (\Sw, \partial \Sw) \rightarrow (\ST ) $
 the boundary condition that $f:\p \Sw \to \amalg {\cal G} $
is in the homotopy class
\eqn\hmtpycl{
\p \Sw {\buildrel \{ \kG\}\over \longrightarrow} \amalg_{\G} S^1
{\buildrel \amalg \G\over \longrightarrow} \amalg {\cal G}.
}

The worldsheet and target are oriented. The worldsheet boundary
has orientation compatible with that of worldsheet.
The maps are orientation preserving.
The first arrow describes a covering of oriented circles by
oriented circles. The second is the homotopy class of the curves
defining the Wilson loops.

Equivalences of such maps are defined as in Def 3.2 .
 Spaces of such maps will be called
Hurwitz-Wilson spaces. More precisely
we expect that Wilson averages are
 generating functions for orbifold Euler
characters of Hurwitz-Wilson spaces.

To compactify the space of branched
covers we will
{ \it allow  branch points to lie on  the
 Wilson graph} \foot{ The idea of letting the branch
locus intersect the Wilson
graph has also appeared in work of Kostov \refs{ \Kos}. } .
For a map with branch points on the graph, the image of
 the boundary $\partial \Sigma_W$
is deformed away from the branch point to the right of
the oriented Wilson loop. The image of the deformed Wilson loops
is an infinitesimal   deformation  $\amalg \cal {G }^{\prime}$  of the graph
 $\amalg \cal { G }$. With these prescriptions, the worldsheets
are { \it always smooth } and
 the lifts of curves in the target space to the  worldsheet
boundary are  unambiguously defined.  $\spadesuit $

The conjecture was proved for non-intersecting
 Wilson loops in \CMROLD .
The first main result of this paper is to prove it for  intersecting loops.

The proof will involve  a derivation of the chiral large $N$ expansion
in the form of a delta function over symmetric groups,
using exact answers of \Witdgt\ and \Ru .   The answers are
expressed in a diagrammatic form, which simply generalises the diagrams
we introduced in sections 2 and 3.  We hope to have given sufficient detail, in
sections 5-9,
 to convince the reader that
the geometric properties alone  determine
the chiral zero area  expansions.

\newsec{ Non-intersecting Wilson Loops}

\subsec{ Covering space geometry}

We describe some implications of conjecture
 3.1 when the Wilson loops
are non-intersecting.
The worldsheets contributing to the Wilson loop
 expectation value
have boundaries which  map to the Wilson
graph according to the
 specified conjugacy classes.
For a map satisfying this condition, there can
 still be in the
inverse image of the Wilson graph, curves
 which are not at the boundary
of the worldsheet $\Sw$ , but  in the interior.

\ifig\fbifi{Locally the covering map looks like this.
Above $\Gamma^r$ there are only interior curves.
Above $\Gamma^l$ there are interior and boundary
curves. }
{\epsfxsize5.0in\epsfbox{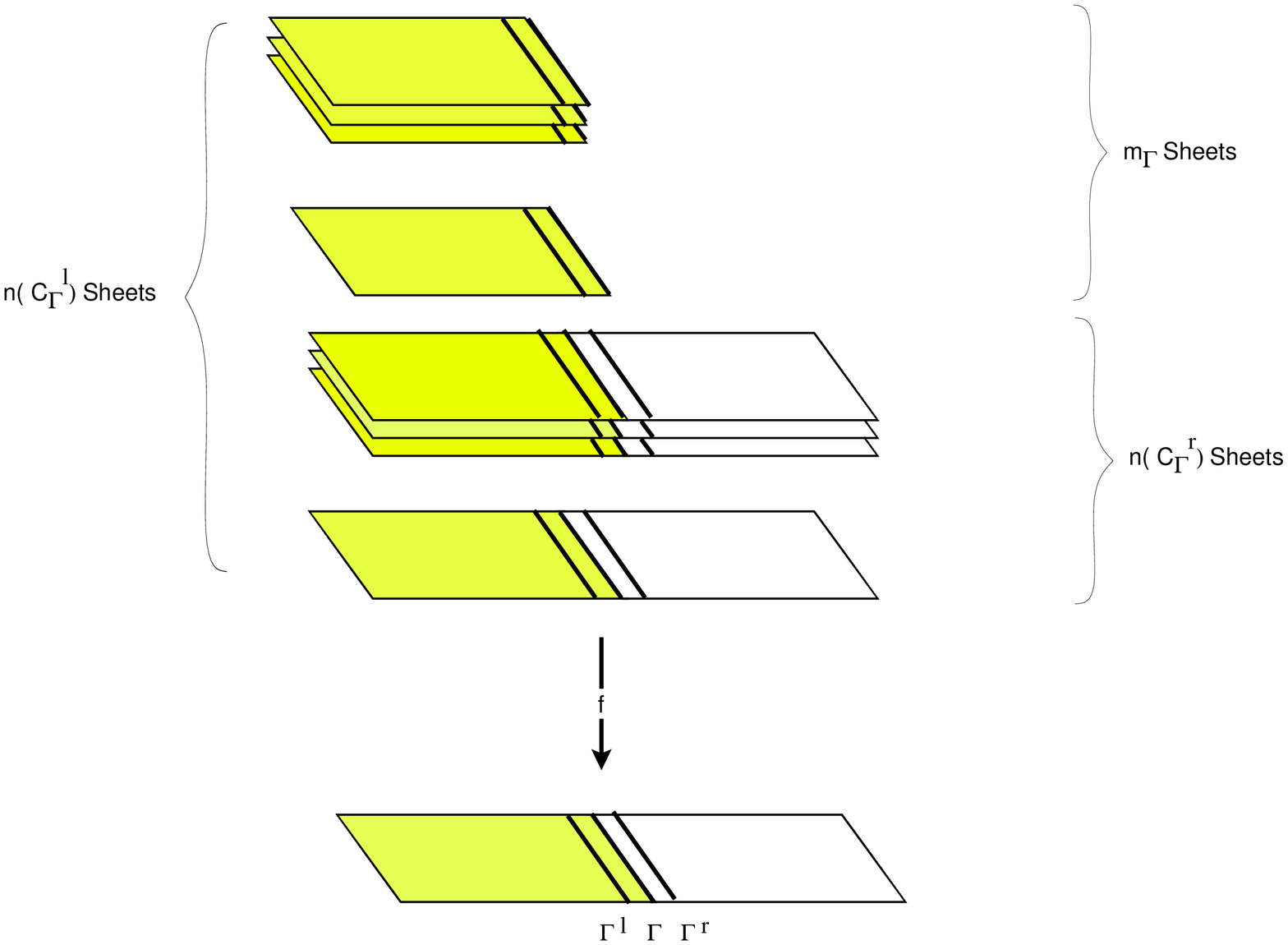}}

  Inside each component $\Sc$,  the maps restrict to branched covers.
 In the inverse image of each  Wilson line there are
  $\vert \vec k_\G \vert = \sum_{j} k_{\G}^{(j)} $  circles which
are worldsheet boundaries. The map from this
union of circles is determined by the vector $\vec k_{\G}$:
there are $ k_{\G}^{(j)} $ circles which
map with winding number $j$, so that the number of
 points belonging to worldsheet boundaries
 in the inverse image of each point on a
Wilson line $\Gamma$, is $n (\vec k_\G)= m_\G$.
There can be an arbitrary number
of circles which are  in the interior of the
worldsheet.
Since the map has to be orientation preserving on the components,
$\Sc$, and on the boundaries,
 the degree of the map is higher to the left of $\G$.
We have:
\eqn\match{
n(\cgp)= n(\cgm) + m_\G,
}
as illustrated in \fbifi .

\ifig\ftrvl{Figure showing  Wilson loop on sphere with handles attached :
 Gluing condition relates permutations
seen from different basepoints }
{\epsfxsize4.5in\epsfbox{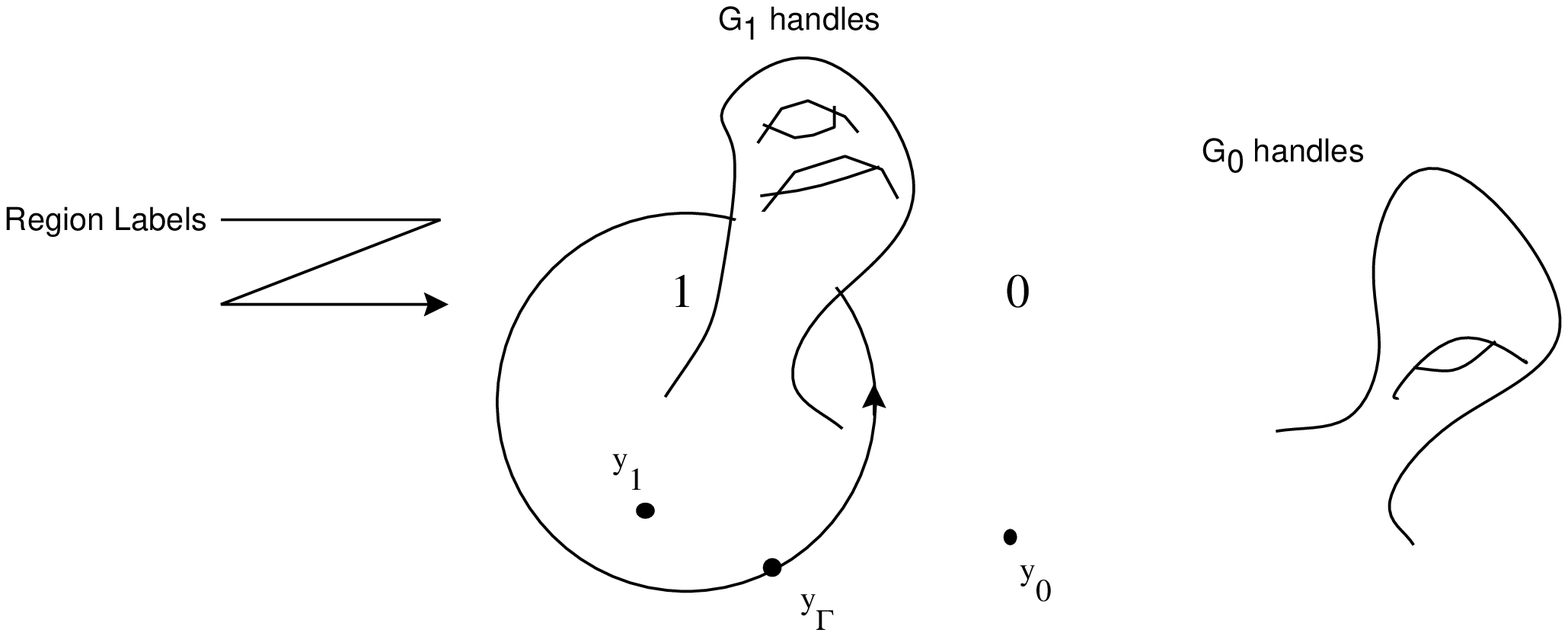}}

\subsec{ Description in terms of homomorphisms into symmetric groups}

We can apply  ideas from the theory of branched covers
 to count equivalence classes of these maps
in terms of homomorphisms into symmetric
groups. This is done
formally and for any non-intersecting Wilson
 loop in  \CMROLD .
We will summarise  the main points here,
as applied to  the example of a single
non-intersecting loop, on a sphere with
  $G_1$ handles attached to  region $1$
on the left  and $G_0$ handles
 attached to region $0$ on the right (\ftrvl ). From \match\
we have
\eqn\matche{ n_1 = n_0 + m_\G.  }

Given any map onto $\Sigma_T$  of the
 kind in Conjecture 4.1 ,
we can construct a sequence of
permutations satisfying some conditions.
Choose a basepoint $ y_c$ in region $\Sc$ , and label the
$n_c$ points in its inverse image.
By lifting  paths in non-trivial homotopy
 classes in each region, we pick up permutations
satisfying a condition:
\eqn\cond{  v_1(c)\cdots v_{L_c} (c)
 \prod_{i=1}^{G_c} [s_i(c),t_i(c)] u_c^{-1}   =1.   }
The $v_i$ are permutations representing paths
 which wrap around branch points
 (at zero area they come from expanding the $\Omega_{n_c}$
factors),
  the $s_i$,$t_i$ represent paths  around the handles;   $u_1$,  $ u_0$ are
permutations along  $\G$ as seen from a basepoint in the regions $0$ or $1$.

 There are additional gluing conditions across each
 Wilson line, since the permutation of all the sheets in the inverse image
of the Wilson line can be described in two ways.
Approaching from the left (with more sheets)
there is a  permutation $u_1 \in  S_{n_1}$ obtained by
 using the basepoint in region $1$.
Choose a basepoint $y_\G$ on the Wilson line. We
 also choose a path
from $y_0$ to $y_\G$ and label the endpoints
 $1$ to $n_0$: then we label
the remaining $m_\G$ points
 (which lie on worldsheet boundaries) $n_0+1$ to
$n_0+m_\G=n_1$.
 We can combine $u_{0}.\sigma_{\G}$
into a permutation in $S_{n_1} $.  These are
equivalent ways of describing the same permutation
 from the point of view of two basepoints, so they
 must be related by a conjugation
in $S_{n_1}$. We have the second gluing
condition at each Wilson line: there must exist
$\gamma \in S_{n_1}$ such that
\eqn\matchii{
u_1 =  \gamma u_{0}. \sigma_\G  \gamma^{-1}.  }
This permutation $\gamma$  is obtained by lifting a path
connecting $y_{1}$  to $y_{\G}$.

Conversely given a set of permutations satisfying the
 above conditions
we can reconstruct the cover. First construct covers
of  the  regions $0$ and $1$,
following \Ez :  Cut open the target
along the generators of
$\pi_1(\Sc  - { \hbox{branch points}} )$ and take $n_c$
labelled copies of this cut surface.
Glue according to the permutations $v_c$, $s_c$, $t_c $ etc.
Next, glue together
regions $0$ and $1$ along the common boundary $\CG$,
and correspondingly
glue together the covering surfaces according to the
permutation $\gamma$. This means that
the  $i  $'th  sheet  on the left is glued to
 the $ \gamma (i)$ 'th on the right.

Equivalent covers which are related by a homeomorphism
 $\phi$ such that $f_1= f_2\circ \phi$, lead to permutations
which are related by conjugations, and there is a $1-1$ correspondence
between equivalence classes of covers and equivalence classes of
permutations describing the covers (see \CMROLD ).

\subsec{Chiral non-intersecting Wilson loops---
 qcd answers and geometric interpretation}

We will now write the  \ymt\ answer for
 this  Wilson loop and then
 express it in a diagrammatic form from
which the covering space geometry can be read off.
This exercise will be useful in understanding the
more complicated intersecting Wilson loops.
\eqn\chwisp{ \eqalign{
 \biggl\langle{   \vert T_{ \kG} \vert  \over {m_\G ! }  }
 \Upsilon(\kG,\G) \biggr \rangle \cr
& =  \sum_{n_0, n_1} \sum_{u_1 \in S_{n_1} }
 \sum_{u_0 \in S_{n_0} }
\delta_{n_1, n_0 + m} \cr
& {1\over {n_1!}} {1\over {n_0!}}
\delta ( F_1 (n_1) u_1^{-1} ) \delta ( F_0 (n_0) u_0^{-1} ) \delta
  \bigl ( \gamma  u_1  \gamma^{-1}  (u_0 .  \sigma_{\Gamma} )  \bigr) \cr
  &= \sum_{ n_1, n_0} {1\over {n_1!}} {1\over {n_0!}}
 \delta \bigl( \gamma F_1 (n_1) \gamma^{-1}  F_0 (n_0 ). \sigma_{\Gamma}
\bigr)
 } }

\ifig\fchnib {Delta function for single chiral nonintersecting Wilson loop. }
{\epsfxsize 4.5in\epsfbox{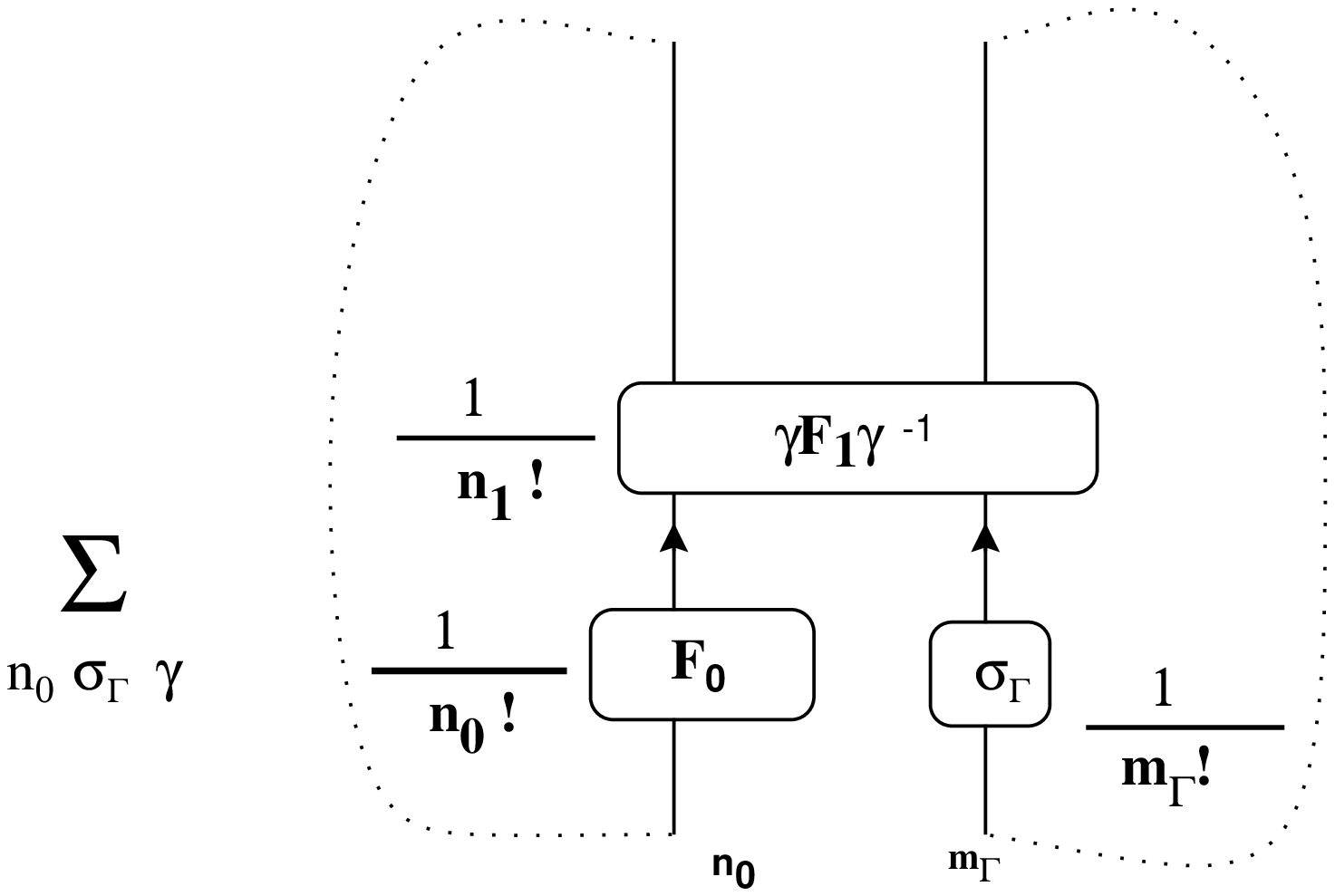}}

We recognize the first expression from  \cond\ and \matchii.  In the second
expression,
we have done the sums over $u_1$
and $u_0$ to get a simple formula,
which still can be interpreted geometrically.
We can express this in diagrammatic
form (\fchnib) as we did for the partition function
 for manifolds with and without boundary.

The $F_0$ factor acts on the first $n_0$ numbers (which correspond to
the inverse image of a point on the Wilson graph).
It gives permutations representing
 branch points or non-trivial loops around
handles in region $0$.  $F_1$
 contains permutations obtained by lifting closed
paths starting from the
basepoint  in region $1$. Since the degree of the
map in region $1$ is $n_1$,  all
the $n_1$ integers in the diagram can be permuted
 by
$F_1$. The conjugated $F_1$ gives
monodromies obtained from the basepoint
on the Wilson loop. Restricting the delta function
 to the first $n_0$ numbers,
we see that they label points lying on sheets
covering the target over both  regions $0$ and $1$.
After  picking up permutations along all the
 generators of $\pi_1$
we get a trivial path, which is mapped to $1$:
following the lifts
starting from each of the $n_0$ points,  of  all the generators of $\pi_1$
associated with cycles  in regions $0$ and $1$, we get
the trivial permutation.
Restricting  the delta function to the
last $m_\G$ numbers we see that they correspond to
sheets covering region $1$ only, as they are only acted on by $F_1$.
The product of paths around the branch points in region $1$ and
 the appropriate commutators for the handles in region $1$ is homotopic to
a path around the Wilson loop.
So travelling along the inverse of the Wilson loop and then along all
the non-trivial paths in region $1$,    we get the identity in $\pi_1$
 which maps to the identity
permutation.   This is exactly the condition we
read off by restricting to the $m_\G$ strands.

Note that we may construct the covers by starting with
 a cover of constant degree
$n_0+m_\G$
of the whole surface $\ST$. In region $0$, the first $n_0$
 sheets can receive arbitrary permutations from the
branchpoints etc.
coming from the $F_0$ factors. The last $m_\G$ are
 permuted by a single branch point
whose cycle structure is  $[\vec k_\G ]$. To get the
 correct degree in each region, we cut out the
 $m_\G$ sheets from the outside. This construction
 gives the same kind of cover as that
discussed in the previous subsection. The role of the
 branch point in the outside region permuting the
 $m_\G$ sheets is to produce the correct winding of the Wilson loop.
These ideas will be useful in the intersecting case.

The above discussion  shows that the
discrete data (sums over permutations) which
enter the Wilson average count covers
predicted by conjecture 4.1. To see that we
get Euler characters, of these spaces we
use the fact that the $F_c$ factors contain
$\Omega_{n_c}^{\chi (\Sc) }$. Thus contributions
with $L_c$ branch
 points in $\Sc$ are weighted by $\prod_{c}
 \chi ( \Sc, L_c)$, the product of Euler characters
of configuration spaces of $L_c$ points in $\Sc$.
Using the fibration of Hurwitz-Wilson spaces
over configuration spaces shows that we
 have Euler characters of these spaces.

\noindent
{\bf Remarks :}
a) In the case of non-intersecting loops
the Euler character of the space of branched
covers is the same whether or not
the branch points can lie on the Wilson graph.
This is because the Wilson graph has Euler
 character zero in this case. A consequence
is that the delta functions contain
$\Omega $ factors only inside the $F$ factors
for the different regions.
This will no longer be true in the intersecting
case, where there will be extra
$\Omega$ factors.

b) { \bf Relation to closed strings. }
If we  multiply the Wilson insertion by
 $N^{ \vert \vec k_\G \vert  - 2m_\G G_0  }$,
the Wilson average has the interpretation of
counting  covers of constant degree with the
permutations having some special structure
 (given by the delta function).
So Chiral Wilson averages compute  Euler
characters of subspaces of
ordinary Hurwitz spaces.

\newsec{\bf  Algorithm for computing the
 chiral expansion for arbitrary Wilson
Loops  }
In this section, we will  show how to
 construct a delta function
when  the Wilson graph  is a connected one living on a sphere
with $G_c$ handles attached to  $\Sc$ .
The generalisation is simple  (see comment at the end of this section).

\ifig\fedco{ Edge contraction does not change the fundamental group. }
{\epsfxsize2.0in\epsfbox{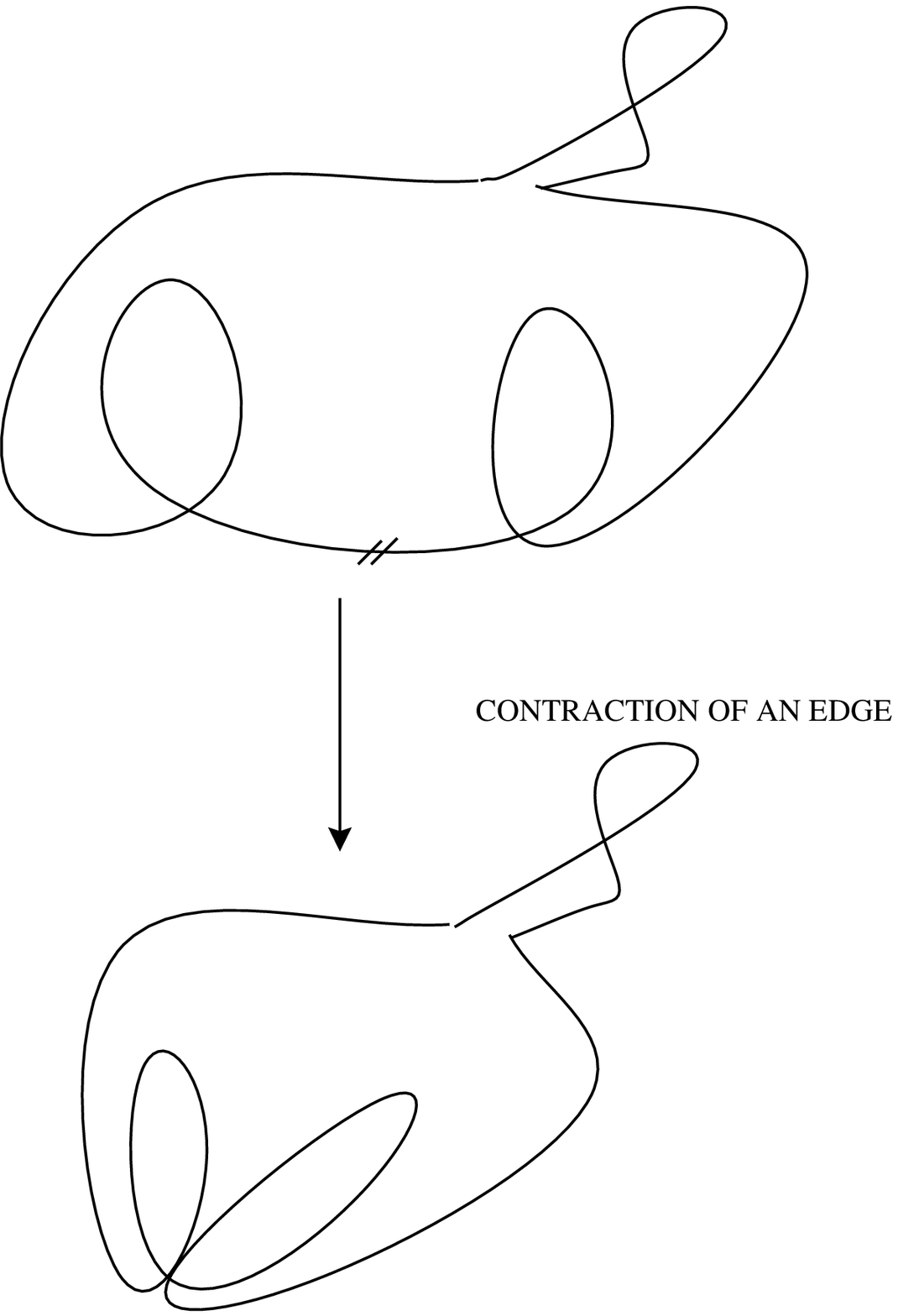}}

\ifig\freg{ Fundamental group of flower is a free group. }
{\epsfxsize1.8in\epsfbox{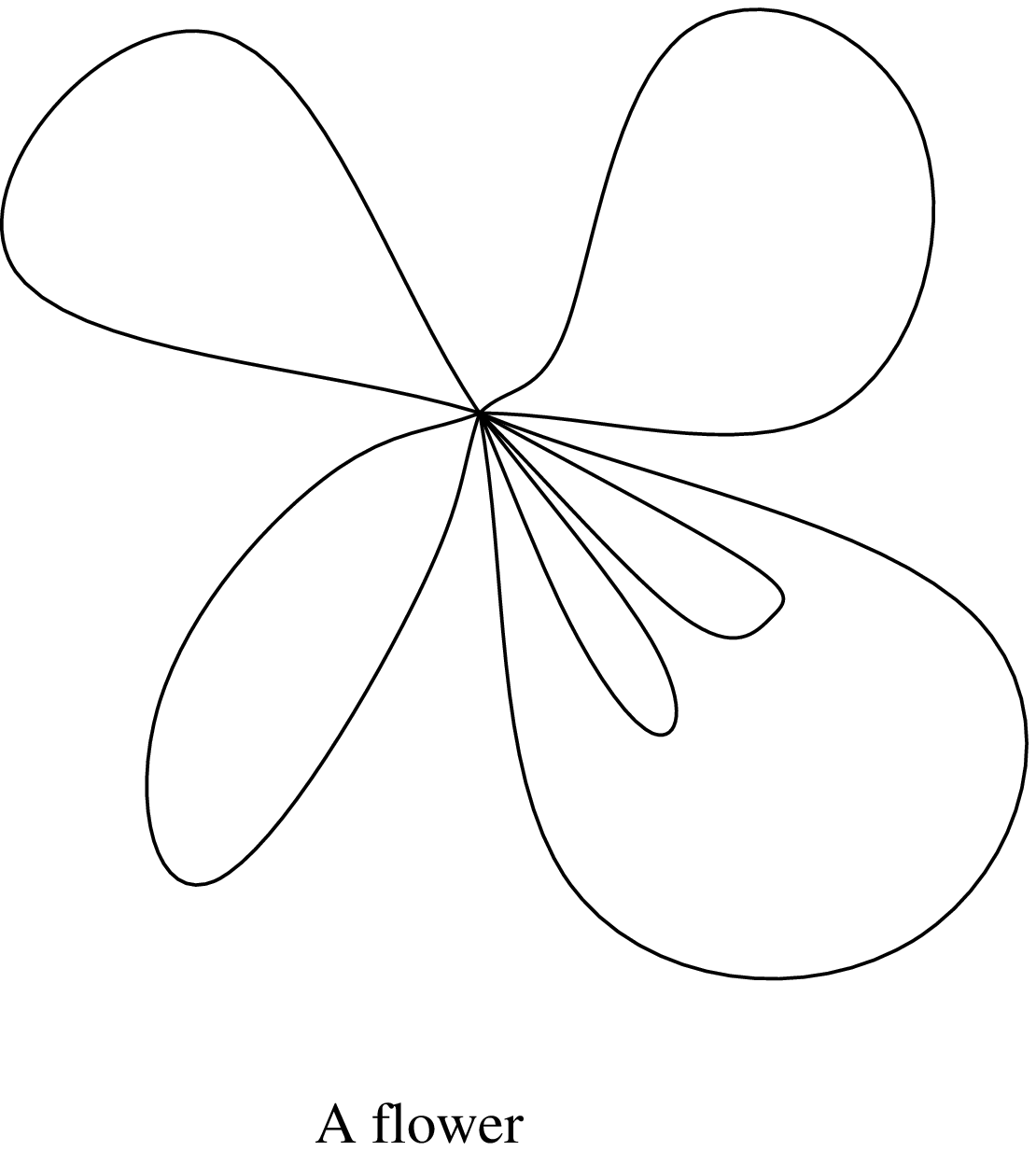}}

We first  express the Wilson loops $\{ \G \}$
 in terms of the fundamental group
of the Wilson graph. A set of generators of the Wilson
 graph can be written down in $1-1$ correspondence
with the edges. In terms of these generators the
 fundamental group is not free.
 The relations can be used to
 set to one any generator corresponding to
an edge connecting distinct points. Thus, for each such edge,
 we can
 define a contraction of the graph, which
 is a map sending it to another graph with
 the open edge removed and the two vertices at the end  of
 the edge identified, see \fedco.
Under this map the fundamental group does not change \seif .
We can continue this contraction process until the graph
 is mapped to a graph with one vertex and all  edges
incident on that vertex. We will call such a graph a {\it flower}.
 Its fundamental group is
a free group on $g$ generators.
In  \freg\ we have drawn an example.
Suppose $c$ in \ccomp\ runs from $0$ to $R_I$.
Choose a region,  say the one labelled $0$,
which we call the outside region, and the remainder
are called interior regions.
A convenient set of generators is the set
of  {\it oriented boundaries of the interior regions}
( $W_c , c=1, \cdots R_I $  )  which we call
the {\it region paths} .
The boundary of the outside region is not independent (proof in sec. 7).
The contraction procedure allows a simple way
 to write the Wilson loop in terms
of independent generators.

\medskip
\noindent
{\bf Step1}:   Choose
some set of edges to be contracted,
which allows us to map the Wilson graph to a flower.
 Then write the remaining edges in terms of the oriented  region
paths. Write each Wilson loop in terms of the chosen generators.
Corresponding to  this word in the fundamental
group, draw  a closed strand,  joining a sequence of generators.
Label the  strand  by an integer $n(\vec k_{\G} )$ called its multiplicity.

\ifig\fdef{ Deforming a  strand carrying generators
of the fundamental group. }
{\epsfxsize2.8in\epsfbox{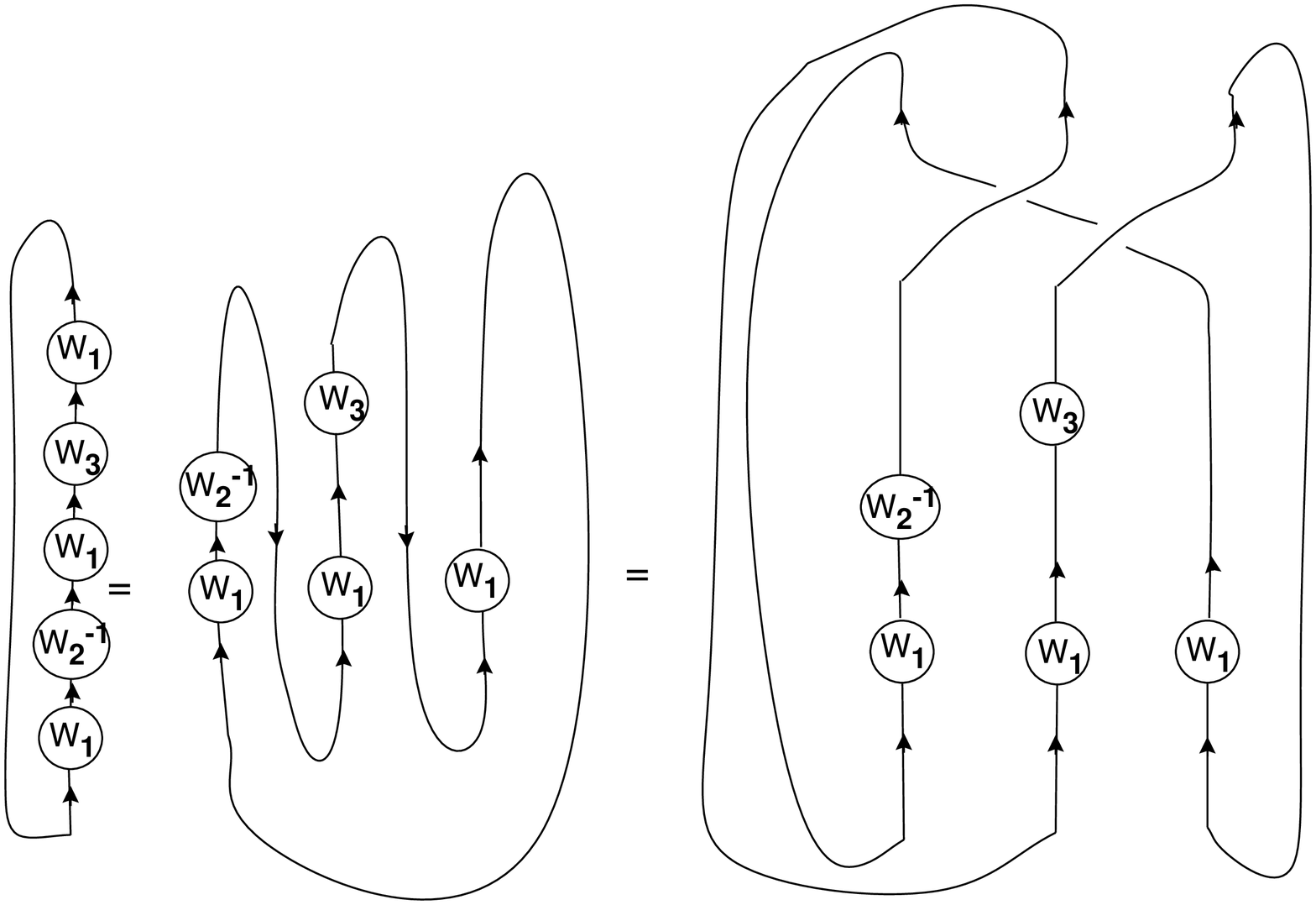}}

\medskip
\noindent
{\bf Step 2 }: Choose an ordering of the generators.
 Set the array vertically with the first generator at the
bottom. If a generator occurs more than once,  then
deform the strand to obtain an ordered set of generators
 together with some cyclic permutation,  see \freg .

\medskip
\noindent
{\bf Step 3} : Repeat the procedure for all the
Wilson loops and the inverse of the oriented boundary
of the outside region    and arrange them
 so that any given generator occurs at the same
 level in all strands.
For the boundary of the outside region the
 label on the strand is equal to some integer
 $n_0$ which is summed over. Insert at the bottom
 of this strand
 the factor  $F_0 (n_0)= F( G_0, A_0, b=1, n_0)$
the $F$ factor for the
 outside region, its explicit form is given in \Fbdy .
The first entry on the strand corresponding to a
Wilson Loop is
${1\over {n(\vec k_{\G} )   ! }} \sum_{\sigma_{\G}
 \in S_{n(\vec k_{\Gamma}) }} \sigma_{\G} $.
 The strands are placed in some arbitrarily chosen
 order  from left to right. At the end of this  step, the central part  of the
diagram
contains upward directed strands incident on $W$'s and the top
 and bottom of these strands are tied as in \fdef.

\bigskip
\noindent
{\bf Comments:  }

\noindent
a) To simplify the diagram it may be
 convenient to change the expressions of the
Wilson loops independently  to others in the same
 conjugacy class of the fundamental group, by
conjugating with some generator.

\noindent
b) It is  convenient to remove redundant
expressions like a  generator followed by its inverse.

\noindent
c) It follows easily by inspection of the flower (see \freg ) ,
that the oriented boundary of the outside region can be
written as a product of the inverse of the all the region
paths  in some order,
each region path occuring once.

Starting with this diagram we will replace the
generators of the fundamental group
of the graph with
sums over  permutations acting on a set of numbers, to get a diagram
generalising     \fonpf , \ftwpf , \fchnib , and \fgnsmb .
All permutations will act on
some subset of the integers ranging from  $1$ to the sum of
all the multiplicities of the strands appearing in the diagram.
A pair of strands with multiplicities $m_1$ on the left and $m_2$ to
the  right
incident on a generator $W_c$ is equivalent to a a single strand
 of multiplicity $m_1+ m_2$ incident on $W_c$. The combined
 strand carries a natural action of $S_{m_1}$ on the first $m_1$
 strands, of $S_{m_2}$ on the second set of $m_2$ strands,
and of $S_{m_1+m_2}$ on the combined strand.

\ifig\wbastwo{The basic weight for the delta function}
{\epsfxsize5.0in\epsfbox{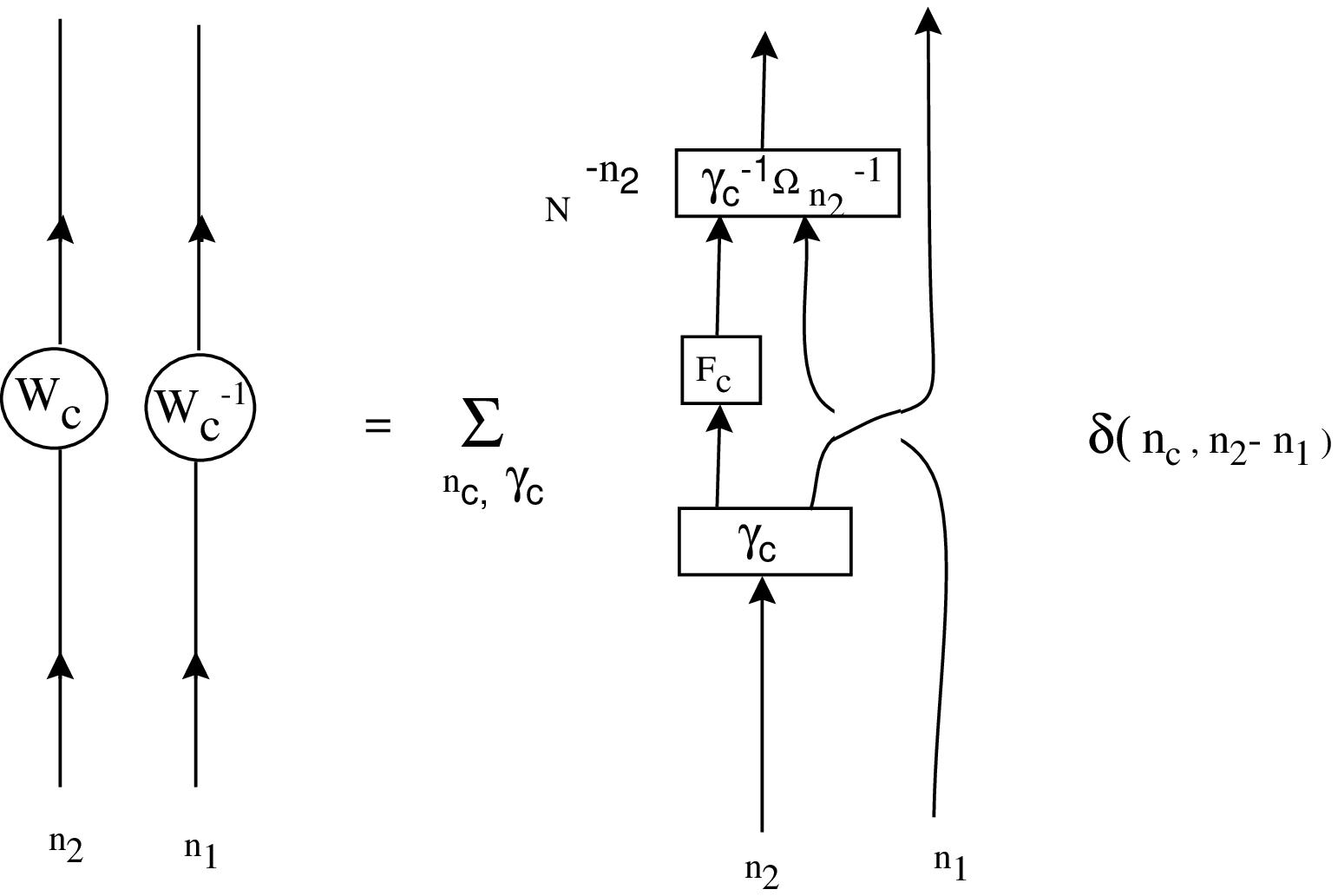}}

\ifig\fbws{Basic weight: simple case}
{\epsfxsize4.0in\epsfbox{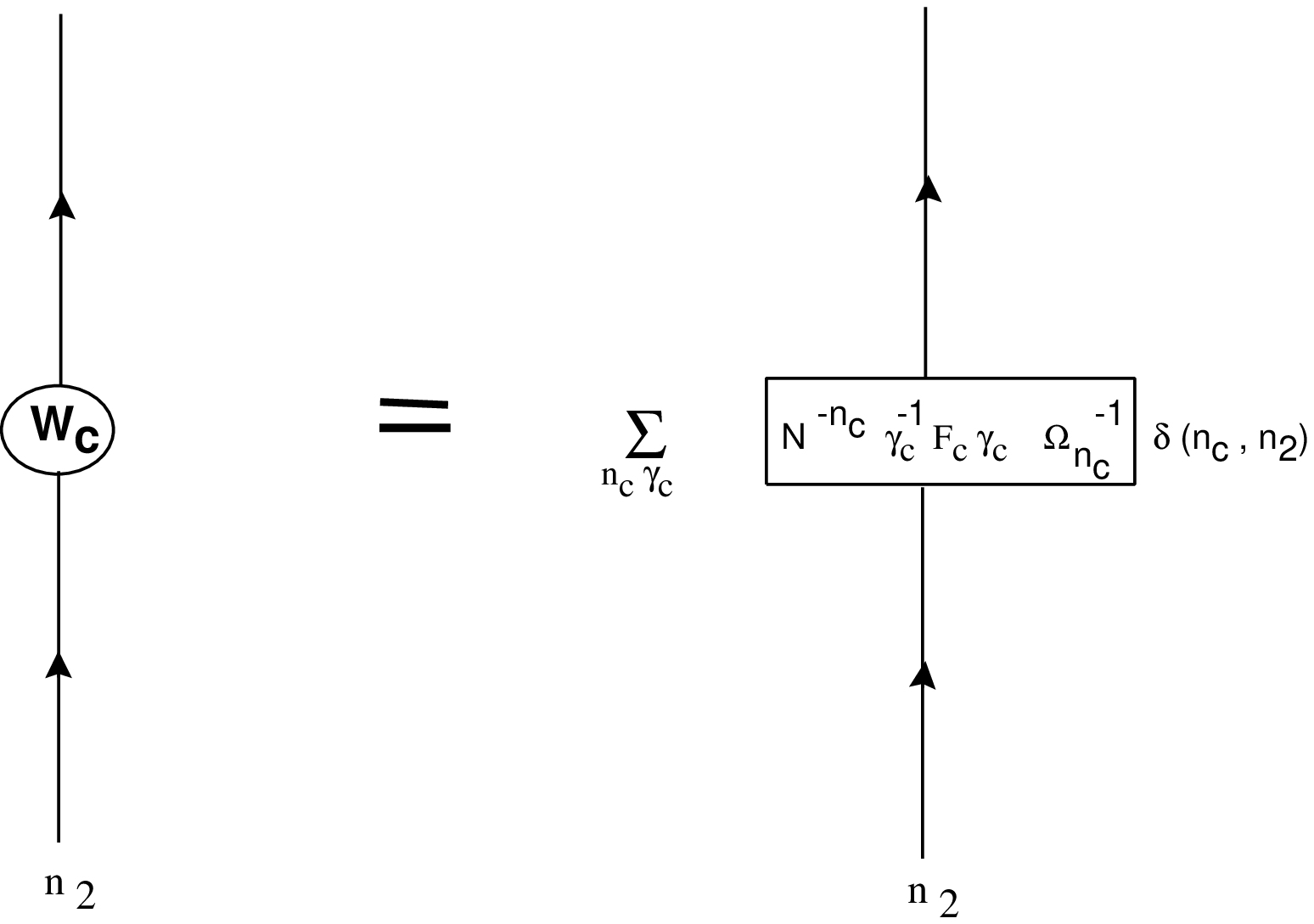}}

Focusing on the occurrences of  a given generator $ W_c$
and its inverse,  the diagram looks
 like the left hand side of  \wbastwo . We have used
 the usual shorthand of combining all the
strands incident on $W_c$ (or $W_c^{-1}$ ) into one strand.

\noindent {\bf  Step 4:}
We will replace this by a sum over permutations acting on
the labels carried by the strands as in the RHS of \wbastwo .
 Because of the
 delta function which imposes $n_2= n_c + n_1$, we have  used the
isomorphism between a strand of muliplicity $n_2$ and a pair of strands
of multiplicities $n_c$ and $n_1$.
The choice of letting the $F_c$ act on the first $n_c$
as opposed to the last say is done for convenience,
and it does not change the result
because of the sum over $\gamma_c$. In  case  there are
only incidences of $W_c$
the vertex simplifies to the one in \fbws .

\noindent {\bf  Step 5: }
At the top of the diagram,  we place a  factor
 $N^n \Omega_{n}$
where $n$ is the total number of strands
going upward counted with multiplicity,
and we replace the solid lines
 joining top to bottom by dotted lines. This
 is the delta function.  It is always a sum over
 homomorphisms from fundamental group
of $\Sigma_T - \{ \hbox{ points} \}$ into symmetric groups,
with certain restrictions on the allowed
 permutations which are  in the diagram.

\noindent
{\bf {Comment:  }}

\noindent
In the general case of  Wilson loops  in a closed manifold, the exact answer
 is constructed by starting with delta functions of
for manifolds with boundary obtained by cutting along the Wilson loops.
These  boundary permutations also  enter a complicated delta function of the
kind we
discussed, with embeddings of symmetric groups into a large one,
determined by the Wilson loops.
By solving for the boundary permutations  we can combine
into a single delta function containing $F$ factors (see appendix).
When we have  Wilson loops in a manifold with boundary, the
 single delta function will contain $F$ factors and  permutations associated
with the real boundaries ( not those associated with cutting along the Wilson
loop).

We will now describe the derivation of the rules
 from the group integrals and outline a proof that
expectation values for chiral intersecting Wilson
 averages compute the Euler character of Hurwitz-Wilson
 space. We will illustrate this in a simple example.

\newsec{ Derivation of algorithm }

We will use the exact results on Wilson averages
 in terms of integrals over the gauge  group described in section 4.

\ifig\fbdrone{vectors and operators.}
{\epsfxsize3.5in\epsfbox{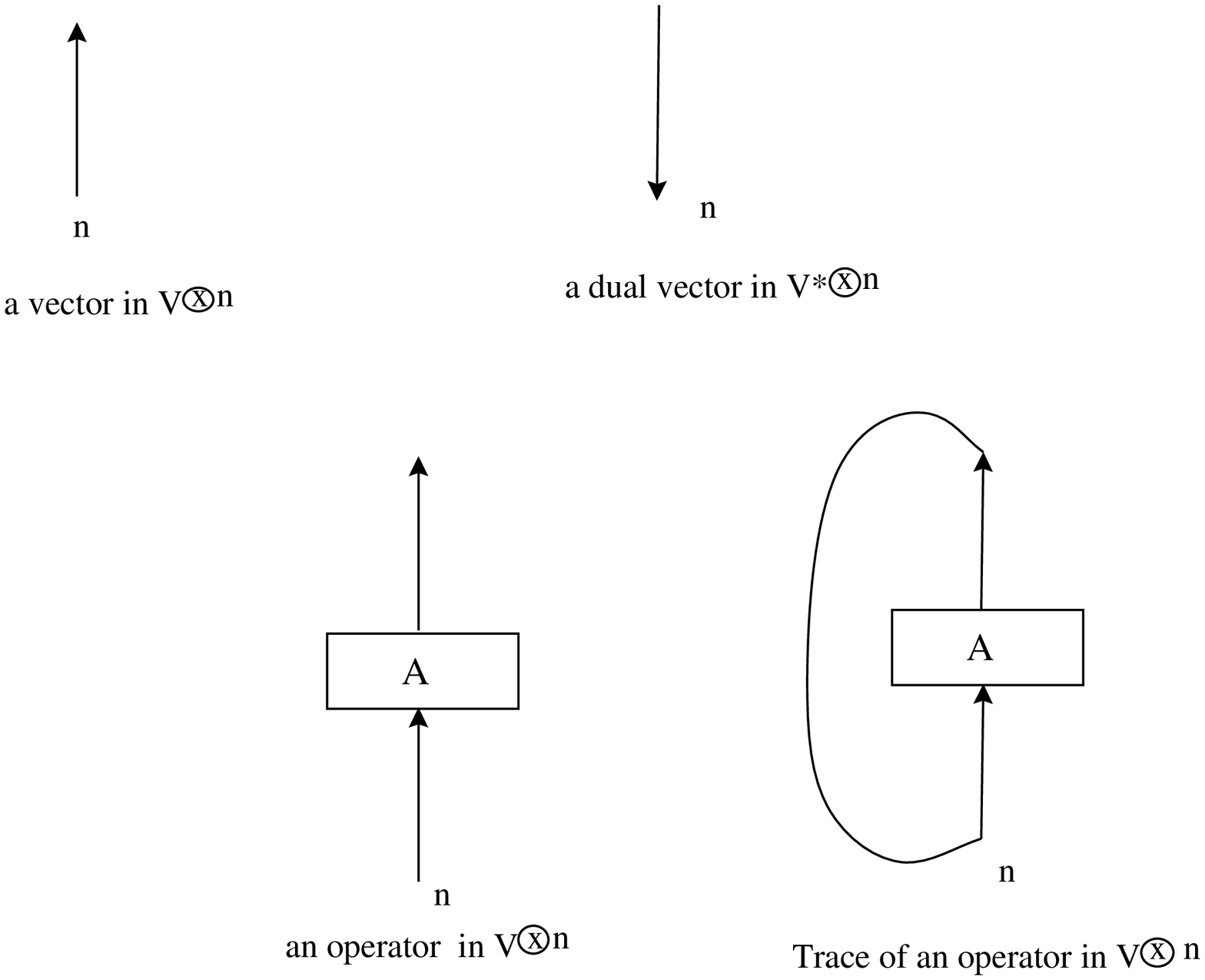}}

\ifig\fbdrtwo{vectors in tensor space: a basic isomorphism }
{\epsfxsize2.6in\epsfbox{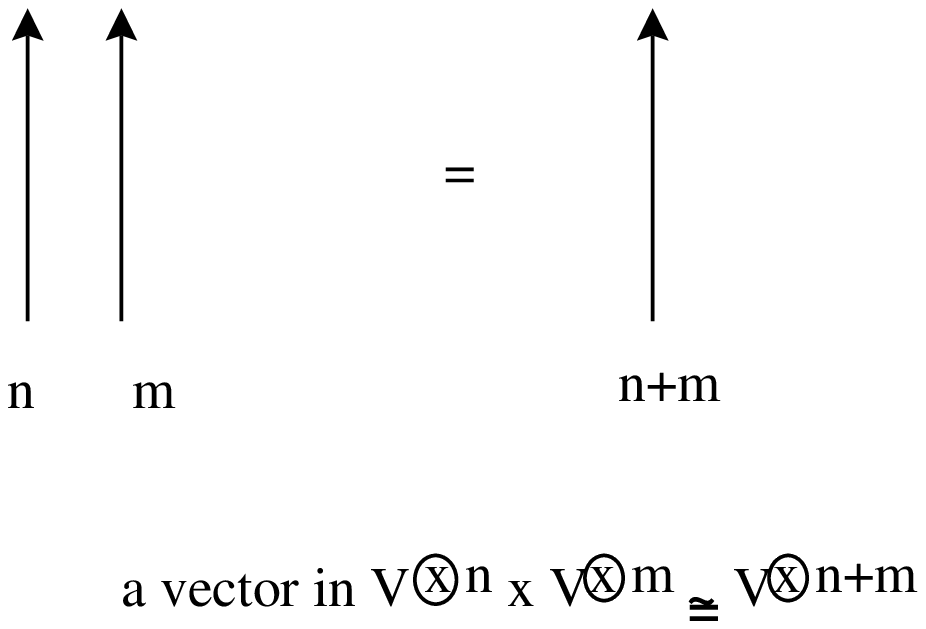}}

\subsec{ From group integrals to braid diagrams}

Three simple ingredients go into the derivation of these rules.
First  we use a basic
relation between gauge invariance and fundamental groups.

Second  we make extensive use of a diagrammatic
 approach to  linear algebra in tensor spaces
 (more generally to Braided Tensor Categories)
 which may be familiar from  many contexts , and
has been useful
before in algebraic constructions of topological
 invariants, e.g knot
and $3$-manifold invariants, and for proving
 properties of  RCFT. \refs{ \mrsb , \RT , \Resh }

Finally we use a few integrals which can be derived from  Schur-Weyl duality.

\ifig\fbdrthr{Linearising Group Multiplication in tensor space. }
{\epsfxsize2.6in\epsfbox{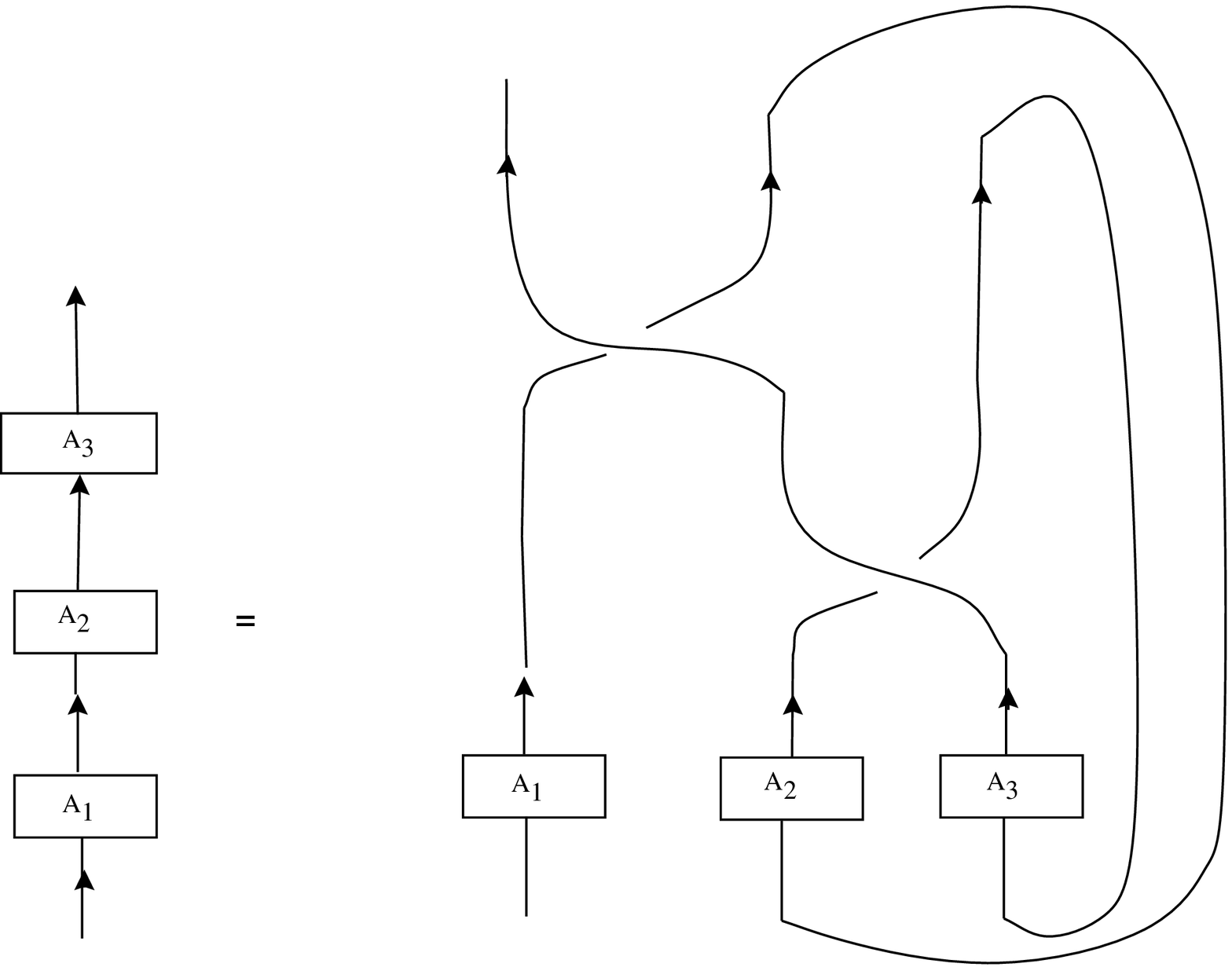}}

The  basic diagrammatic presentation of vectors,
endomorphisms and traces
is shown in \fbdrone . A basic isomorphism
between tensor spaces
is shown in \fbdrtwo . A very useful identity
is shown in
\fbdrthr . Writing it out, using summation
 convention for repeated indices,
we have
\eqn\ident{\eqalign{
 RHS &=  (1\otimes tr \otimes tr ) P_{(123)}  A_1
\otimes A_2 \otimes A_3 \vert e_{i_1}> \otimes
  \vert e_{i_2} > \otimes \vert e_{i_3}> \cr
 &= (1\otimes tr \otimes tr ) P_{(123)}
(A_1)^{j_1}_{i_1}(A_2)^{j_2}_{i_2}(A_3)^{j_3}_{i_3} \vert e_{j_1}>
\otimes  \vert e_{j_2} > \otimes \vert e_{j_3}>\cr
&= (1\otimes tr \otimes tr )
 (A_1)^{j_1}_{i_1}(A_2)^{j_2}_{i_2}(A_3)^{j_3}_{i_3}
 \vert e_{j_3}> \otimes  \vert e_{j_1} > \otimes \vert e_{j_2}> \cr
&= 1 \otimes <e^{i_2}\vert  \otimes <e^{i_3}\vert
(A_1)^{j_1}_{i_1}(A_2)^{j_2}_{i_2}(A_3)^{j_3}_{i_3}
\vert
e_{j_3}> \otimes  \vert e_{j_1} > \otimes \vert e_{j_2}>\cr
&= (A_1)^{i_2}_{i_1}(A_2)^{i_3}_{i_2}(A_3)^{j_3}_{i_3}
\vert e_{j_3}>\cr
&= (A_3A_2A_1)^{j_3}_{i_1} \vert e_{j_3}>= LHS.\cr } }
The identity has an obvious generalisation to a
 product of $n$ elements:
\eqn\identwo{
(1\otimes tr \otimes \cdots  tr ) P_{(12 \cdots n )}
 (A_1 \otimes A_2 \cdots A_n)
= (A_n \cdots A_2 A_1 ).
}

We perform for each $Z_c (U_1^{(c)}, U_2^{(c)}, \cdots  , U_{b_c}^{(c)} ) $
 the chiral expansion, replacing the sum over all
 representations by a double sum,  over the
number of boxes $n$ and the set
of Young diagrams with $n$ boxes. This will
 yield
\eqn\partint{\eqalign{
Z_c ( U_1^{(c)},U_2^{(c)},\cdots, U_{b_c}^{(c)} )
= \sum_{n_c=0}^{\infty}  {1\over {n_c!}} \delta \bigr
 ( \quad F(A_c, G_c,b_c,n_c) u_1^{-1} \cdots  u_{b_c} ^{-1} \quad  \bigl ) \cr
 \Upsilon_{u_1} (U_1^{(c)}) \cdots \Upsilon_{u_{b_c}} (U_{b_c}^{(c)})   . \cr
}}
The $F$ factor is obtained from \Fbdy .
The loop functions  will be integrated over.
Let us now restrict to the case where each
region $\Sc$ has one boundary. This will bring out the key
features which make the intersecting case
 different from the non-intersecting case.  In this
 case the delta function
will set $F_c$ equal to $u_c$, and the $u_c$ enters
a braid diagram, so we can do the sum
over $u_c$ producing $F_c$ in the braid diagram.

So we now have a product
\eqn\wils{ \eqalign{
<W> = \int
\prod_c \sum_{n_c} {1\over{ n_c!} }
\delta _{n_c} (F_c u_c^{-1} )  \Upsilon_{u_c} ( V_c)
\prod_{\Gamma}\sum_{\sigma_{\Gamma} \in [ \vec k_{\Gamma} ] }
 {1\over {n(\vec k_{\Gamma} )!}}
\Upsilon_{\sigma_\Gamma}  (U_{\Gamma}^{\dagger} ) \cr} }
The integral is over all the edge variables.
 The $V_c$ are expressed
in terms of the
edge variables.

Now the first observation is that following a
standard argument in lattice gauge
theory we can set to $1$  group  variables
 associated with edges  connecting  distinct points.
Recall that a gauge transformation in lattice gauge
 theory can be specified by giving a group variable
 $g_i$ for the $i'th $ vertex. The holonomy $U_{ij}$
is transformed to $g_i U_{ij} g_j^{-1}$.
Consider an edge joining two distinct points, $k$
and $l$ with holonomy $U_{kl}$.
 Let the group variable $g_l$ be related to $g_k$ by
$g_l=g_k U_{kl}$ . This gauge transformation sets
to $1$ the variable $U_{kl}$.
 This specifies a conjugation  for
the holonomies along  the remaining edges which
can be absorbed into the Haar measures for these
variables. We are left then with an integral
 over $U_{kl}$ where the integrand is
 independent of $U_{kl}$. The integral  gives  $1$
 (the Haar measure is normalised such that the group volume is $1$).

This shows that the integral  is unchanged if such
an edge is contracted  until $k$ and $l$ become
a single point. If this deformation is done
 while keeping the areas of all the regions fixed,
the Wilson average is unchanged, e.g
the Wilson loops in \fedco\ have the same
expectation value if the areas are left
 unchanged in the deformation.
Note  that as the contraction is done
 the {\it Euler character of the Wilson
 graph  is not changed}. The contraction
 procedure can be repeated until the only variables left are closed loop
variables.
The process of contracting the edges does not change the fundamental   group
of the graph. In fact this is used in \seif\  to show that
 the fundamental group of a graph is a free group.

Having set to $1$ some set of edges such that the
reduced graph is a flower, we
find it convenient  to choose one region which
we call the
 outside region, $\ST ^0 $,  and  use as independent
variables, the holonomies $V_c$ around the $R_I$
interior regions. The choice of these variables is
convenient for an intuitively clear geometrical
interpretation, but other choices can be made
and have to be made, e.g
 for higher dimensional models. In the case of interest here, let us prove
 that our chosen variables form a complete set . This may be seen by computing
the Euler character of the target in two ways:
\eqn\varnum{\eqalign{  \chi (\ST) &= \chi ( \hbox{graph} ) + \sum_{c=0}^ {R_I}
 (1 - 2h_c)   \cr
&= 2- \sum_{c=0}^{R_I} ( 2h_c) . \cr
 }}
Now use the fact that $\chi ( \hbox{ graph}  ) =1- g$ to find that $R_I = g$.
The holonomy around the outside, can be
written $\prod_c V_c^{\dagger}  $ in some order.
The Wilson averages are also expressed in terms
of the $V_c's$. Using the conventions of
\fbdrone\ we can write the
$\Upsilon_{u_0} ( V_0) = tr_{n_0} (u_0 V_0 ) $ as a
vertical  strand containing  $u_0$ at the bottom
and a sequence of interior  $V^{\dagger}$'s, and
with the upper end of the strand tied back to the
 lower end.  We do the same thing with the Wilson
loops. We arrange the diagram such that
all the $V_c$'s and $V_c^{\dagger }$'s
for a given $c$ occur at the same level.
To arrange this we may need to use the
 fact from linear algebra \identwo , which
has the diagrammatic interpretation \fbdrthr . So we have accounted for steps
$1$ to $3$ of the algorithm in section 7.
At the end of this we have a sequence of
 operators in tensor space, consisting of
permutations, and integrals of the form
 $\int dV_c \Upsilon_{u_c} (V_c)
 \rho_{n_2} (V_c^\dagger) \rho_{n_1} (V_c)$.

\ifig\fbasint{ Basic integral }
{\epsfxsize2.4in\epsfbox{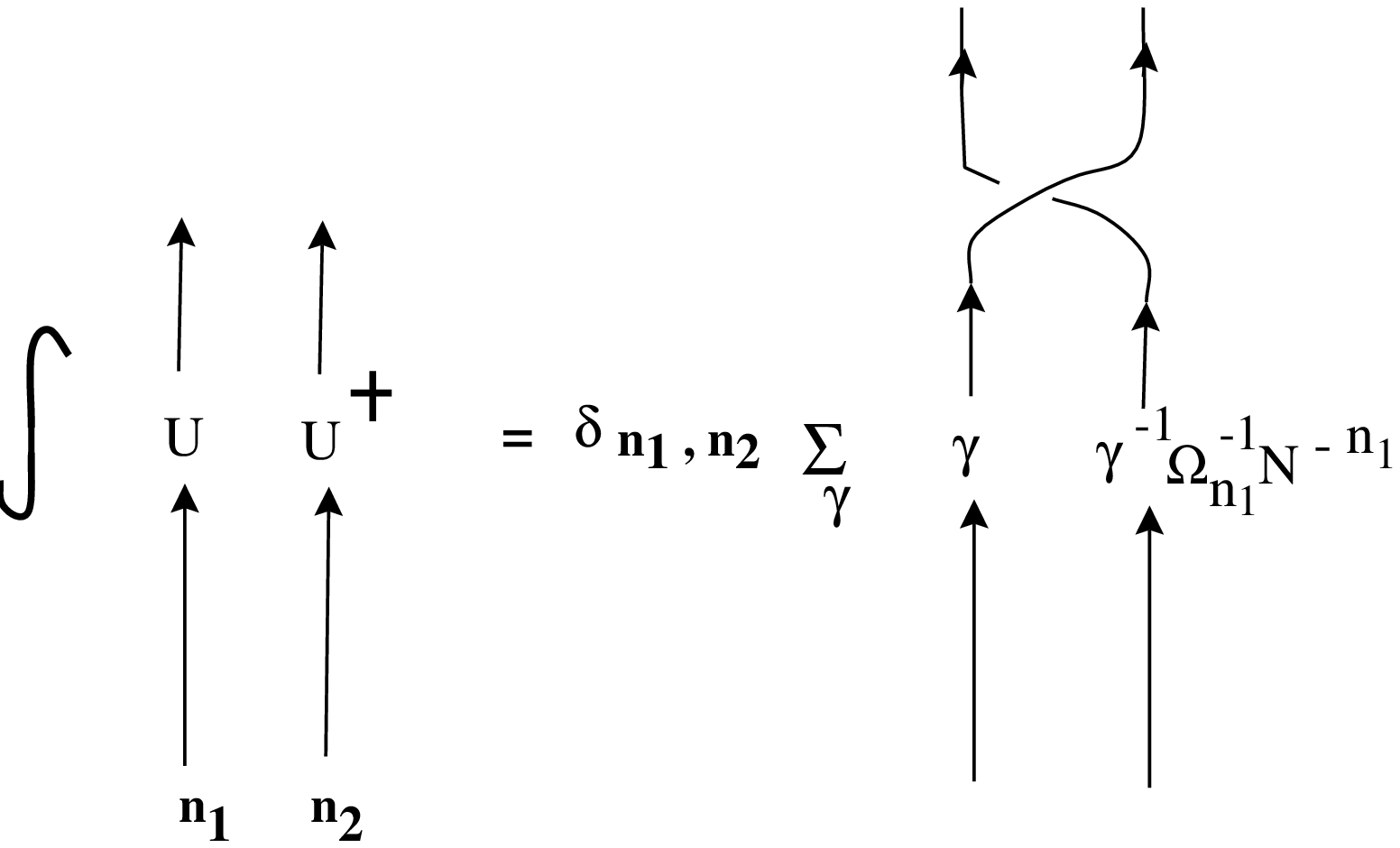}}
\ifig\fbasinti{  Integral derived from basic integral . }
{\epsfxsize2.4in\epsfbox{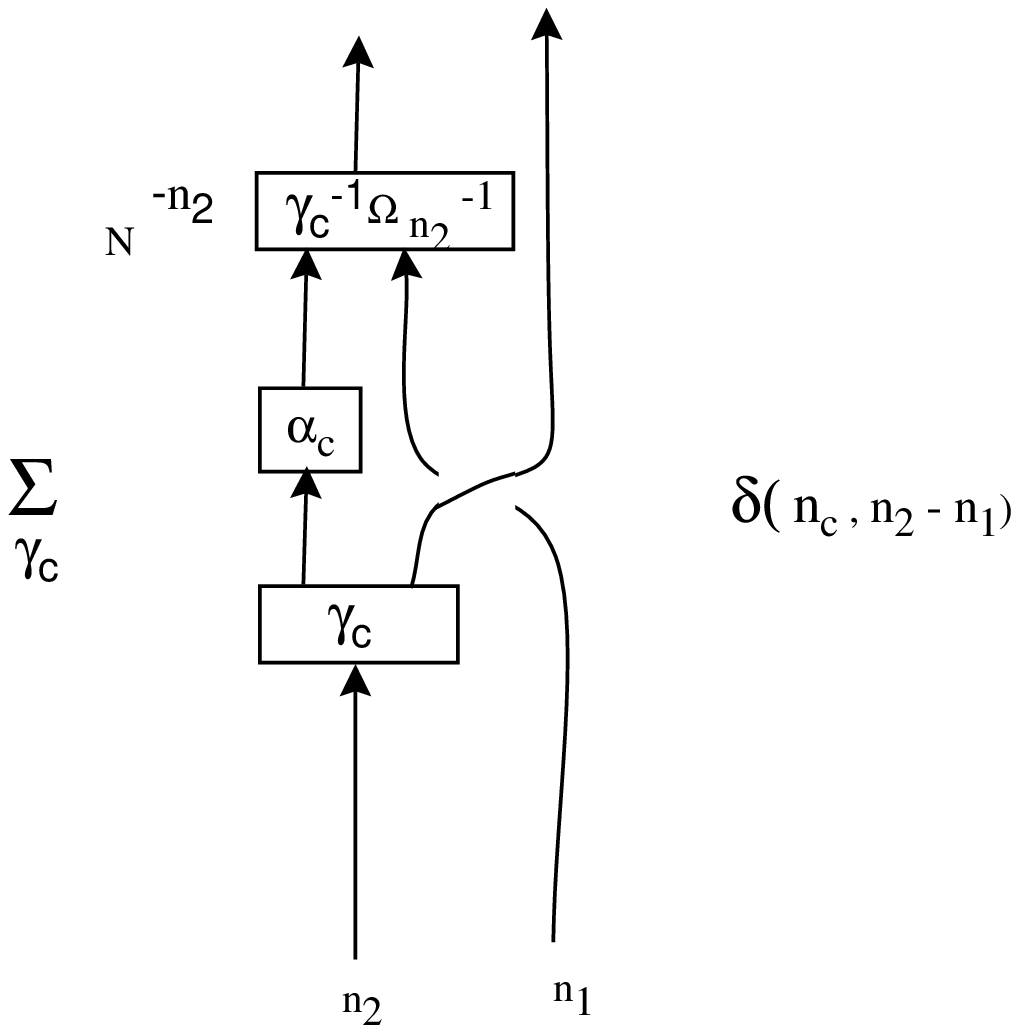}}

This is derived by using  the integral $\int dU
 \rho_{n_2} ( U^\dagger )  \rho_{n_1} ( U)$.
For large $n_1,n_2 \le N$,  it is expressed
 in terms of $\Omega $ factors in \GrTa .
Then  the integral is proportional to
 $\delta_{n_1,n_2}$,
and is expressed in diagrammatic
notation in \fbasint .
The result for $\int dV_c \Upsilon_{u_c} (V_c)
  \rho_{n_2} (V_c^\dagger)\rho_{n_1} (V_c)$
an operator on $V^{\otimes n_2} \otimes
 V^{\otimes n_1}$ is shown in
diagrammatic notation   in \fbasinti .
Note we have used the isomorphism
$V^{\otimes (n_c + n_1)} \cong V^{ \otimes n_c}
 \otimes V^{\otimes n_1} $.
Now $u_c$ also appears in $\delta ( F_c u_c^{-1})$, so after
doing the sum over $u_c$, we recover
  step 4 of section 7.
After doing all the integrals using this formula
 we have a trace in tensor space
of certain permutations. This is converted to a
 delta function using \deltatr . This is step 5 of section 7.

\noindent
{\bf {Comment:  }}
In the general case we have delta functions of the type
we get for manifolds with boundary from the $Z_c$ factor as in \partint .
The permutations entering here also enter more complicated delta functions.
The latter come from doing  group integrals. All the integrals needed can be
expressed
in diagrammatic form using the integrals presented in this section.
Combining the delta functions into a single one proceeds by  simple operations
on braid diagrams, e.g connected sums (see appendix).

\newsec {   \bf  Braid Diagrams  and Branched covers  }

To relate the large $N$ expansion to branched covers,
we will show how to construct branched covers from
the data in the delta function, using  cutting and pasting techniques
of Hurwitz.
We will sketch briefly the converse :
how the symmetric group data can be extracted from the covers.

We will then explain how all the key features
 of the general construction can be understood
in terms of  Euler characters of Hurwitz -Wilson spaces.
 This will involve  showing  that the Gross-Taylor rule,
 and a generalisation,
 has a simple interpretation in terms of Euler characters
 of Hurwitz-Wilson spaces.

The result of applying the  algorithm  of  section 6 to an example
 of intersecting Wilson loops will be written down
 in section 10.
 Its interpretation
 in terms  of branched covers and Euler characters
 of spaces of branched covers will be demonstrated.

 \subsec{  \bf From  Braid diagrams to Branched covers }

The braid diagram  is
just a more elaborate version of \fonpf ,
with the allowed permutations
 having some special structure. Exploting this
observation,   we see
that the easiest way to construct the covers
starting from the braid diagram is to start with a
branched cover of constant degree. Then we
cut along the boundaries of the various regions
 $\Sc$   to obtain the right degrees
(as dictated by the powers of $1/N$, the string coupling constant),
 in the various regions.
The degrees are easily read off from
the braid diagram by inspecting the number of strands
that the $F_c$ factor  acts on.
They also follow from inspection of the
 group integral for each edge, which
 imposes some selection rules on the difference
 in degree between regions
on  the left and right of the edge \GrTa.  We will illustrate
such a construction which starts with a
branched cover of constant degree using usual techniques
of  construction of branched covers
and makes cuts as instructed by the braid diagram to produce
 a cover satisfying the condition \hmtpycl .

\ifig\fexcb{a  Wilson loop and its braid diagram }
{\epsfxsize4.0in\epsfbox{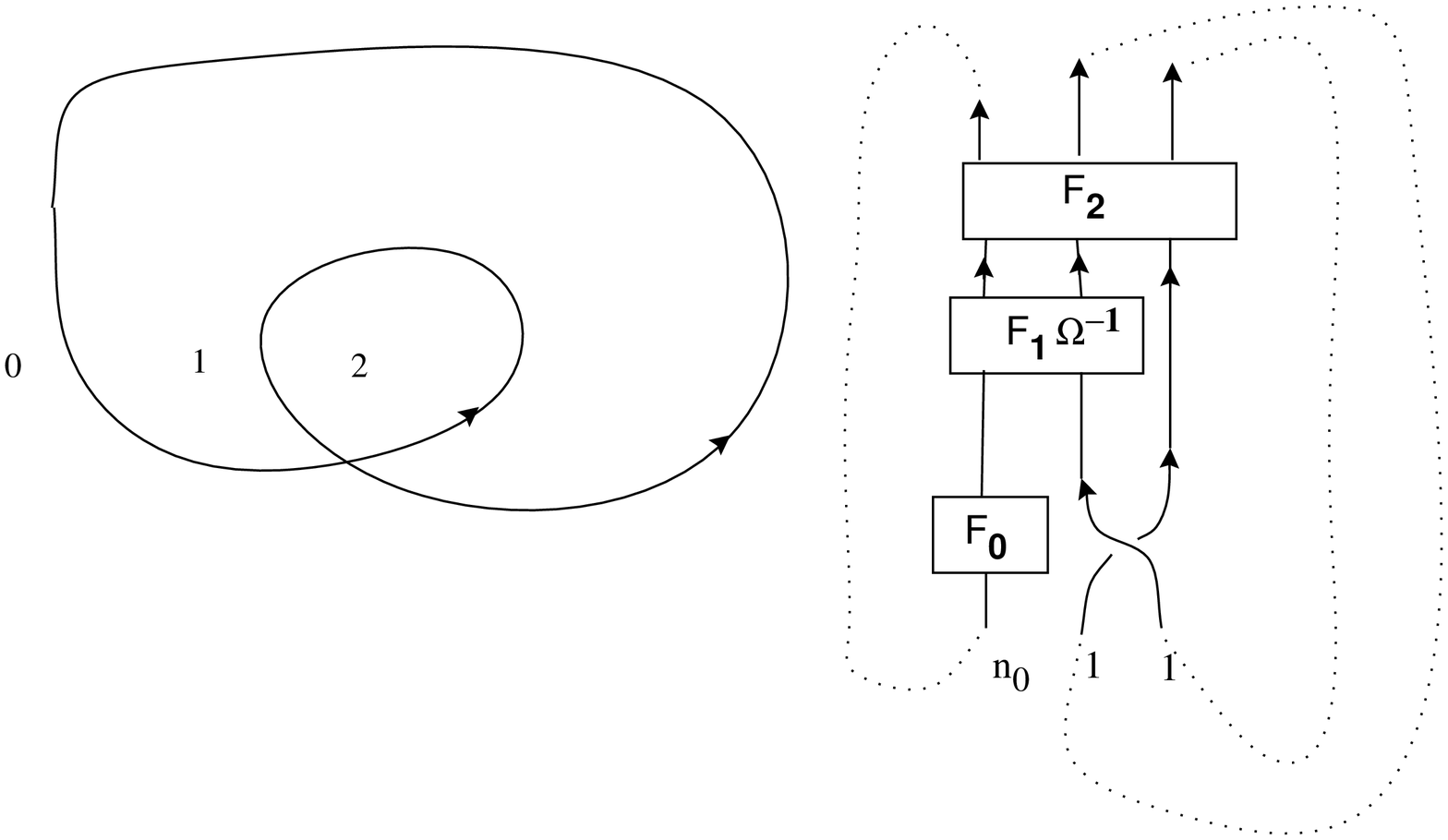}}

\ifig\fec{The cut and paste construction. }
{\epsfxsize2.5in\epsfbox{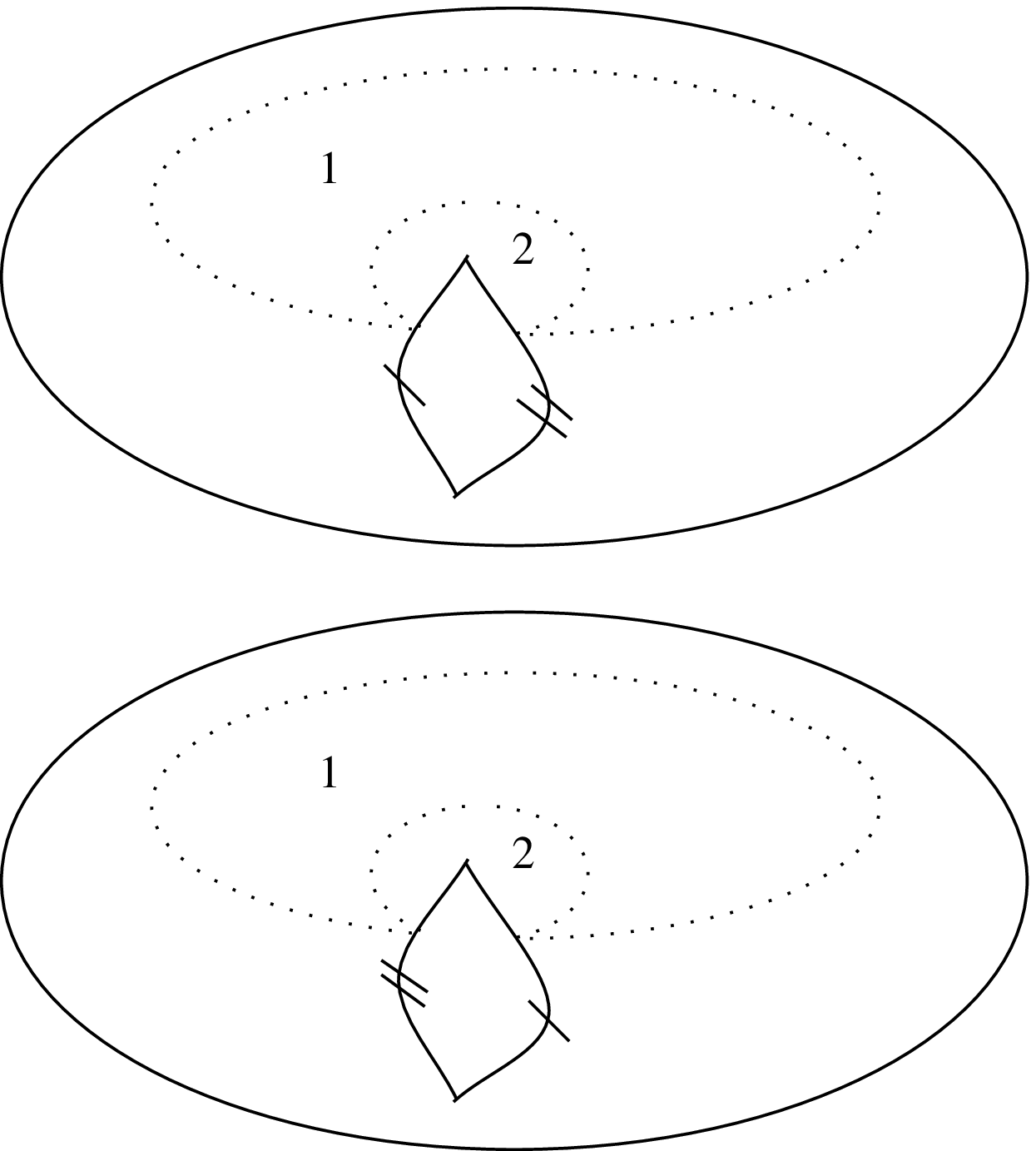}}

Consider the example Wilson loop shown in \fexcb .
We have drawn a sketch of the braid diagram
 obtained by following the  rules of section 6 . For
 simplicity we are considering a Wilson loop in
fundamental representation,
but  the delta function keeps the same form  in
the general case.
We have also cancelled the conjugating
 $\gamma$ factors ( we can do this since
 they commute with the $F's$ in this case )
which are unnecessary since we are
 not concerned with the detailed counting at
 this point, but rather in construction
techniques for branched covers.
One solution to the delta function is obtained
 by setting $F_1= F_0= 1$, and $F_2 $  a
 transposition acting on the two strands
coming from the Wilson loop. The extra
 winding present from the construction
of the braid diagram, is like a simple branch
point, but does not carry the usual factor of
 $1/N$ for simple branch points. Opting for
the simplest cut and paste construction of covers with
 two simple branch points,  we make
 two slits joining the branchpoints and then identify
 the edges as indicated in   \fec . Then we get the correct
degree (namely $1$ for region $1$ and $2$ for region $2$,
 and $0$ for outside region) by cutting both top and bottom
sheets along the larger dotted circle and dropping the part
covering the outside, and only the top sheet along the smaller
dotted circle throwing the part covering region $1$.
The fact that the simple branch point which gets
associated with the outside doesn't have the usual
$1/N$ is that this permutation acts on sheets that
are eventually removed; its  role is to produce the
 correct winding of the Wilson loop. This basic
construction generalises to arbitrary braid diagrams
and shows that, in the chiral theory,
{\bf the only singularities  we need are branch point
 singularities }.

The permutations in
$F_c$ are associated with branch points and handles
 in region $\Sc$, as viewed from a basepoint in $\Sc$ .
The $\gamma_c$  instructs us on how to tie
 the covers of the separate regions along $\G$ .
  There may also be extra branch points that have
to be inserted in region $c$, but which permute
 sheets that will be cut out in the final step.
 We saw this already  in the non-intersecting case, and
 in the example above.
 Extra branch points are needed in sheets
that will be cut out because they produce
 the correct monodromy along the Wilson lines
for the maps from the worldsheet boundaries to the Wilson graph.

\noindent
{\bf Remark: Relation between Hurwitz-Wilson spaces
and ClOSED strings.}

In fact for a non self-intersecting Wilson loop  which
 has on its right
 a surface with $G$ handles, an
 insertion of $N^{\vert  \vec k_{\G}  \vert  - 2G n(\vec k_{\G} ) } $ times
 the usual insertion would count covers where one
does not need to cut along that Wilson loop. For
self intersecting loops the renormalisation needed
would also depend on the  windings
of the Wilson loop and can be read off from the
 delta function
constructed according to the rules above.
With these renormalisations,  Wilson averages
 count  Euler characters of
 subspaces of ordinary Hurwitz  spaces for closed
 worldsheets. For example,
for  a general Wilson loop on the sphere the renormalisation needed is
$N^{\vert \vec k_{\G} \vert }$. For higher genus targets different
renormalisations
may be needed according to different ways of
embedding Hurwitz-Wilson spaces inside Hurwitz spaces.
These correspond to different choices that can be made in
constructing the delta functions. These choices appear, on the one hand,
in how one writes the integrals, and
geometrically in how one chooses generators
of the fundamental groups, and basepoints, etc.
to describe the  covers in Hurwitz-Wilson spaces.

\subsec{ \bf  From branched covers  to Braid diagrams }

Choose a region which we call the
outside region and a basepoint in it. Choose one
 basepoint
on the edge and one in the interior
 of each region cut out by the Wilson loops.
Choose a set of paths each homotopic to a path
around  $c$'th region
 traversed in a direction compatible with the
 orientation of that region.
If the boundary of the $c$'th region has a
branch point, deform the path
to the right of the Wilson loop.

Suppose the degree of the map
is $n_0$ on the outside. Label from $1$ to $n_0$ the
 inverse images of the basepoint, and draw points corresponding to
 these in a row which will be the base of the braid
diagram.
Complete the cover as discussed before. There may
be more than one prescription for  completing the cover,
these correspond to different ways of writing the braid diagram ,
which are related by identities in tensor space.

 Permutations in the $F_c$ factors will be picked up, by
lifting paths based in the respective regions
in the usual way as  for manifolds with boundary and
 for non-intersecting Wilson loops.
The conjugating permutations $\gamma_c$
will be read off by lifting paths connecting the basepoints
in the region to basepoints on the Wilson loop.
 We will see this in detail in an example in the next subsection.
\ifig\fez{ Ordered set of generators  }
{\epsfxsize2.5in\epsfbox{ 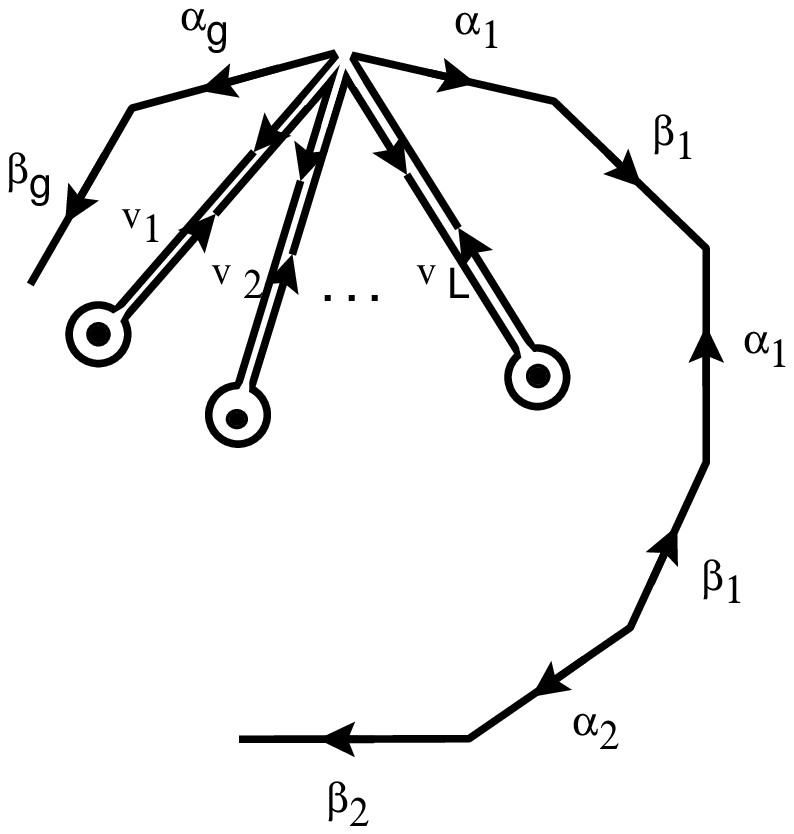  }}

 { \bf Remark: Ordering the generators.}

The fact that  the braid diagram contains all regions
 in a chosen order
 from down to up corresponds to the fact that the
 product of permutations produced in all regions is $1$.
This  just follows from the $1$ relation  on the generators
 of the
fundamental group of $\pi_1 ( \ST - \hbox{branch points})$.
 It is clear geometrically that
we should be able to construct the braid diagram by
 choosing any ordering of the regions.   This ordering
corresponds
to the fact that the relation satisfied by the generators
 of the fundamental group of a Riemann surface with
punctures are expressed by choosing some ordering
of the
generators. Geometrically, having made a choice
of representatives of the non trivial classes of the
 fundamental group
there is a definite cyclic order in which  we have
to take the product of these paths to get  a path that spans
 a $2$-cell and is therefore trivial, see \fez .
That the ordering of the regions in the braid diagram can
 be changed arbitrarily
 can be recovered as a consequence   of Weyl duality
and cyclicity property of traces.

\subsec{  Equivalences and Automorphisms }

A combinatoric definition of equivalence  is  an action
of the symmetric
 groups involved which preserves the delta function.
We can  read off from the braid diagram an action
 of the symmetric groups which leaves the
 delta function invariant, and  indicate the relation  to
an automorphism of the cover constructed from the braid diagram.
To give an equivalent braid diagram  we specify
 permutations $x_c$  for each  component,
which act by conjugation on all the permutations in the  $F$ factor.
These correspond to
 relabelling the points in the inverse image
of the basepoint in the regions $\Sc$.
We also specify permutations $x_{\Gamma}$
associated with the Wilson loops  for the basepoints
on the boundary of each region.
Let $\beta_c$ denote the winding that occurs when
the vertex contains both positive and negative powers
of a generator \wbastwo .
The action on the $\gamma_c$ and $\beta_c$ is
determined by requiring that the product
of the $$ \gamma_c^{-1} F_c \beta_c \gamma_c, $$   is conjugated
by the product of $x_0$ and $x_{\Gamma}$'s for the
 strands acted on by the $\gamma_c,  F_c $ and $\beta_c$.
 Let us call $x_a $ the combination that relabels the first two
 strands in \wbastwo, and $x_b$ the one that relabels the last one.
We find that the action
\eqn\equiva{\eqalign{
&\gamma_c \rightarrow (x_c.1.  1 ) \gamma_c (x_a. 1)^{-1} \cr
                         & F_c \rightarrow (x_c.1.1 ) F_c (x_c.1.1)^{-1}  \cr
                          &\beta_c \rightarrow (1.1. x_b) \beta_c  (1.1.
x_b)^{-1}  \cr
}}
guarantees the conjugation.
These relabellings correspond to restrictions of the map $\phi$ to the various
basepoints $y_{c}$ in the regions $\Sc$
and those on the Wilson graph ,  used in relating
 maps to the homomorphism into permutation groups.
The equation \equiva\ does not lead to a unique way
of setting up symmetric group actions which
 preserve the delta function. Different choices
should correspond to different ways of setting up the
description and counting of the covers.

Once we know that equivalence classes
of maps  are in $1-1$ correspondence with
equivalence classes of homomorphisms
under the combinatoric
equivalence just defined, it is clear that the
centraliser subgroup of an equivalence class
 of homomorphisms is isomorphic to the group
of automorphisms of the branched cover.
Counting the homomorphisms represented in the
braid diagram with weight equal to the orders of the
permutation groups involved is the same as
counting equivalence classes of homomorphisms
with weight equal to the inverse order of the
centraliser subgroup . This shows that
the symmetric group data in the
braid diagram  counts equivalence classes of maps
with weight equal to the inverse order of the automorphism group.

 \newsec{ \bf Euler Characters of Hurwitz-Wilson spaces }

For the partition functions we saw that the power
of $\Omega$ factors is equal to
the Euler character of the target space, and
for maps of degree $n$ the $\Omega$ factor has degree $n$. After
expanding the $\Omega $ factors this led to
the result that the number of branched covers
counted with inverse
order of  automorphism group, was weighted by
the Euler character of
the configuration space for motion of the branch points
on the target. This was, for fixed degree, branching number
and fixed number of branch points, the Euler character
of Hurwitz space, a discrete bundle over configuration
space. Here the base space of the bundle will be the
product of configuration spaces for the separate regions
labelled by $c$  and for the graph defined by the Wilson
 loops. The number of $\Omega $ factors approriate for
each region appears in the $F$ factors. Then there are
 inverse omega factors for each generator of $\pi_1$
and an $\Omega$ factor.  We prove for arbitrary Wilson loops

\medskip
\noindent
{\bf Theorem 9.1}
The  net number of $\Omega$ factors (outside the $F$ factors ) is equal to
 the Euler character of the Wilson graph .

\noindent
 {\bf Proof}

This follows from the fact that the contraction
procedure does not change the Euler character,
 since each contraction
reduces the number of edges by $1$ and the number
of vertices by $1$.
The Euler character of the reduced graph is clearly
 the $1 - g$ where $g$ is the number of generators
of $\pi_1$. In our integration scheme we choose
 these $g$ generators to be the interior region paths. Each integration
gives an inverse omega factor  $\Omega^{-1}$, and going from traces
to delta functions using \deltatr\ we get an $\Omega$  factor.
$\spadesuit$

{\bf Remark : }
This  property has a generalisation beyond
Wilson loops in two dimensions,
and motivates a  conjecture on the relation
 between the Euler characters
of an appropriate space of branched covers
and higher dimensional Lattice gauge theory.
We return to this briefly in the final section.

\ifig\fomegpl{ Omega factors and branch points on Wilson graph.   }
{\epsfxsize5.0in\epsfbox{ 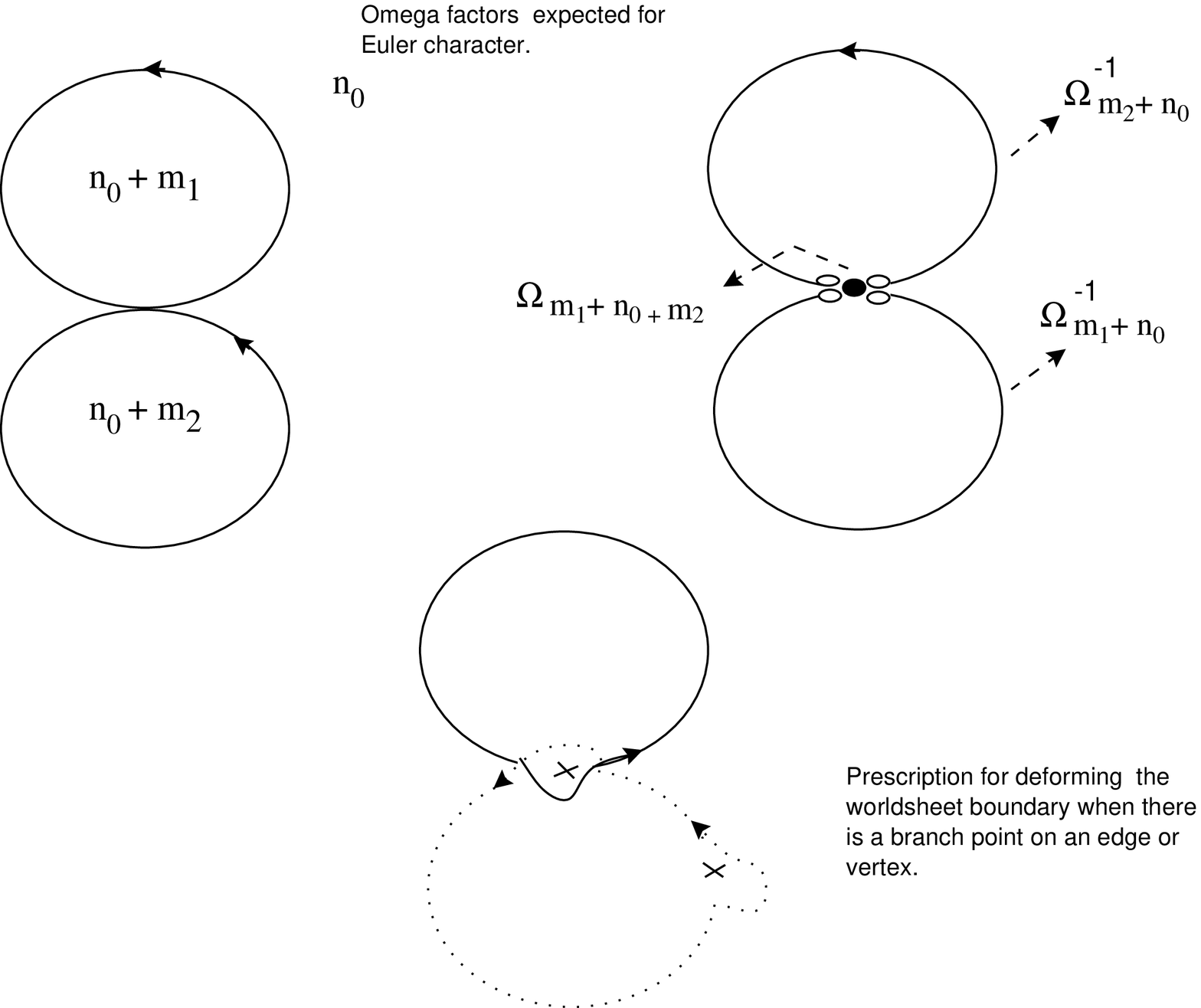  }}

\subsec {Detailed placing of Omega factors,
 Euler characters and the Gross Taylor prescription.}
Having discussed the net number of Omega factors in
complete generality,
 we now turn to the detailed nature of the Omega factors
 which appear, in particular, the degree  $d$ of $\Omega_d$
and the power with which it appears.
We have a space of branched covers
where the branch points are allowed to be anywhere
on the Wilson graph
and an important point is that the {\it {degree of the map is
 not constant  over the graph}} .
To calculate the Euler character of such a space of
branched covers ,
we should decompose the Wilson
graph into edges, vertices and for convenience we
 will allow circles, such that
 the degree of the map restricted to any such component
 is constant.

 {\bf  Distributing Omega factors : }

We then assign to each component a factor
$\Omega_{d}^{\chi} $ where $d$ is the
degree of the map restricted to that component
and $\chi$ is the  Euler character of that component.

This predicts which $\Omega$ factors appear in the
delta function. It is equivalent to the Gross Taylor rule
when the latter is directly applicable (transversal double intersections)
and generalises it to  arbitrary  Wilson loops.
 The Gross Taylor approach instructs
us to perturb the inverse Omega factor from
 each edge into the side to its left, which
 predicts the degree of the Omega factor .
For a  vertex  where two edges intersect transversally,
it tells us to perturb the $\Omega$ factor to the left of both edges.
Again this predicts the same degree as the above .
Our  interpretation however is different. We allow the
 { \it branch points to lie on the graph} , and give a prescription
for { \it deforming the paths}  along which the monodromies are measured.

 When the omega factors are expanded out they
produce branch points  which are interpreted to  lie on the graph.
Reconstructing the delta function from the map
 proceeds as before. The monodromies are well
defined using the prescription of deforming the
image of the worldsheet boundary away from the
 branch point, to the right of the oriented Wilson loop.
When we restrict  the delta function to
the  strands corresponding to the Wilson loop, we are
taking a path which is the image of the deformed worldsheet
boundary, i.e it is deformed to the right of the branch point.
 \fomegpl\   illustrates our prescriptions
with a graph that has  a tangential intersection.

We comment on the distribution of Omega factors
 that comes from the rules of integration we described.
 It is most clearly understood in terms of covering
 spaces of the reduced Wilson loop.
The omega factor is naturally associated with the
vertex  of the reduced graph.
The inverse omega factors are associated with edges
according to the number of sheets they act on.
For reduced graph and unreduced graph,
the rule based on Euler characters of Hurwitz spaces,
gives different distributions of Omega factors.
These are expected to related by identities in the algebra
$Z (S_{\infty}) \times \IR (1/N) $, or equivalently
in the group algebra of arbitrarily large symmetric
 groups tensored with power series in $1/N$.
Typically the identities involve equation 11.3 and its generalisations.

The final step in  the proof of Theorem 9.2 proceeds as before.
\medskip
{\bf Theorem 9.2}
 The Chiral Wilson average counts Euler characters of
Hurwitz-Wilson spaces.

  We expand the $\Omega $ factors.
The $F_c$-factors contain $2-2G_c-1$ omega factors
so these generate Euler characters of configurations
 of points in the respective regions $\Sc$.
The extra  Omega factors associated with the
Wilson graph  generate Euler characters
of configuration spaces of points on the graph.
So each term in the Wilson average is a  product
 of Euler characters of configuration spaces  times
discrete data counting, with inverse automorphisms,
Hurwitz Wilson branched covers  for fixed branch locus.
This completes the proof.

\newsec{ \bf  Example  }

\ifig\fbrings{ A pair of  Borromean rings }
{\epsfxsize3.0in\epsfbox{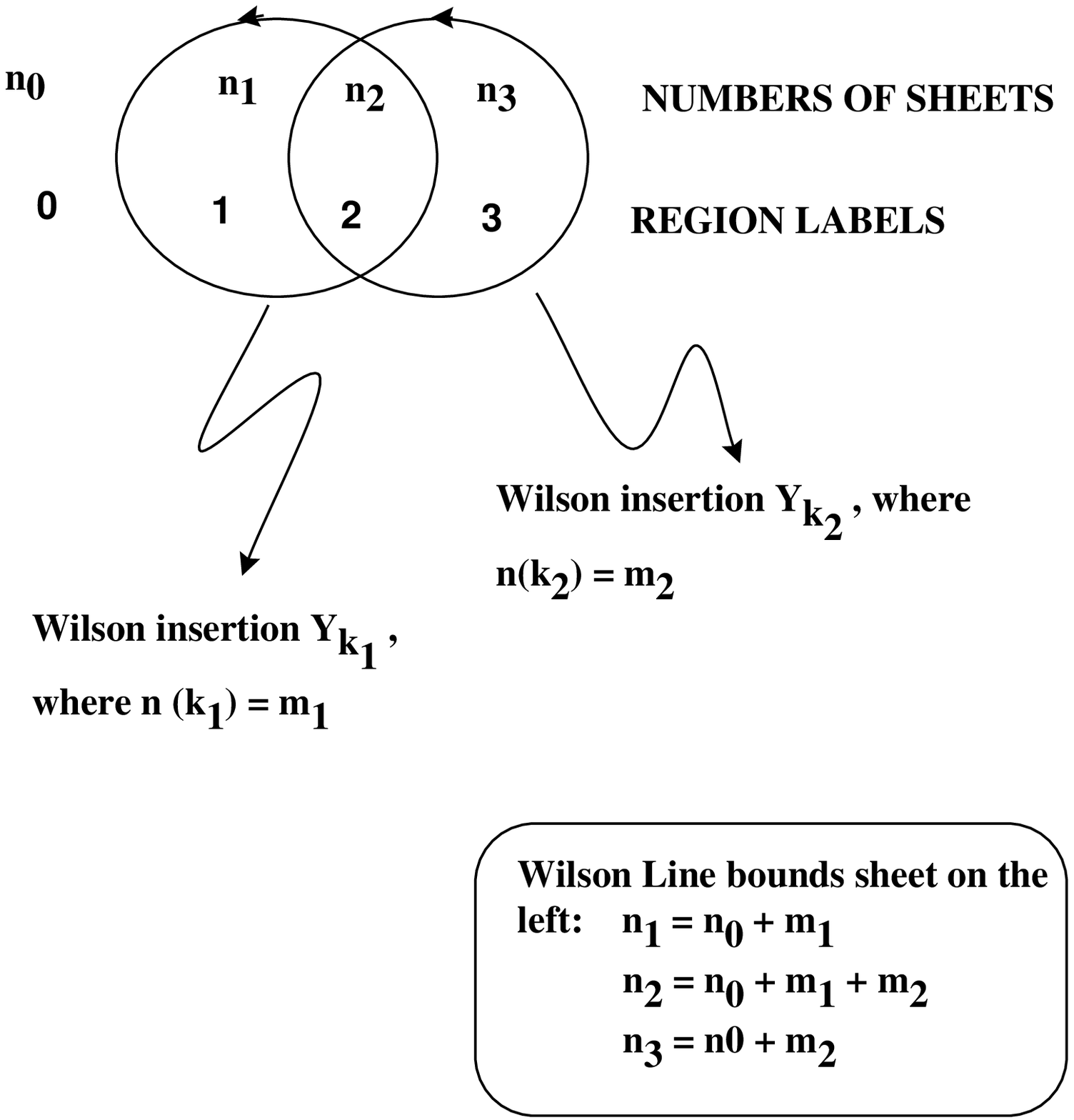}}

\ifig\fbrbd{ Braid diagram for Borromean rings }
{\epsfxsize5.0in\epsfbox{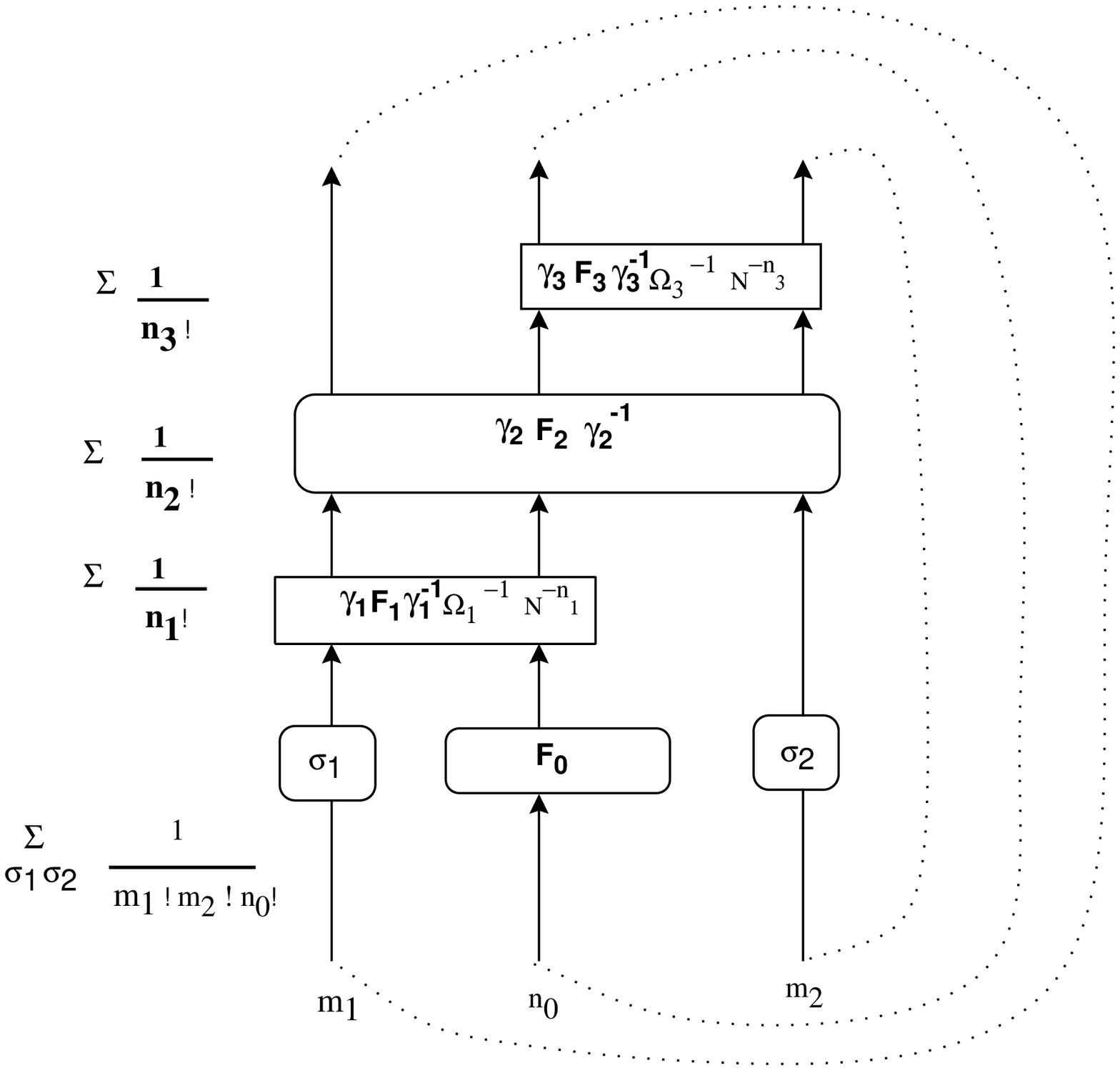}}

We discuss here a simple example (  \fbrings)
 which illustrates how
 the chiral large $N$ expansion generalises
to intersecting Wilson loops,
and admits an interpretation as a generating
function for Euler characters of Hurwitz-Wilson spaces.
We consider a pair of Wilson loops
 which intersect twice . The Wilson loops
$1$ and $2$ are labelled by the
vectors $\vec k_1$ and $\vec k_2$, with $n(\vec k_i)
 = \sum_{j} j k_i^{(j)} = m_i$.
The answer is illustrated diagrammatically in \fbrbd .
The sums over $n_i$ are restricted by
the conditions written in \fbrings .
The permutations $\sigma_i$  $(i= 1, 2)$ are
summed over the conjugacy classes
specified by $\vec k_i$ in $S_{m_i}$.

\subsec{ From Braid diagram to maps, and automorphisms}
The discussion of how to go from braid diagram to maps has
already been given in complete generality.
We just state the main points as applied to this example.
 Start with a covering of constant degree $n_0+m_1+m_2$
by the same techniques that were used for covering
surfaces counted by
the partition function, and cut out a number of
sheets to the right of each Wilson loop at the end.
 The only difference here is that
there are specified  monodromies along the Wilson loops,
which restricts the kind of branch points that are allowed.
Branch points coming from expanding the $F_c's$ are
clearly to be placed in the respective regions $c$.
In the example shown we also need to add branch
points giving monodromies $\sigma_1$ and
$\sigma_2$ on the $m_1$ and $m_2$ sheets outside.
These sheets are going to be cut out from the outside
along the Wilson loops $1$ and $2$ respectively  but they
guarantee
that the right monodromy is recovered from paths along the
Wilson loops.

We discuss here in detail the  issue of
equivalences and automorphisms.

In the example,
the $n_0$ points  at the base of the braid
 diagram can be relabelled
according to permutations
 $ x_{\Gamma_1}. x_0. x_{\Gamma_2}  \in  S_{m_1}
\times S_{n_0} \times S_{m_2}$, to give maps related
 by a homeomorphism $\phi$. The basepoints in the $\Sc$ can
be relabelled by permutations $x_c$.
$x_c$ acts on the permutations in $F_c$ by
conjugation, which leads to
\eqn\conjF{
 F_c \rightarrow x_c F_c x_c^{-1}.    }
 $x_{\Gamma}$ acts
 on the permutations $\sigma_{\Gamma}$
 by conjugation.
The action on the $\gamma's$ needed to
 leave the delta function invariant can be
read off from the braid diagram. For example $\gamma_1$
 is acted on the left by $x _1$, and from the right by
$( \sigma_1 .x_0 )^{-1} $.
These relabellings correspond to restrictions
of the map $\phi$ to the various
basepoints used in going from the map to the
 homomorphism into permutation groups.
Applying a theorem  from the theory of branched
covers \Ma\ ,  this shows
that the data $x_c, x_{\Gamma_i}$ is equivalent to
 a geometric equivalence $\phi$.  For ordinary branched covers
the argument is reviewed in \Ez\ and briefly in section 3 of \CMROLD.

\subsec{  From Maps to Braid diagrams  }

It follows immediately  from conjecture 4.1,  that the degrees
 of the maps
in the various regions are related as shown in the diagram.
The Wilson average contains a sum over  $n_0$ ,
which  is the degree of the map for the region outside.
The cover may be  completed by adding
$m_i$ sheets to the outside of
$\G_i$, which cover the outside, with just   one branchpoint
in the class specified by $\vec k_i$.

Let us ignore
 for a moment the omega factors outside the $F_c$.
The $F_c$ describe  permutations
associated with branch points and handles
 in region $c$.
We can understand how a sequence of
permutations satisfying
the delta function  can be obtained from a map
of type described in section 8.5.1,
 {\it by restricting it, in turn, to
 the three sets of strands of multiplicity
 $n_0$, $m_1$ and $m_2$ respectively }.

Choose a basepoint $y_0$ on the outside.
 Label the inverse images  from $1$ to $n_0$, and
correspondingly draw the $n_0$ middle points at
the base of the braid diagram.
 Follow the lift of a path from the outside joining the
 basepoint outside
to some point $P_1$ on
$\Gamma_1$. The endpoints of the lifts
will be points in the interior of the worldsheet.
There will be $m_1$ extra points on the outside
which lie on the boundary of the worldsheet.
These can be mapped to $m_1$ points in the inverse
 image of $y_0$ for the
{\it completed } cover.
Repeat the procedure by choosing a basepoint $P_2$ on
the second Wilson line.
With the completed cover we have $n_0+m_1+m_2$ points
 in the inverse image of $y_0$ which correspond to points
 at the base of the braid diagram in direct analogy to
 the partition function.

Also choose basepoints in the interior of regions
 $1$ through $3$,
and read off permutations around non-trivial cycles
 in these regions.
These permutations enter the factors $F_1
 $ through $F_3$.
 The permutations $\gamma_c$ $(c=1,2,3) $ are
obtained by following paths from regions $c$
to the point $y_0$. Using the conjugations all
permutations are described with respect to one
 basepoint $y_0$.

Now consider the lifts of curves which start from
 any of the $n_0$ points on the outside.  Consider a
product of paths  around
all the branch points and the appropriate commutators for the
handles in region $0$,
$1$, $2$ and $3$. It is
homotopic to the trivial path. This is reflected in the
braid diagram by the fact
that the result of the windings
of the $n_0$ strands produced successively by
 $F_0$, $F_1$, $F_2$
and $F_3$ is a trivial permutation.   Restricting to
 the $m_1$ sheets, which only cover regions $1$ and $2$,
we see that the image in $S_{n_0+m_1+m_2} $
of the path round region $1$ times the image of the path
 round region $2$ which is homotopic to the oriented
path described by the Wilson loop $\Gamma_1$ is in the
class of $\vec k_{1}$, as in
Conjecture 9.1. The same argument applies to
$\Gamma_2$ which is homotopic to the product
of paths around regions $2$ and $3$.

The same discussion goes through when we consider branch points
coming from the extra $\Omega$ factors, by using the
deformation of paths described in section 4.2 and 9.1.

\ifig\fbrde{Decomposing 2 intersecting circles into components
 where map has constant degree.}
{\epsfxsize4.0in\epsfbox{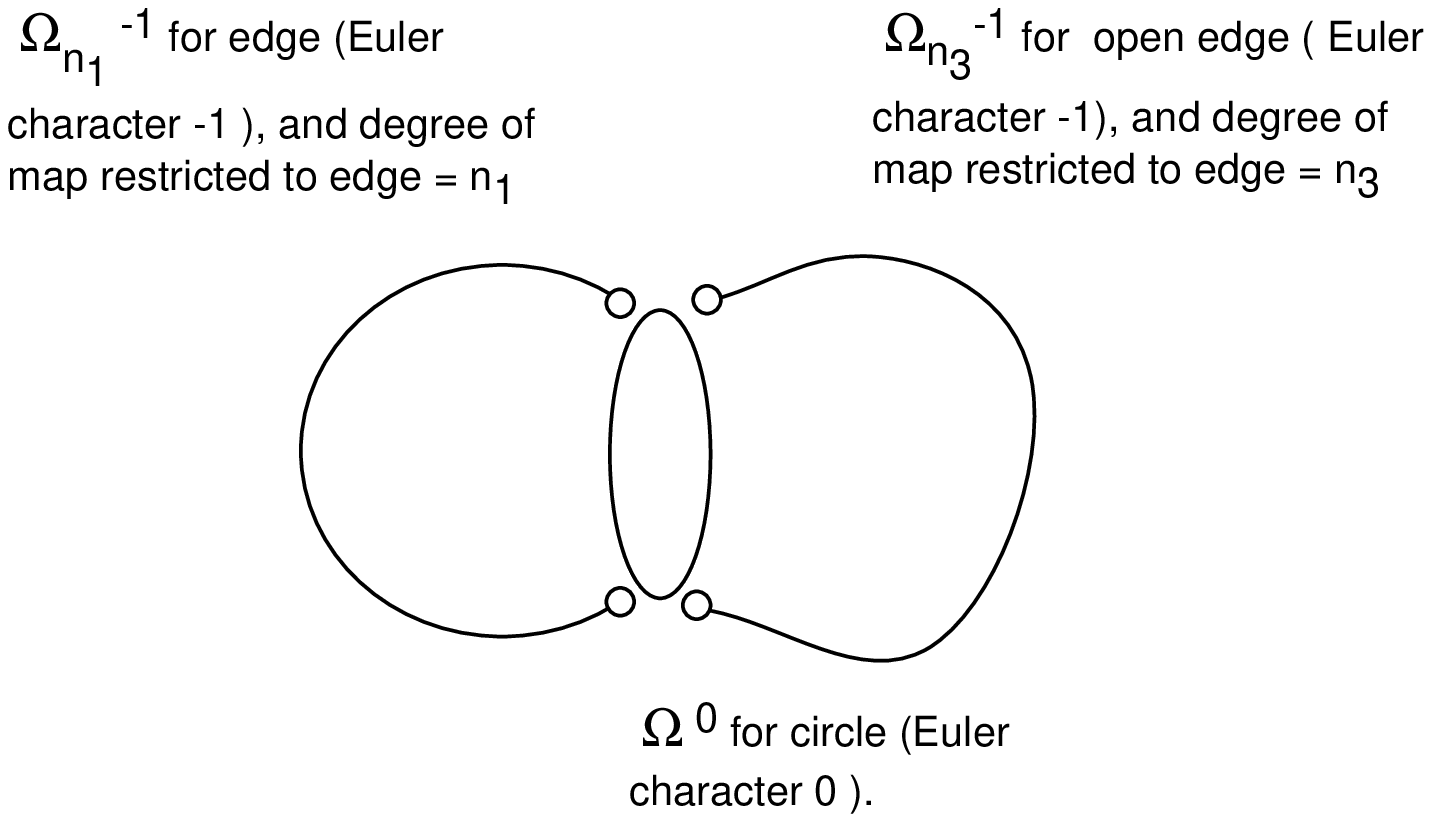}}

\subsec{ Euler characters}
 In this example a decomposition of the graph
  satisfying the above conditions is
shown  in \fbrde . The degree of the map on the
left edge is $n_1$ and on the right it is $n_3$ .
This explains why the delta function contains
$\Omega_{n_1}^{-1}$ ,  $\Omega_{n_3}^{-1}$,
$\Omega_{n_2}^{0}$. Proof that we have Euler
character of Hurwitz-Wilson space proceeds as in section 9.
\ifig\fphmi{ Proof of Loop Equations.}
{\epsfxsize5.5in\epsfbox{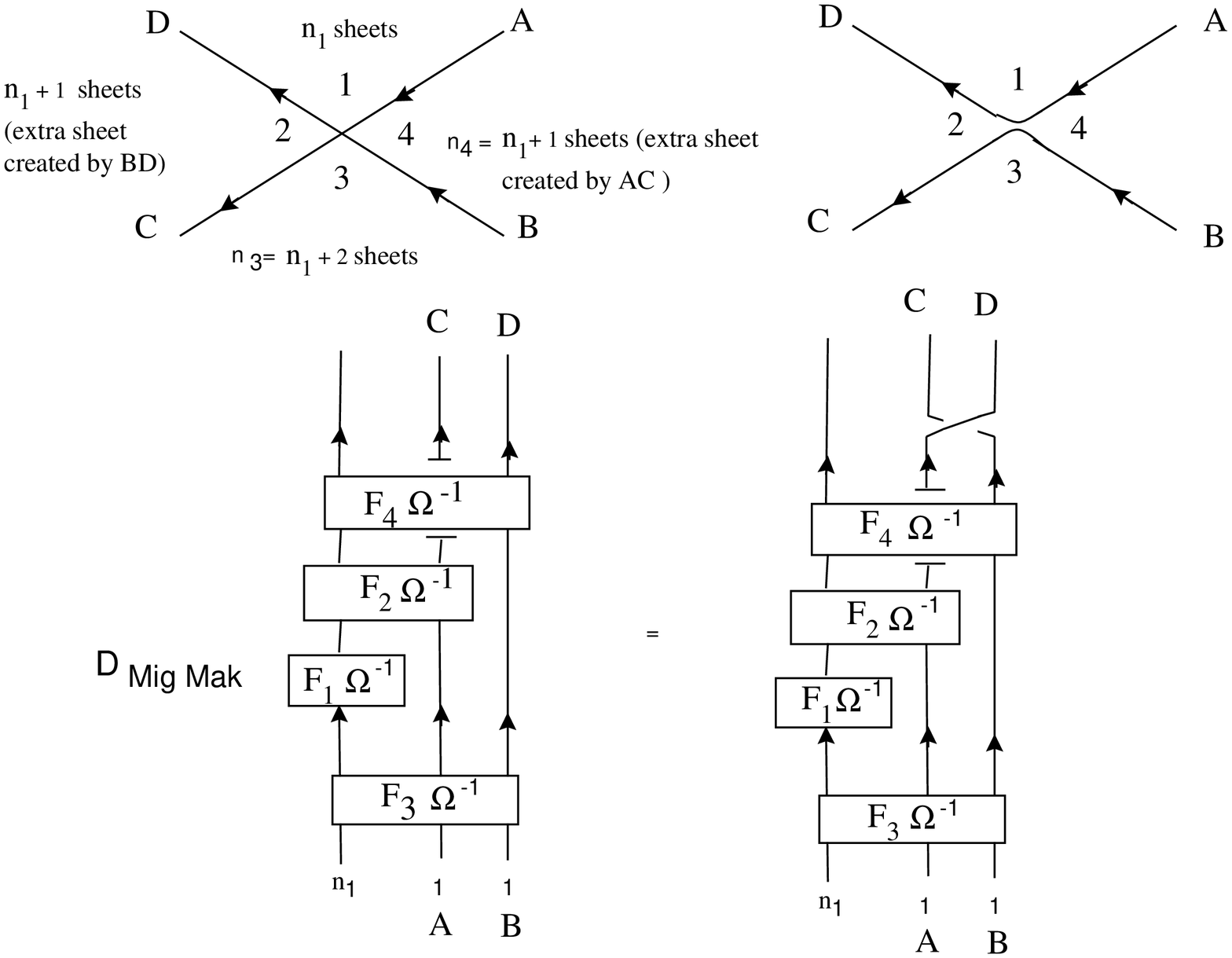}}

\newsec{ \bf Chiral Loop Equations  \foot{ I thank G. Moore for collaboration
on this section.}}
In this section we  use the diagrammatic formalism  that we
developed in the previous sections to prove that
 the Migdal-Makeenko loop equations
are valid for Wilson loops in the chiral expansion
\foot{ The observation that the Migdal-Makeenko Loop equations are
valid for the chiral theory has been made independently by W. Taylor. }.

The figure explains  a proof
of the Migdal-Makeenko equations for the chiral theory. These
are differential equations giving the result of acting by a first order
diferential operator
on  the expectation value of Wilson loops (in the fundamental
 representation)
with an intersection as in the top left of \fphmi\ to the Wilson
average with the
intersection replaced  by  the top right of \fphmi .  The differential operator
is a combination of derivatives
in the areas on different sides of the intersection.

Since the Wilson average is a sum over coverings of the target, such that
the oriented worldsheet boundaries map to the oriented Wilson
 lines by orientation
preserving maps,
the degrees of the maps in regions $1$
 through $4$ are related as in  \fphmi .
This corresponds in the delta function
 for the Wilson average,   to the
pattern of embeddings shown.
The sheet bounded by the side $BD$
 covers regions $2$ and $3$
and the corresponding strand is acted
 on by the $F$ factors for these regions.
The sheet bounded by the side $AC$
covers
 regions $3$ and $4$ and so the corresponding
 sheet is acted on by  $F_3$ and $F_4$. Now if
we act by the differential operator
\eqn\difop{ D_{Mig Mak} = { N\partial \over
 {\partial {A_2}} }+  { N\partial \over
{\partial {A_4}} } -{N\partial \over {\partial {A_1}} } -
  {N \partial \over {\partial {A_3}}}  + { 1\over N}. }
( If we are doing the chiral zero charge sector
 of the $U(N)$ theory,
 we can drop the $1/N$ to  give a simpler
differential equation).
It is an easy check, using the form of the $F$ factors in \Fbdy ,
 that the only contribution comes from
$T_2^{(n_3)} +T_2^{(n_1)} -T_2^{(n_2)} -T_2^{(n_4)}  $,
 where the $T_2's$ are sums of permutations
in $S_{n_1}$ to $S_{n_4}$, all  embedded in $S_{n_1+2}$
 as shown in the figure, e.g $S_{n_4}$ acts on the
 first $n_1$ and the last strand. This difference of
operators is just equal to the transposition which
 switches the last two sheets:
\eqn\cantra{\eqalign{ &T_2^{(n_3)} +T_2^{(n_1)} -
T_2^{(n_2)} -T_2^{(n_4)} \cr
 &= \sum_{i<j=1}^{n_1+ 2 } (ij)
+ \sum_{i<j=1}^{n_1} (ij) - \sum_{i<j=1}^{n_1+1}
  (ij) - \sum_{i < j \in \{ 1\cdots n_1 \} \cup \{n_1+2\}  }
(ij) \cr
& = (n_1+1 ~ n_1+2).   \cr}}
Transpositions acting on the first $n_1$ sheets
occur twice with a plus sign from $T_2^{(n_1)}$
and $T_2^{(n_3)}$ , and twice with a minus sign
 from $T_2^{(n_2)}$ and $T_2^{(n_4)}$. Transpositions
 that mix the strand labelled $A$ (the $(n_1 + 1)$' st  strand)
with the first $n_1$ cancel because of positive contribution
 from $T_2^{(n_3)}$ and
negative contribution from $T_2^{(n_4)}$.
 Transpositions that mix the strand labelled $B$
with the first $n_1$cancel due to contributions
from $T_2^{(n_3)}$ and $T_2^{(n_4)}$. So we
are left just with the transposition mixing the
 last two strands from $T_2^{(n_3)}$. But  the resulting
 braid diagram is
exactly what one would construct for the  Wilson
average on the upper right of \fphmi .

\ifig\flgen{ First order differential operator acting  on
 Wilson loop
the left gives the Wilson loop on the right  }
{\epsfxsize3.5in\epsfbox{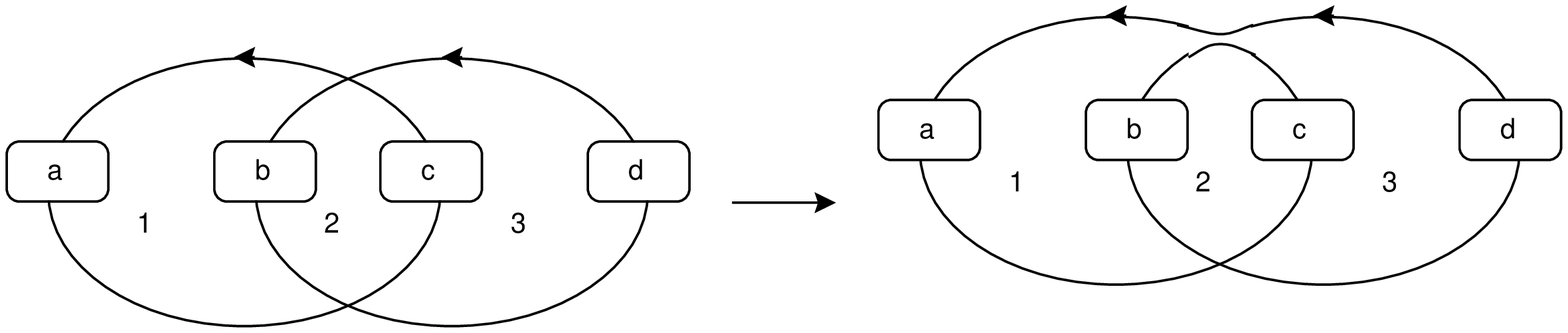}}
In making an identification between certain
points on the Wilson graph
and certain strands of the braid diagram, we
have used some physical intuition.
We have seen from the construction of the braid
 diagram that
it contains strands which start off being
diagrammatic
 representations of the path traced by the
 Wilson line. Physically
 (as is obvious in a Hamiltonian quantization
on a cylinder
\refs{  \MiPoetd , \dgcrg  } ) the Wilson observable
creates a
worldsheet boundary which propagates to form part
of a covering of the manifold.
To make the argument completely mathematical, we
 can analyze the structure of the braid diagram for
some very general classes of  Wilson loops.
 For example suppose we
 have two Wilson loops which intersect at the
 upper intersection of \flgen . Since they are two
 distinct closed loops, they must intersect at least one
more time.
So quite generally we
have the structure of \flgen\ where another intersection is
shown as the lower one.
The boxes indicate some arbitrary windings. Note that
 in each region there may be
some handles attached. For such a class of Wilson loops,
 one constructs the basic structure of the braid diagram, and
shows that the above
argument goes through :  namely that the pattern of embeddings
is the one shown in \fphmi , and that the remaining transposition
obtained after acting with $D_{MigMak}$ is exactly the
 one needed to give the braid diagram  for the Wilson loop on the
right of \flgen .

\noindent
{\bf  Remarks : }

1. There are also loop equations involving Wilson loops which
 have tangential intersections \KaKo.
These can also be derived for the chiral theory
 using properties of the delta functions.
The derivation for the case when the intersection
 is transversal
 involved the additive structure of classes of symmetric
 groups embedded in different ways in $S_{\infty}$ ( the group of permutations
of all the natural numbers).
The derivation in the case of tangential intersections
  involves the multiplicative structure  in
the algebra $\IZ (S_{\infty} ) \otimes \IR (1/N)$.
For example if $\Omega_{n_0}$ acts on integers
 $1$ to $n_0$, and $\Omega_{n_0+1}$ acts on
 integers $1$ through $n_0+1$, then we have:
\eqn\omegconv{ ( 1+ \sum_{i=1}^{n_0} { (i ~ n_0+1)\over N} )
 \Omega_{n_0} = \Omega_{n_0+1}. }

2. The Loop equations are of course true for
the correct large $N$ expansion of 2d qcd (the coupled expansion).
This does not contradict the fact that  the Wilson average at large $N$
 is determined by the loop equations
{\it given the  correct boundary conditions}. The
 boundary conditions,
i.e the Wilson averages for the non-intersecting
 non-overlapping Wilson loops,
are in general  different between chiral and
 non-chiral theory.
Consider non-chiral Wilson Loops on the plane. In the
non-chiral theory, for a collection of non-overlapping
non-intersecting Wilson averages, the expectation
value is just the product of
$e^{-A_i/2} $ where the $A_i$ are areas enclosed
by the Wilson loops. In the chiral theory,
 there is also {\it a simple way to characterize the
boundary conditions} :  we have $e^{-A_i/2}$ for Wilson loops
with orientation compatible with that induced by the region inside,
which we call the plus orientation,  and zero
for Wilson Loops with the opposite orientation.
If a Wilson loop can be reduced by the Loop equations to
a collection of circles which all have the plus orientation,
then chiral and non-chiral theory have the same boundary condition,
and give the same answer.
 This can be understood  from the string picture by
 using the rule  \GrTa \ that, in the non-chiral theory, each Wilson loop
can bound an orientation preserving sheet
 on the left or  an orientation reversing sheet on the right.
We see that only orientation preserving sheets are allowed,
 since the right of a loop  is always the
outside which has infinite area and therefore  cannot have any sheets.
 So it is not surprising that for this class of Wilson loops
chiral and non-chiral theory give the same answer.

\ifig\fhat{Loop equations as constraints }
{\epsfxsize5.0in\epsfbox{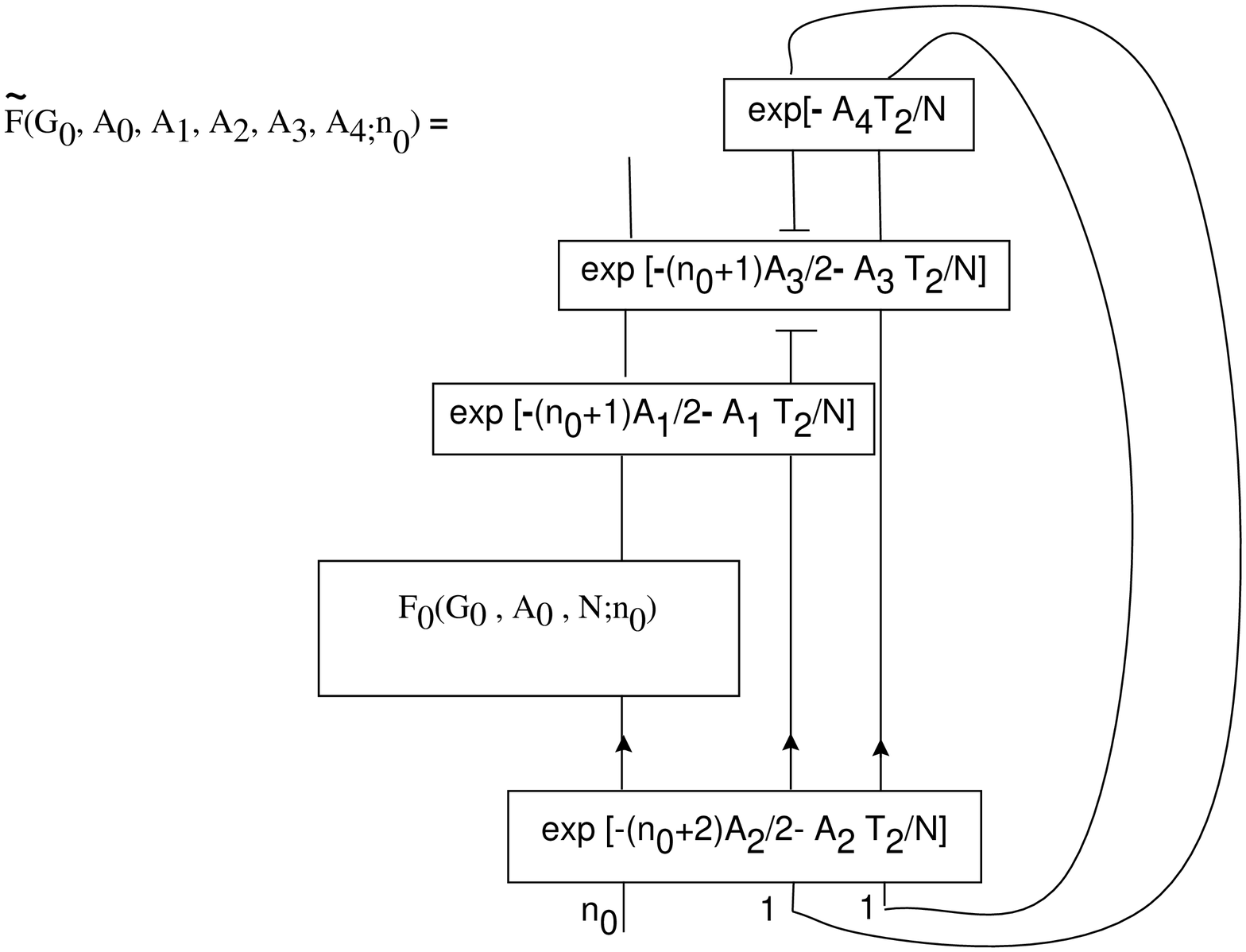}}
\newsec { \bf Loop equations as linear
constraints on the partition function }

We  have  an expression
for the loop equation as a relation between
 sets of operators
acting on tensor spaces. We now show that
the loop equations can be used to write
first order differential equations for perturbations
of the partition function.

Consider
${1\over {N^2}} \tilde F ( A_0, A_1, A_2, A_3, A_4;n_0) $  of \fhat\ as
 an operator on $V^{\otimes n_0}$.
For any $A_i$ it can be proved that this operator commutes
with $U(N)$ acting on $V^{\otimes n_0}$  and also with $S_{n_0}$,
so it must be possible to expand it  in terms of casimirs acting
on $V^{\otimes n_0}$,
which would be the more usual way of describing deformations
of the partition function \refs{ \CMRlsch , \GaYaSo }.

Consider
\eqn\Zpert{
Z(G_0, A_0, A_1, A_2, A_3, A_4)= \sum_{n_0} Z (G_0, A_0, A_1, A_2, A_3, A_4;
n_0),  }
where
\eqn\zno{ Z (G_0, A_0, A_1, A_2, A_3, A_4; n_0) =
 tr_{n_0} ( {\tilde F( G_0, A_0, A_1, A_2, A_3, A_4; n_0)\over {N^2} }  ) .}
If  $A_i=0$ for $( i = 1, \cdots ,  4)$, $Z$ is equal to the
term in the partition function
for a  target of area $A_0$, with $G_0$ handles. For $A_4=0$ (other $A_i \ne 0$
) the
 trace of the operator is the expectation value of the Wilson
average for a pair of rings intersecting in two points as in \fbrings\
and carrying the fundamental representation ( $m_1=m_2=1$).  The derivative
$ {- \partial Z(G_0, A_0,A_1,A_2,A_3,A_4) \over {  \partial A_4}}
\vert_{A_4=0}$
gives the Wilson loop resulting from applying the Migdal-Makeenko equation
at the upper  intersection of \fbrings. The derivative brings down the
transposition
shown alongside $F_0$ in \fexcb.
The key cancellation
of transpositions
responsible for the Migdal-Makeenko equation can be expressed
as a differential equation for
$Z(G_0, A_i; n_0)$
\eqn\diffeq { \bigl [ -{\partial  \over {\partial A_0}} -{\partial
  \over {\partial A_2}}
+  {\partial  \over {\partial A_1}} + {\partial  \over
 {\partial A_3}}  + {\partial  \over {\partial A_4}}\bigr ]  Z(G_0, A_i; n_0 )
= 0,  }
 which results in the same equation for $Z(G_0,A_i) $.
By introducing one extra area,  like $A_4$ here,
we can always
write the Migdal-Makeenko equation as a
differential operator annihilating a braid diagram.
In general the braid diagram will not reduce to
the partition function when the areas go to zero,
but rather to the partition function with
 insertions of some local operators.

So we find a close analogy to  2D gravity :  loop equations
are equivalent to differential operator constraints on perturbations
of  the partition function.
 This was conjectured in section 9 of  \CMROLD . The surprise
 is that this is true for the {\bf chiral } theory
where the partition function has a geometrical interpretation as the
generating function of orbifold Euler characters of
classical Hurwitz spaces.
It was further conjectured that the algebra satisfied
by the constraints is $W_{\infty}$ .
 The only algebra we have used is $\IZ (S_{\infty} ) \otimes \IR ({1/{N}} ) $.
It would be interesting if this algebra could be
 related to $W_{\infty}$ in a meaningful way.

While one  side of the Migdal-Makeenko
 equations is of course
linear, the other  side contains a product of Wilson loops.
 So, expressed in the space of Wilson loops,
the Migdal-Makeenko equations are non-linear.
 But we  find that replacing  ``the   space of Wilson loops''
 by   ``the space of delta functions over $S_{\infty}$, parametrised by
area perturbations'' , linearises the loop equations completely.
It is an interesting question whether this can be done
 in higher dimensions.

\newsec{ General  remarks }
\subsec{ \bf Finite N}

We have focused on the  chiral large $N$ expansions
 for $U(N)$ and $SU(N)$. For finite $N$ one can still use  the
same diagrammatic form
to express the exact answer : only the exact formulae for converting
the integrals into symmetric group data are different
and haven't been fully worked out. In addition,  there are some
 extra projection operators introduced in \refs{ \BaTa}.

If $n_0$ is kept small
and fixed, (e.g by having outside area
equal to infinity which sets $n_0=0$ or by
choosing to work with a manifold where one region
 has an extra boundary where the boundary condition
 is chosen to ensure fixed $ n_0$ with   $N \gg n_0 $), and if  the
 windings and $n(\vec k_\G)$  of the Wilson loops
 are less than $N$, then the  formulae we have already
derived can be  directly applied.

\ifig\fsuex{an example  $su(N)$ integral at finite $N$. }
{\epsfxsize2.5in\epsfbox{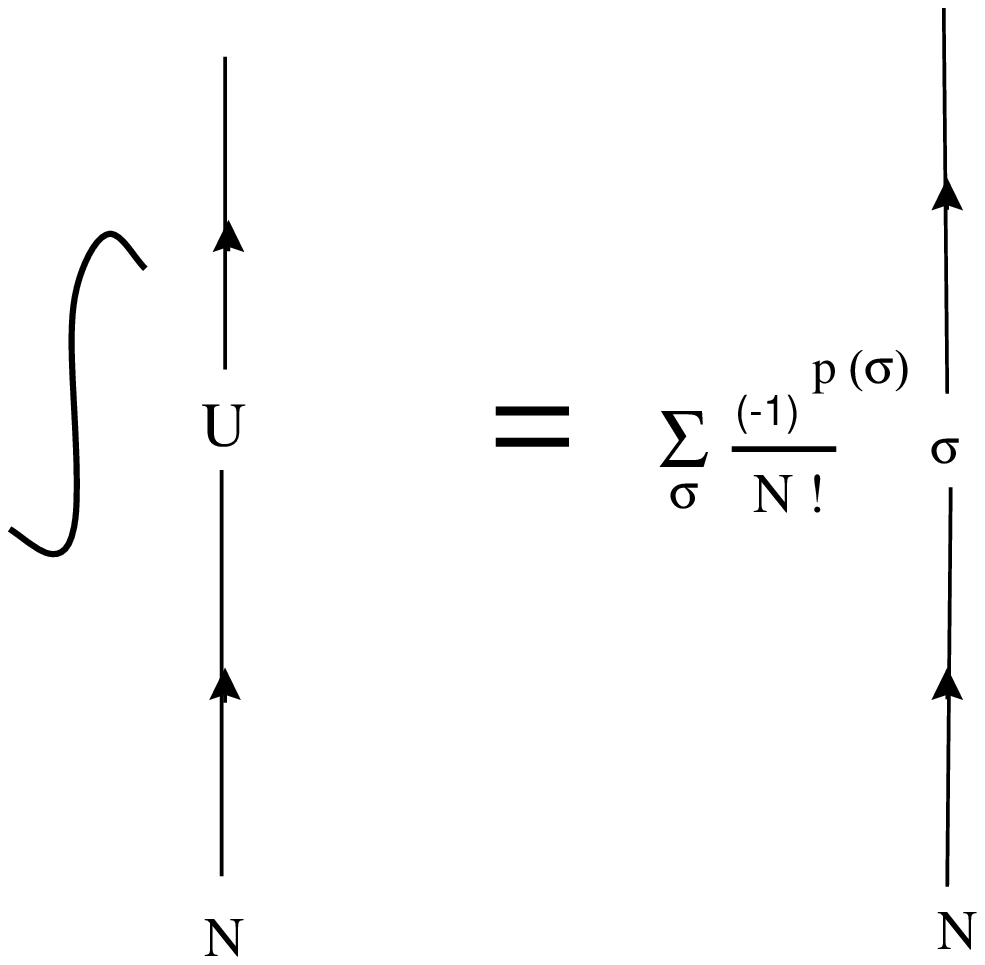}}

For $SU(N)$ at finite $N$, there is another subtlety  in addition to projection
operators.
For $n_1$ and $n_2$ not necessarily small compared to $N$,  the integral
is no longer proportional to $\delta_{n_1, n_2} $ but rather to
 $\delta_{n_1,n_2 {\hbox{mod N}}}$ (see  \DrZu ).
 For example if $n_1=N, n_2=0$ we
can compute the integral by using the fact that the
 integral is the projector onto the identity representation.
 But the  Young diagram  with $1$ column of length $N$ is
isomorphic to the identity representation, so we can use
the formula for
a projector onto a Young diagram $Y$
\eqn\proj { P_Y = d_{r(Y)} \sum_{ \sigma \in S_n }
 {1\over n!} \chi_{r(Y)} ( \sigma^{-1}) \sigma . }
For the relevant $Y$ we have  $d_{r(Y)} = 1$,
 $\chi_{r(Y)} ( \sigma^{-1}) = (-1)^{p(\sigma)} $
 where $p(\sigma) $ is the parity of the  permutation $\sigma$.
This leads to the formula in \fsuex\ where the permutations
summed are in $S_N$, in agreement with eq. 21 of  \Creu .
It should be possible to
 write a general formula for this integral in the case of $SU(N)$,
in terms of a sum over  permutations, for arbitrary $n_1$ and $n_2$.
This might be useful
in the calculation  of correlation functions in  interacting fermion
systems
\refs{\MiPoifs} .

Since the basic structure of the braid diagram is unchanged,
similar proofs to the one we described for the loop
 equations should be applicable at finite $N$.

\ifig\fcontop{ Contraction operators }
{\epsfxsize0.3in\epsfbox{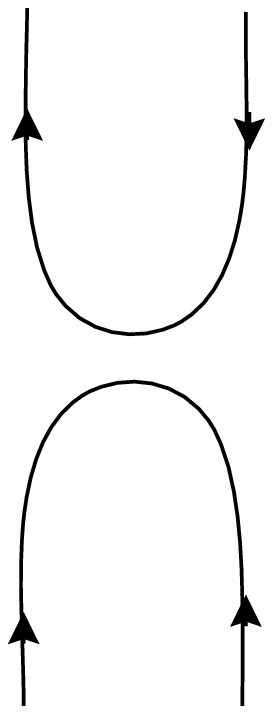}}

\noindent
\subsec{ \bf Non-chiral Theory}

The  method for writing and interpreting  Wilson averages
 developed in this paper should generalise to the non-chiral theory.
So far the main ingredient in the diagram has been
 a sequence of upward directed strands, which are
acted on by permutations. The reason why we have only permutations is that
the commutant of $SU(N)$ in $V^{\otimes n} $ is $S_n$. In the non-chiral theory
we have to deal with $V^{\otimes n }\otimes \bar V^{\otimes m} $.
Now the commutant is not just
$S_n \times S_m$. Rather it includes the dual
pairing of a vector and its dual in the complex conjugate representation.
A generalisation of our construction would
 contain  upward and downward
strands, acted on by permutations and with
 operators like  those in \fcontop .
It is clear that these contraction operators
will be needed:  for example they allow us to
write the coupled Loop functions of \GrTa\ as
  a trace of $U$ acting
 in $V^{ \otimes n} \otimes \bar V^{ \otimes m} $. However the
 combinatorics of these contraction  operators
needs to be related to that of tubes in the
 coupled omega factor in order to solve the
problem of describing the covering space geometry
of Coupled  intersecting
Wilson loops. For arbitrary  non-intersecting Wilson loops the
details of the contraction  operators can be bypassed,
to write a general formula in terms of sums over symmetric groups.
This formula is given and its relation to Euler characters
 of spaces of branched covers is discussed in  \SR\  ( see also \NR ).

\noindent
\subsec{\bf Higher dimensions}

Reformulations of  strong coupling expansions
of higher dimensional lattice gauge theories
in terms of surfaces have been made
\refs{ \ObZub ,  \Kazkos}.
It was observed that the  resulting statistical
 mechanics of surfaces is rather peculiar,
not all surfaces are weighted by positive signs,
 and the
weights are generally rather complicated. One source
 of  minus signs
is the $\Omega_n^{-1}$ factor which appears in
 \fbasint . In the
context of the chiral expansion this leads to the $\Omega_n^{2-2G}$ for
 the genus $G$ partition function
(the chiral Gross Taylor expansion can be
 obtained rather directly from  the heat kernel
 plaquette action for  the standard  cell
 decomposition of the Riemann surface into $1$ two cell,
one $0$-cell and $2G$ 1-cells, and the integral of \fbasint ) .
But the story of \ymt\ shows that all the weights
and signs, at least in the chiral expansion,
 can be understood by thinking about the
 topology  (Euler characters) of the space
of branched covers. So we may naturally
ask  if the weights
 for surfaces in higher dimensional lattice
gauge theory can be understood
in terms of Euler characters of a space of
branched covers.

It is probably best to approach this question
 by using the
heat kernel action \refs{ \menof , \Mig} as the plaquette action,
and try to imitate the kind of expansion
that was done in 2D QCD.   This will involve the
 coupled expansion which comes from
putting together chiral and antichiral in the right way.
Here we will discuss the chiral sector
which gives part of the full large $N$ expansion. A
 delta function  over symmetric groups
can be derived
 for any coupling,  giving the chiral expansion in
terms of symmetric group data,
but the hope of relating to Euler chracters
exists when we can ignore everything but
the $\Omega $ factors:  whether there is
 a sensible physical regime (analogous to the
zero coupling (area)
  limit of \ymt\ ) where this is justified is an
important question
which we will not address here, rather we will proceed
formally
 taking as our starting point the heat kernel action with the
exponent set to zero.
 The techniques for integration which we used in section 7
 can be used in higher dimensions. We make some comments
 on the properties
of the delta functions over symmetric groups  which can be
derived from such an approach.

 For concreteness we have
 in mind the $D$-dimensional Eguchi-Kawai model (in which
 spacetime has the topology of a $D$-torus).
 Our remarks  also apply
to an arbitrary cell complex, where the plaquette action is put on the
two-cells
and group variables are attached to the $1$-cells  ( this rather
 general set-up for relating  lattice gauge theory to topology has been
discussed in a different context by Boulatov \refs{ \Boul}).
 The analog of the Wilson graph
of previous sections is now the $1$-skeleton  ( set of $0$ and $1$-cells ) of
the EK complex, consisting
of  one vertex  and $D$ edges.
By expanding the plaquette actions we have sums over  integers
 $n_{ij}$ for the plaquettes ( $i<j$
running from $1$ to $D$ ).  These sums can be converted to symmetric group data
in the standard way, giving $1$ $\Omega$  factor each.
The actual integral is  over $D$ independent variables, each giving
an $\Omega^{-1}$ factor according to \fbasint .  After doing these
integrals we are left with a trace
in $V^{\otimes \sum_{i,j} 2n_{ij}} $ ( the two comes from the operation of
\fbdrthr ) .
It can be converted to a delta function by inserting an $\Omega$ factor using
\deltatr .
So we have an $\Omega$ factor for each two-cell, an $\Omega^{-1}$ for each
1-cell,
 and an $\Omega$ factor which may be associated with the  0-cell.  So the net
number of $\Omega $ factors is equal
to the Euler character of the 2-skeleton of the EK complex. As we saw for the
partition function
and Wilson loops the 2D analog of this fact was
 one of the important ingredients in showing that
the zero coupling partition function in YM2 computes Euler characters
of a space of branched covers.
For more general complexes we have to contract  the 1-skeleton,
 (as we did for Wilson graphs in previous sections) by using
gauge invariance to one that has a single vertex ( For EK we don't  have to do
this step
since it is already a  ``reduced model'' ).
We have seen that this contraction process does not change the Euler character
of the $1$-skeleton. It also does not change the number of $2$-cells (closed
loop variables
 bounding $2$-cells cannot be gauged away). So it does not
change the Euler character of the $2$-skeleton.
By the same argument as for the EK model then,
the net number of $\Omega$ factors is equal to the
 Euler character of the two-skeleton of the cell complex.
 This motivates the following

\noindent
{ \bf Conjecture :}
The partition function for Lattice gauge theory with
 the zero area heat kernel action for a $D$-dimensional
 cell complex is the generating function for
 orbifold Euler characters of a space of branched
 covers of the $2$ skeleton of the cell complex.

To account for the structure of the  delta functions completely
 in terms of covering spaces, and to define precisely the relevant
class of branched covers appears  rather
subtle:   the  mathematics of  { \it branched} covers of arbitrary
$2$-complexes
 is not well-developed  like that of ordinary surfaces. We may however
expect many techniques
to carry over : this is indicated  for example  by the fact that the chiral
expansion
 for the EK complex (which is orientable) contains only even powers of $N$.
Moreover one wants to explain from the geometrical point of view,
not just the net number of $\Omega$ factors, but also their degrees.
These issues are under investigation.

\bigskip
\centerline{\bf Acknowledgements}
I thank G. Moore for many discussions and for collaboration on section 11.
I  thank S. Cordes, I. Frenkel, M. Henningson, R. Howe,
R. Plesser, R. Puzio, S. Shatashvilli, W. Taylor for  useful discussions.
This work is supported by DOE grant DE-AC02-76ER03075,
 DOE grant DE-FG02-92ER25121.
\bigskip

\vfill\eject
\appendix{ A} { Some properties of Braid Diagrams. }

\ifig\fgenp{ generic operator }
{\epsfxsize2.5in\epsfbox{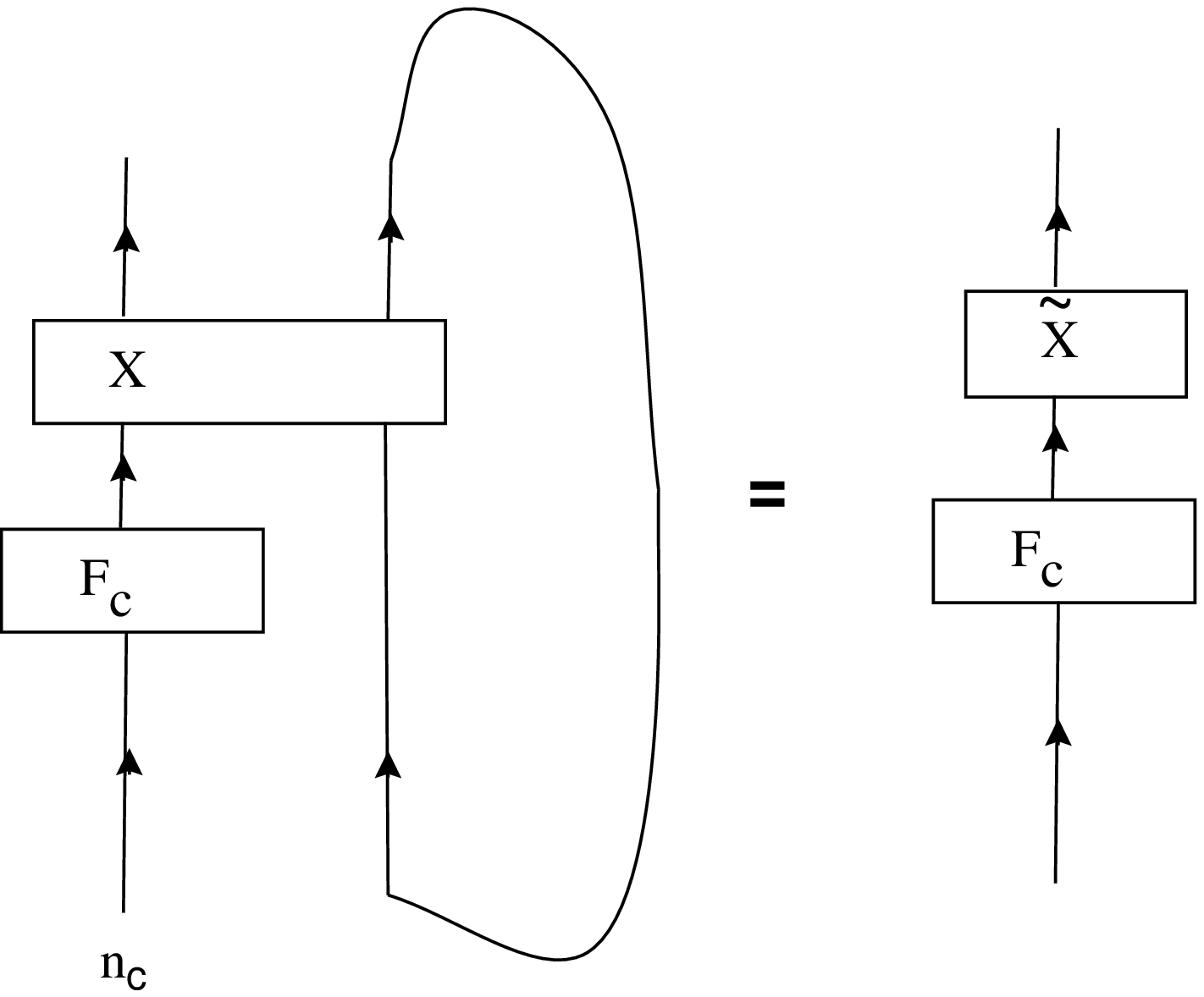}}
We discuss some properties of braid diagrams which are useful in manipulating
them.
Consider the $n_c$
 strands entering  a given $F_c$.
The remaining strands have been lumped,
using the isomorphism shown in \fbdrtwo ,  into the
second strand of multiplicity $l$. The braid diagram
appears as in \fgenp\ where the $X$ operator
 is the sequence of permutations, coming from
other $F$ factors,
their conjugating factors, and  from the windings
 of the Wilson loop. Having traced away the indices for
 the second strand, we are left with an operator on the first strand $  \tilde
X$ .
This operator commutes with the action of $SU(N)$ on $V^{\otimes n_c} $:
\eqn\com{\eqalign{&(1 \otimes tr ) ( U\otimes 1 ) X \cr
& =(1 \otimes tr ) ( 1\otimes U^{-1} )( U\otimes U )  X \cr
& =(1 \otimes tr ) ( 1\otimes U^{-1} )  X ( U\otimes U ) \cr
&= (1 \otimes tr )   X  ( U \otimes 1 ). \cr } }
Here we have used the fact that on
$V^{\otimes ( n_c + l ) } $ the standard action of $SU(N)$
commutes with permutations. In the last line we used cyclicity of the trace.
So $\tilde X$ commutes with $SU(N)$. By Schur-Weyl duality
it must be a permutation.
More intuitively the strands from the first $n_c$ wander
 through the second set of strands and come back with
a permutation. It is often useful in proving properties of
braid diagrams to use these commutativity properties,
and to simplify the braid diagram to make manifest
certain properties which do not use the full bulk of the
braid diagram e.g for loop equations
or for giving evidence,  from this point of view,  for the
 Gross-Taylor rule on the placing of Omega points.

\ifig\fconsumi{ }
{\epsfxsize3.0in\epsfbox{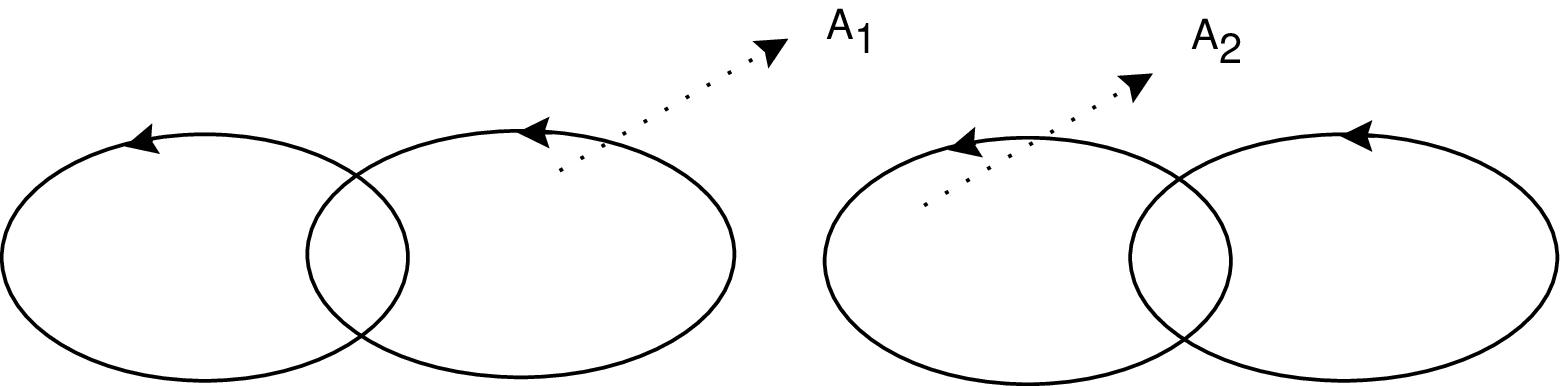}}
\ifig\fconsumii{ connected sum of plane curves }
{\epsfxsize3.0in\epsfbox{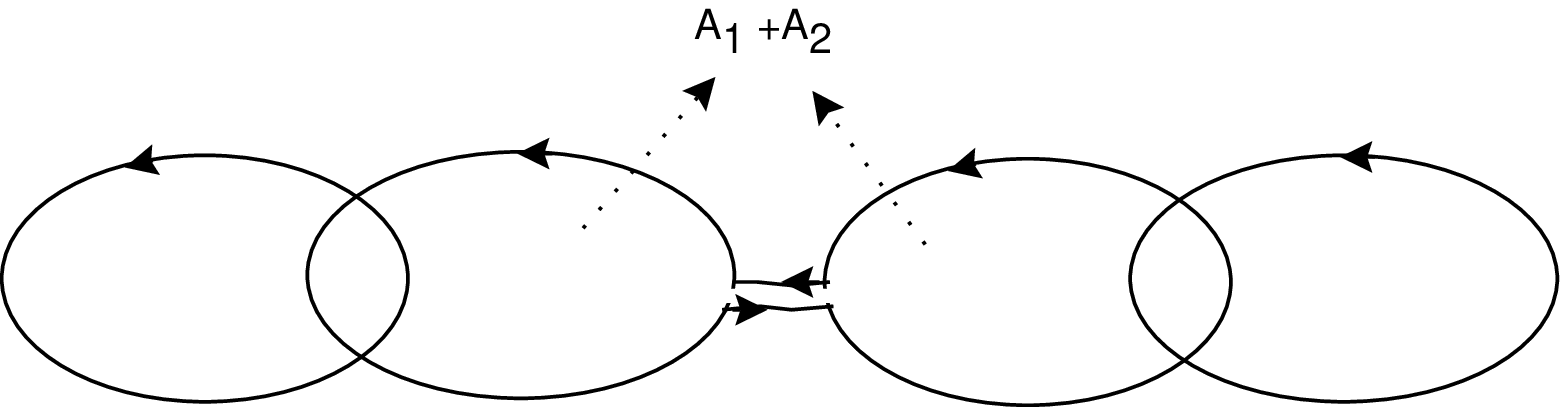}}

Simple operations on braid diagrams are both
natural and useful in expressing properties of Wilson loops.
We can always  combine delta functions for
a Wilson average into $1$
delta function even if the
Wilson graph is disconnected.
Suppose it has $2$ components, as in the example in   \fconsumi\ .
Using the group integrals and the techniques we have described, one derives
sums with $3$ delta functions.

\ifig\fonedel{ Lower braid diagram has structure of connected sum of the  two
on
upper right }
{\epsfxsize4.0in\epsfbox{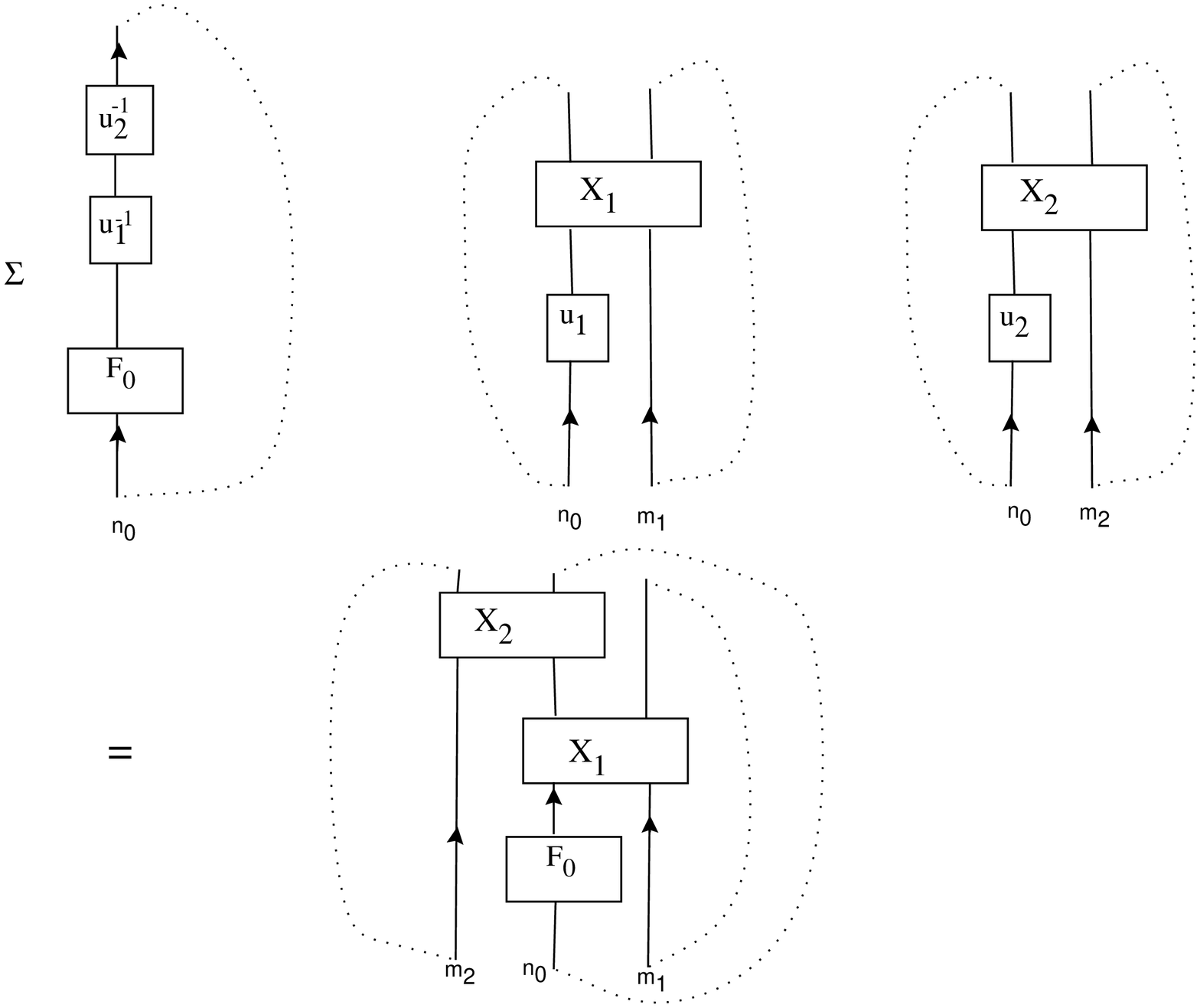}}
The form of the answer is shown in \fonedel .
There are sums over $n_0$ and permutations
$u_1$ and $u_2$ in $S_{n_0}$.
Solving the delta functions gives a single
delta function, which is a {\it connected sum}
of the braid diagrams we would have for the
 two Wilson loops separately.
 In case $n_0$ is zero we have a trivial
concatenation into a single delta function on
of permutations of $m_1 + m_2$ integers,
where the permutations have the property of lying in
 the subgroup $S_{m_1} \times S_{m_2}$.
It is clear from
 the covering space geometry that we should
always be able to write a single delta function, with different
 permutations associated with different paths on the same
manifold. This point is important
in expressing both left and right hand side of
Migdal-Makeenko equations as linear operators on
a single delta function, as we did in section 12.

Connected  sums of braid diagrams  (these are studied
in knot theory
 e.g  \refs { \Kauff} )
  are also useful in
proving relations between different Wilson loops.
For example consider the set of  Wilson loops  in  \fconsumii ,
which is a connected sum of those in \fconsumi\ (here
we are using connected sums
of curves in the sense used for example in \Ar ). Suppose we
are working in the representation basis and  that the
outside area is infinite. Then this  Wilson average of \fconsumii\
 equal to the Wilson average  for \fconsumi\
divided by  dimension
 of the representation  $R$ carried by the middle
circle of \fconsumii .
The $ \dim ~ R$ factor is simply understood from the
 Euler character point of view
by counting the Euler character of the space over
 which the branch points are
allowed to move. This relation between the expectation value of
connected
sums of Wilson loops
and the original Wilson loops (when outside area is infinite)
 is very  general and the proof based on forming connected sums of braid
diagrams can also be applied at finite  $N$.

\listrefs

\bye